\documentclass[graybox,secnum]{svmult}

\usepackage{mathptmx}       % selects Times Roman as basic font
\usepackage{helvet}         % selects Helvetica as sans-serif font
\usepackage{courier}        % selects Courier as typewriter font
\usepackage{type1cm}        % activate if the above 3 fonts are
\usepackage{makeidx}         % allows index generation
\usepackage{graphicx}        % standard LaTeX graphics tool
\usepackage{multicol}        % used for the two-column index
\usepackage[bottom]{footmisc}% places footnotes at page bottom
\usepackage{hyperref}        %for hyperlinks
\usepackage{soul}            % for high-lighting of text
\hypersetup{colorlinks=true,urlcolor=blue}
\usepackage[square,numbers]{natbib}
\usepackage{amsmath}	% Advanced maths commands
\usepackage{amssymb}	% Extra maths symbols
\usepackage{xspace}
\makeindex

\newcommand{\Lo}{L_{\rm UV}}

\newcommand{\Lx}{L_{\rm X}}

\newcommand{\aox}{\alpha_{\rm ox}}

\newcommand{\om}{\Omega_{\rm M}}
\newcommand{\ol}{\Omega_\Lambda}

\newcommand{\msun}{M_\odot}
\DeclareRobustCommand{\ion}[2]{%
\relax\ifmmode
\ifx\testbx\f@series
{\mathbf{#1\,\mathsc{#2}}}\else
{\mathrm{#1\,\mathsc{#2}}}\fi
\else\textup{#1\,{\mdseries\textsc{#2}}}%
\fi}

\newcommand{\chandra}{\textit{Chandra}\xspace}
\newcommand{\xmm}{\textit{XMM-Newton}\xspace}
\newcommand{\funits}{\,\mathrm{erg\,cm^{-2}\,s^{-1}}}
\newcommand{\ang}{\textrm{A\kern -1.3ex\raisebox{0.6ex}{$^\circ$}}}

\newcommand{\EL}[1]{#1}

\makeindex             % used for the subject index
\begin{document}
%\tableofcontents{}
\title*{The dawn of black holes} 
% - The dawn of black holes
\titlerunning{The dawn of black holes}
\author{Lusso Elisabeta\thanks{corresponding author; \email{elisabeta.lusso@unifi.it}}, Valiante Rosa\thanks{corresponding author; \email{rosa.valiante@inaf.it}}  and Vito Fabio\thanks{corresponding author; \email{fabio.vito@inaf.it}} }
\authorrunning{Lusso, Valiante, Vito}
\institute{Lusso Elisabeta %\email{elisabeta.lusso@unifi.it} 
\at Dipartimento di Fisica e Astronomia, Universit\`a di Firenze, via G. Sansone 1, I-50019 Sesto Fiorentino, Firenze, Italy,\\ INAF -- Osservatorio Astrofisico di Arcetri, L.go Enrico Fermi 5, I-50125 Firenze, Italy
\and 
Valiante Rosa \at INAF-Osservatorio Astronomico di Roma, via di Frascati 33, I-00078 Monteporzio Catone, Italy\\
INFN, Sezione di Roma I, P.le Aldo Moro 2, I-00185 Roma, Italy %\email{rosa.valiante@inaf.it}
\and Vito Fabio
%\email{fvito.astro@gmail.com}
\at INAF-Osservatorio di Astrofisica e Scienza dello Spazio, via Gobetti 93/3, 40129, Bologna, Italy}
\maketitle
\abstract{In the last decades, luminous accreting super-massive black holes have been discovered within the first Gyr after the Big Bang, but their origin is still an unsolved mystery. We discuss our state-of-the-art theoretical knowledge of their formation physics and early growth, and describe the results of dedicated observational campaigns in the X-ray band. We also provide an overview of how these systems can be used to derive cosmological parameters. Finally, we point out some open issues, in light of future electro-magnetic and gravitational-wave astronomical facilities.}

\section{Keywords} 
%(max 10 keywords)
Early Universe; Galaxies: active; Galaxies: high-redshift; Quasars: supermassive black holes; Quasars; X-ray astronomy; Cosmology: observations

\section{Introduction} \label{sec:intro}

\begin{figure}
    \includegraphics[width=\columnwidth]{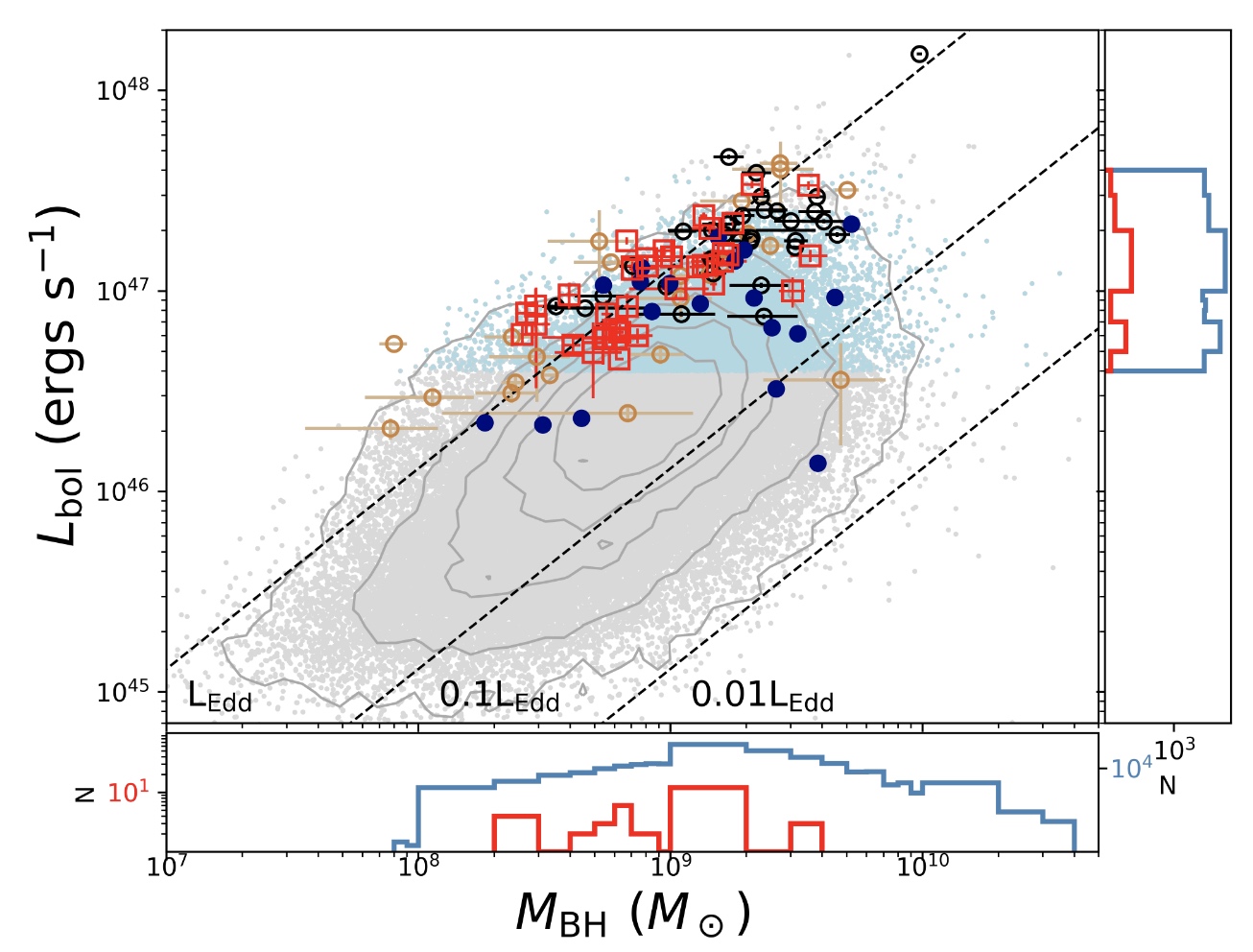}
    \caption{Bolometric luminosities and black hole masses of %known quasars     at $6.3<z\leq 7.64$ \cite[red open squares from][]{yang2021}), and at $z\sim 6$ (the black open from \cite{Schindler2020}, the blue solid from \cite{Shen2019} and light brown open circles from \cite{Willott2003, kurk2007, willott2010, deRosa2011, Mazzucchelli2017}) 
    a compilation of $6.0\lesssim z \lesssim 7.6$ QSOs (red, black, and brown symbols)
    compared with the SDSS lower-redshift ($0.40<z<2.1$) sample (grey dots).%; from \cite{Shen2011}). 
    %All $z\geq 6$ BH masses in this figures are derived from the same virial BH mass estimator from \cite{vestergaardOsmer2009}, based on MgII line observations in near-infrared (NIR) spectra.
    The light blue dots represent a sub-sample from the SDSS lower redshift survey with bolometric luminosity comparable to the highest-redshift sample presented by \cite{yang2021} (red). The right-hand and bottom histograms show the comparison of the luminosity and black hole mass distributions, respectively, in the two aforementioned samples, with the same colour-coding. The figures show that a large number of high-redshift QSOs are located close to (or above) the line of Eddington luminosity (upper dashed line). %suggesting that the SMBHs are accreting close to the Eddington limit.
    Figure is taken from \cite{yang2021}. \textcopyright AAS. Reproduced with permission.}
    \label{fig:quasarSample}
\end{figure}

%\FV{{\bf (Io ristrutturerei l'intro cosi', mixando l'intro di Rosa e il mio).}} \RV{Concordo con Fabio. Ho già inserito e sistemato un pochino la versione ristrutturata e ho spostato la mia precedente in un file separato (deletedSections.tex). Restano da inserire alcune referenze}\\ 
In the last two decades, the availability of wide-field (i.e., $\approx10^3-10^4\,\mathrm{deg^2}$) optical/near-infrared (NIR) surveys, such as the Sloan Digital Sky Survey (SDSS; e.g., \cite{Jiang2016}), the UKIRT Infrared Deep Sky Survey (UKIDSS; e.g., \cite{Mortlock2011}), the Canada-France
High-redshift Quasar Survey (CFHQS; e.g., \cite{willott2010}) and the Panoramic Survey
Telescope \& Rapid Response System 1 (Pan$-$STARRS1; e.g., \cite{Banados2016}), led to the discovery of $>200$ quasars (QSOs) at $z\gtrsim 6$, when the Universe was $\lesssim 900$ Myr old. Currently, eight of these objects have been identified at $z>7$ %\cite[][]{Mortlock2011, Banados2018,  Wang18b, matsuoka2019, Matsuoka2019b,Yang2020}, 
\cite[][]{Mortlock2011, Banados2018,  Wang18b, Matsuoka2019b,Yang2020}, with the redshift record holder recently identified at $z=7.642$, during the reionisation era \cite[][]{wang2021}.  
The relatively shallow coverage provided by wide field optical/NIR surveys implies that only high-redshift QSOs with the strongest rest-frame UV continuum emission can be detected, such that their selection is strongly biased towards rare (about 1 Gpc$^{-3}$ at $z\sim 6$; \cite{ deRosa2014}) and luminous ($ \log\frac{L_{bol}}{L_{\odot}}\gtrsim13$, or $M_{1450}\lesssim-24$, where $M_{1450}$ is the QSO absolute magnitude at rest-frame $1450$\AA; Fig.~\ref{fig:quasarSample}) systems. Only recently the Subaru High-z Exploration of Low-Luminosity Quasars project %(SHELLQs; e.g. \cite{Matsuoka2016, Matsuoka2019b}) 
(SHELLQs; e.g., \cite{Matsuoka2019b}) discovered the first moderate-luminosity (down to $M_{1450}\approx-22$) QSOs at $z\gtrsim6$, which are thus still poorly 
known in terms of basic demographics, birth, and growth. 

Moreover, all of the confirmed $z\gtrsim6$ QSOs are optically classified as type-1 systems, i.e., objects with detectable blue rest-frame UV emission and typically with broad emission lines. For these objects, single-epoch virial black hole (BH) mass estimates can be derived from the Mg II broad emission line (e.g., \cite{Mazzucchelli2017}) and suggest that most of the high-redshift QSOs are powered by already evolved BHs with masses log$\frac{M_{BH}}{M_\odot}\approx9-10$ (e.g., \cite{wu2015}) which are nonetheless still accreting efficiently with Eddington ratios $\lambda_{Edd}=L_{\rm bol}/L_{\rm Edd}\approx0.1-2$ (see Figure \ref{fig:quasarSample}), where $L_{\rm Edd}$ is the Eddington luminosity limit for a BH of mass $M_{\rm BH}$: 
%(i.e. the maximum luminosity that an accreting BH can achieve, as a result of the balance between the radiation pressure and the gravitational force)
\begin{equation}
 L_{\rm Edd}=\frac{4\pi G M_{\rm BH} m_p c}{\sigma_{\rm T}} \simeq 1.26 \times 10^{38} \bigg(\frac{M_{\rm BH}}{\rm M_\odot} \bigg) \rm erg/s = 3.2\times 10^4 \bigg (\frac{M_{\rm BH}}{\rm M_\odot} \bigg) \rm L_\odot,
 \label{eq:EddingtonLum}
\end{equation} 
where $G$ is the gravitational constant, $c$ is the speed of light, $m_p$ is the proton mass and $\sigma_{T}$ is the Thomson scattering cross-section.
%The bolometric luminosity, $L_{\rm bol}$, and SMBH mass of high redshift QSOs are shown in Figure \ref{fig:quasarSample} (taken from the recent work of \cite{yang2021}). %% Specificare da dove sono prese le Lbol? 
To date, the most massive and luminous QSO detected at $z>6$ is \textsc{SDSS J0100+2802}, at $z \sim 6.3$ \cite{wu2015}, which is powered by a BH mass M$_{\rm BH} =(1.2 \pm 0.19) \, \times \, 10^{10} \rm \, M_{\odot}$, and yet still accreting at $\lambda_{\rm Edd}\approx1$, whilst the most distant source is J0313$-$1806, recently detected at $z = 7.64$, with an estimated mass of M$_{\rm BH} = (1.6 \pm 0.4) \times 10^9 \, M_\odot$ and $\lambda_{\rm Edd}=0.67\pm 0.14$ \cite{wang2021}. 

The theoretical implications of the observations of these super massive black holes (SMBHs) as central engine of $z>6$ QSOs are discussed in Sections~\ref{sec:seeds} and \ref{sec:seedGrowth}. X-ray observations of $z>6$ QSOs as well as the physical properties derived from such datasets are discussed in \ref{Sec_selection}. Large statistical QSO samples at redshifts $z\gtrsim6$ are also pivotal standardisable candles to test cosmological models in a poorly explored cosmic epoch. This topic is discussed in Section~\ref{Quasars as cosmological probes}. Section~\ref{sec:unexploredBHs} focuses on the high-redshift BH population which is missed by current astronomical observatories, while Section~\ref{sec:futureProspects} reports the prospects linked with future facilities.

%%%%%%%%%%%%%%%%%%%%%%%%

%textit{\bf Remove or keep the following discussion about overmassive/undermassive BHs at high-z? } \RV{non saprei, ci ragiono un po ma ti lascio carta bianca a riguardo.}\FV{Se non serve per la parte di seeds e early growth, la toglierei!} Using the dynamical mass as a proxy for the stellar mass of the host galaxies, several works found $z>6$ QSOs to be overmassive with respect to the local $M_{BH}-M_{bulge}$ relation (e.g., Pensabene et al. 2020), implying a ``catch-up" scenario during the early galaxy evolution phases, according to which the host galaxies grew subsequently around already formed massive BHs (e.g., Volonteri et al. 201?), as suggested by some theoretical models \textit{(?, link with heavy seeds?)}. However, it is unclear if these findings can be extrapolated to the overall population of QSOs at high redshift, or if they are driven by selection effects, as the faintest ($M_{1450}\approx[-22,-24]$) known $z>6$ QSOs are often found to be less massive (log$\frac{M_{BH}}{M_\odot}\approx8-9$) and to lie on, or sometimes even below, the local $M_{BH}-M_{bulge}$ relation (e.g., Izumi et al. 2019).

\section{The earliest black holes} \label{sec:seeds}
The observations mentioned above place important constraints on the formation and the evolution history of QSOs at high redshift %(i.e. on the seeding and accretion conditions required to form these systems) 
and represent a major challenge for theoretical models 
%\cite[see e.g. the recent reviews by][]{valiante2017, Woods2019, inayoshi2020}.% Volonteri2021}.
(see, e.g., the recent review by \cite{inayoshi2020}).
The main question is how these extremely massive and luminous sources form and evolve during the first billion year of cosmic history ($z>6$), soon after the formation of the first (pre-galactic) structures in the Universe ($z\sim 20-30$).

From a theoretical point of view, a SMBH is the evolutionary product of a less massive progenitor BH, called \textit{seed}, which subsequently grows via accretion of gas from their surroundings (as the gas settles onto a disc-like configuration, the accretion disc, around the BH) and mergers with other BHs. The study of possible formation modes of massive BHs starts way back in time to explain the luminous emission observed in active galactic nuclei (AGN; see, e.g., \cite[][]{Rees1978} and references therein). However, how numerous and massive these seeds must have been (i.e. what are their birth mass function and number density) as well as the physical mechanisms regulating their formation and rapid growth in the early Universe are still debated astrophysical problems. %\cite[see e.g.][for thorough reviews]{volonteri2010, Natarajan2011, Natarajan2014, valiante2017,inayoshi2020}. 
%\cite[see e.g.][for thorough reviews]{volonteri2010, Natarajan2011, VolonteriBellovary12, Natarajan2014, JohnsonHaardt2016, valiante2017,inayoshi2020, Trakhtenbrot2020review}.
%are not known yet
%
%In what follows we review the seed formation scenarios proposed in the literature so far. 
%and refer the reader to recent reviews, such as \cite{LatifFerrara2016, inayoshi2020}, for detailed discussion of the formation mechanisms.

\subsection{Light seed black holes} \label{sec:lightSeeds}
We classify as light seeds the population of BHs with masses ranging from few tens to few hundreds/thousands solar masses, predicted to be the end-products of the first generation of stars, the so-called Population III (Pop~III) stars.%, the first sources of light in the Universe. %marking the transition from the dark ages to the cosmic dawn.

Pop~III stars are expected to form at $z\sim 20-30$ in the first collapsed dark matter structures, the minihalos, with virial temperatures $T_{\rm vir}<10^4$K ($\sim 10^5-10^6~\msun$) where the collapse of primordial gas is driven by molecular hydrogen ($\rm H_2$) cooling %\cite{OmukaiNishi1998, Abel2002, Heger2003, Yoshida2008} and see \cite{BarkanaLoeb2001} for a comprehensive review). 
(see, e.g., \cite{Bromm2013} for a review).
%At such early epochs, the collapse of primordial gas is driven by molecular hydrogen ($\rm H_2$) cooling. 
In the absence of heavy elements (metals) and solid molecules (dust), or if their abundance %(the metallicity and dust-to-gas mass ratio) 
(often indicated by the metallicity, $Z$, the ratio between the mass of metals and gas)
is below a critical threshold (e.g., \cite{Schneider2003}),
%\cite[e.g.][]{Bromm2001, Schneider2002, Schneider2003} 
the cooling process is inefficient and fragmentation, i.e. the decomposition of the gas cloud in smaller agglomerates (fragments) that will become single stars, is limited/prevented. Thus, stars forming under these conditions are expected to be more massive compared to stars forming in chemically enriched environments at later epochs. 
Indeed, as the gas is enriched with metals (produced by the first stars), above a given metallicity threshold ($Z > Z_{\rm cr} = 10^{-3.8}Z_{\odot}$) gas fragmentation is induced and the formation of lower mass stars, in the range $[0.1 - 100]~\msun$ (Population~II/I stars) can start, mediated by efficient metal fine-structure line cooling (e.g., \cite{Omukai2005}, and references therein).
%Indeed, once the gas has been enriched above a given metallicity (and/or dust-to-gas mass ratio) threshold %(with the nucleosynthetic products of the first stars), 
%more efficient gas collapse is driven by metal fine-structure line cooling \cite{Omukai2005, Schneider2012}, leading to the formation of less massive, Population II/I stars in the range $[0.1 - 100] \msun$.

Unfortunately, the initial mass function (IMF) of Pop~III stars is still poorly constrained as a consequence of the lack of observational information of these kind of sources and of the large uncertainties in the Pop~III stars formation process (fragmentation, proto-stellar evolution and stellar radiative feedback; see, e.g., \cite{Greif2015} for a review)
%In particular our current understanding of the role of disc fragmentation, protostellar evolution and stellar radiative feedback, prevent to put solid constraints of the initial mass function (IMF) of Pop~III stars (see e.g. \cite{Greif2015} for a review).

Theoretical studies suggest that the IMF of Pop~III stars is top-heavy (i.e. biased towards more massive stars with respect to the IMF typical of Pop~II stars), with masses from few tens to few hundreds solar masses %\cite{HaimanLoeb2001, MadauRees2001, Heger2003}
(e.g., \cite{Heger2003})
or even up to $1000 \msun$ (e.g., \cite{Hirano2015}), forming as single, binary or multiple systems %\cite[e.g.][]{Sugimura2020, Latif2021a}). 
(e.g., \cite{Sugimura2020}). 

In general, the efficiency of the gas cooling process that determines the fraction of gas that can be converted into stars, and thus the number of Pop~III stars (and remnant BHs) that can form, depends on the dark matter halo virial temperature, $T_{vir}$, on its redshift, on the metallicity of the collapsing gas and on the level of ultraviolet (UV) radiation illuminating the halo (e.g., \cite{Omukai2012}). In metal-poor minihalos, $\rm H_2$ cooling, and thus star formation, may be suppressed or inhibited by the illuminating Lyman Werner (LW) field (photons emitted in the $[11.2–13.6]$ eV band) that can easily dissociate the $\rm H_2$ molecules %\cite[e.g.][]{Haiman1997, Omukai2001})
(e.g., \cite{Omukai2001}), the key ingredient to form stars and the main cooling agent in metal poor environments.
%(e.g. Haiman & Loeb 1997; Haiman, Rees & Loeb 1997; Omukai & Nishi 1999; Machacek, Bryan & Abel 2001; Omukai 2001). 
Even a moderate level of LW flux can lead to an increase of the minimum mass required for dark matter halos to host star formation (see, e.g., \cite{Valiante2016}).

Indirect constraints on the Pop~III IMF can be inferred from stellar archaeology studies (see the review by \cite{frebel2015}). The large carbon-to-iron abundance ratio observed on the surface of Galactic halo stars and the low-metallicity tail of their metallicity distribution function are consistent with models where Pop~III stars form in the range $[10-300]~\msun$, with a characteristic mass of $20~\msun$ \cite{deBennassuti2017} and where the chemical enrichment at early times is dominated by faint supernovae (SNe) explosions \cite{Marassi2014}.
In this mass range, less massive Pop~III stars, in the range $[10-40]~\msun$, explode as SNe, enriching the host galaxy with metals and dust, while stars with initial masses of $[40–140]~\msun$ and $>260~\msun$ directly collapse into BHs. In this latter case, non-rotating stars of primordial composition are expected to leave remnant BHs (our light seeds) as massive as their progenitors, as no mass loss is expected (weak stellar winds) during the stellar evolution (e.g., \cite{Heger2003}).
Finally, in the mass range $[140 - 260]~M_{\odot}$, Pop~III stars explode as Pair Instability SN (PISN) and leave no remnant.

\subsection{Medium-weight seed black holes} \label{sec:mwSeeds}
Medium-weight seeds, namely seed BHs of $\sim 10^2-10^4 \msun$ %\cite[e.g.][and references therein]{Devecchi2012} %Reinoso2020, Sakurai2017}
(e.g., \cite{Reinoso2020}, and references therein)
can form in the core of ultra-dense nuclear clusters (of $\sim 10^5 \msun$) %originating in metal-poor protogalaxies, 
via dynamical interactions (runaway collision) of \textit{(i)} stars, leading to the formation of a very massive star, $>260~\msun$, eventually collapsing into a BH %\cite[e.g.][]{BegelmanRees1978, Portegies2002, Omukai2008, Devecchi2009, katz2015, Reinoso2018, Das2021a}, 
%\cite[e.g.][]{BegelmanRees1978, Omukai2008, Devecchi2009, katz2015}, 
(e.g., \cite{Omukai2008}), or of \textit{(ii)} stellar-mass BHs embedded in dense gas clouds %\cite[e,g,][]{Davies2011, Lupi2014}. 
(e.g., \cite{Lupi2014}). 
%Collisional processes were indeed mentioned as an important pathway in the seminal paper by Rees (1984), and subsequently taken into account for instance in semianalytical models by Devecchi et al. (2010, 2012) and Lupi et al. (2014). In N-body simulations employing cosmological initial conditions, Katz et al. (2015) and Sakurai et al. (2017) have shown that black holes with masses of ∼103 M can be formed.
%
%As mentioned in the previous Section, the formation of massive objects is favoured in metal-poor gas clouds, where the cooling process (relying on $\rm H_2$ molecules) is inefficient \cite[e.g.][]{Omukai2005}).
%In particular, the formation of dense/compact nuclear star clusters is possible in metal-poor ($Z<Z_{cr}$) atomic cooling halos, i.e. structures with virial temperatures $T_{vir} \geq 10^4$ K (more massive than minihalos, $\geq 10^7-10^8 \msun$, thus containing a larger gas reservoir) in which the gas is able to cool and collapse thanks to atomic hydrogen.

Numerical simulations suggest that in primordial/metal-poor environments ($Z<Z_{cr}$) fragmentation occurs when the gas density reaches high values ($\geq 10^{9} \rm cm^{-3}$ (e.g., \cite{Latif2013b}) %Greif2011}
), enabling the formation of dense, compact, cluster (with radii $\leq 0.1 \rm pc$).
In mildly enriched halos, and in particular in the presence of small amounts of dust grains (above a critical dust-to-gas mass ratio threshold $\mathcal{D}>\mathcal{D}_{cr}\sim 4.4\cdot 10^{-9}$; \cite{Schneider2012}), dust cooling may drive gas fragmentation at even higher densities (i.e. in the late phases of the collapse; e.g., \cite{Omukai2008}), providing ideal conditions for a very dense cluster to form.
%Early enrichment, with small traces of metals and dust, may be a consequence of prior Pop~III star formation \cite[e.g.][]{Schneider2006, Clark2008}.
%Early enrichment, with small traces of metals and dust, may be triggered by the explosion of the first SNe, i.e. metal-free Pop~III stars with initial mass in the range $[10-40] \msun$ and $[140-260] \msun$)  \cite[e.g.][]{Schneider2006, Clark2008}.

The conditions mentioned above are usually met in atomic cooling halos, i.e. structures with virial temperatures $T_{vir} \geq 10^4$ K ($\geq 10^7-10^8 \msun$, more massive than minihalos and thus containing a larger gas reservoir) in which efficient gas cooling is activated by atomic hydrogen (often referred to as Ly$\alpha$ cooling).
%the gas is able to cool and collapse thanks to atomic hydrogen. 
In the standard structure formation paradigm, more massive systems likely form %at later epochs 
from the growth (via accretion and mergers) of less massive ones (i.e. the minihalos) that collapsed at earlier epochs and that may have experienced prior star formation. Thus, an early enrichment of their gas, with small traces of metals and dust, may be triggered by the explosion of the first SNe, i.e. Pop~III stars with initial mass in the range $[10-40]~ \msun$ and $[140-260]~ \msun$  (e.g., \cite{Schneider2006})
%\cite[e.g.][]{Schneider2006, Clark2008} 
formed in their progenitors.
%In these systems, early enrichment, with small traces of metals and dust, may be triggered by the explosion of the first SNe, i.e. metal-free Pop~III stars with initial mass in the range $[10-40] \msun$ and $[140-260] \msun$)  \cite[e.g.][]{Schneider2006, Clark2008}.
%Evolution of stellar systems via dynamical interactions in these dense environments leads to the formation of BHs \cite[e.g.][]{BegelmanRees1978, Portegies2002}.
%
%In a recent study, \cite{Fragione2021} show that medium-weight seeds can result from repeated mergers of smaller stellar-mass BHs, with larger final masses (merger products) forming in lower metallicity clusters.

As alternative possibilities, stellar-mass BHs in dense nuclear star clusters may efficiently/rapidly grow up to $10^4-10^5$ via supra-exponential accretion of the interstellar gas (e.g., \cite{Natarajan2020}) or via tidal capture and tidal disruption of stars
%\cite[e.g.][]{Stone2017, Sakurai2019}.
(e.g., \cite{Sakurai2019}).

Finally, recent simulations (e.g., \cite{Latif2021b}) show that seeds of $\sim10^3 \msun$ may be the remnants of Pop~III stars of $(1800-2800)~\msun$. Such very massive stars form in halos illuminated by a uniform LW background flux of $(100-500)\times 10^{-21}\rm erg/s/Hz/cm^2/sr$ 
\footnote{The intensity of the LW flux is often indicated in units of $J_{21}=10^{-21}\rm erg/s/Hz/cm^2/sr$.} that is turned on at $z=30$.
%Finally, in a recent simulation \cite{Latif2021b} suggest that medium-weight seeds of few $10^3 \msun$ can results as remnants of very massive Pop~III stars of $(1800-2800)~\msun$, forming (together with a few normal Pop~III stars) in halos exposed to a uniform LW background flux of $(100-500)\times 10^{-21}\rm erg/s/Hz/cm^2/sr$ \footnote{The intensity of the LW flux is often indicated in units of $J_{21}=10^{-21}\rm erg/s/Hz/cm^2/sr$.}, turned on at $z=30$. 
In this scenario, such a level of LW radiation enables minihalos to grow in mass reaching the mass of atomic cooling halos (a few $10^6-10^7 \msun$) and virial temperatures $T_{vir}\sim 10^4$ K, before the collapse but without completely photodissociating their $\rm H_2$ content (shielded in the halo core).
%and self-shielding avoids the complete suppression of their $\rm H_2$ content (in the halo core).

\subsection{Heavy seed black holes}\label{sec:heavySeeds}
Heavy seed BHs up to $10^5 - 10^6 \msun$ (e.g., \cite{LodatoNatarajan2007, Ferrara2014}) are expected to result from the rapid collapse of a massive gas cloud mediated by the formation of a supermassive star (SMS) of $\sim 10^{4-6} \msun$ that directly collapses into a BH of similar mass. This channel is often referred to as direct collapse BH (DCBH) scenario.
The DCBH formation mechanism, under different conditions, has been widely explored in the literature (see, e.g., the recent review by \cite{inayoshi2020}). %via both analytic methods \cite[e.g.][]{Omukai2001, BrommLoeb2003, VolonteriRees2005, Begelman2006, Lodato2006, Ferrara2014} and dedicated high-resolution simulations \cite[e.g.][]{Wise2008, Regan2009, Hosokawa2012, Latif2013b, Inayoshi2014, Regan2014, Chon2016, becerra2018, Wise2019, Maio2019}.

Similarly to dense nuclear star clusters, the ideal birth places for a rapidly accreting SMS are atomic cooling halos
where gas cooling is suppressed (the gas remains at temperatures close to $\sim 10^4$ K) and fragmentation is prevented so that gas can undergo rapid collapse, at high rates (up to $1 \msun/ \rm yr$; %e.g. \cite{Wise2008, Regan2009})
\cite{Regan2009}), a key requirement for the formation of a massive seed.
%Similarly to dense nuclear star clusters, the ideal birth places for a rapidly accreting SMS are atomic cooling halos at virial tempratures $T_{vir} \sim 10^4$ K  where gas cooling is activated by atomic hydrogen (often referred to as Lyman$\alpha$ cooling) that drives the rapid collapse of gas at high rates (up to $1 \msun/ \rm yr$; e.g. \cite{Wise2008, Regan2009}), a key requirement for the formation of a massive seed.
%In most models, the key requirement for a heavy seed to form is a rapid infall of gas towards the central regions of the collapsing gas cloud, promoted by gravitational instabilities. 
For the gas to collapse monolithically into a single massive object (without fragmenting into smaller clumps and/or forming a dense stellar cluster), both metal and dust cooling must not operate. In other words, the gas must be sufficiently metal poor (e.g., \cite{Regan2020}) and dust poor (see Section~\ref{sec:mwSeeds}).

In this scenario, if the newly formed (proto-)star accretes gas at high rates, $0.01-10 \msun/ \rm yr$ %\cite[e.g.][]{OmukaiPalla2003, Haemmerle2018},
(e.g., \cite{OmukaiPalla2003}), negative radiative feedback (ionising radiation from the evolving star itself) is not able to contrast/limit its growth and, eventually, the SMS forms and promptly collapses into a BH via general relativistic instabilities (e.g., \cite{Sakurai2016}). %(but see also \cite{Chon2020}). 
%Eventually, the SMS form and promptly collapses into a BH via general relativistic instabilities %%\cite[e.g.][]{Haemmerle2018, Hosokawa2013}. 
%\cite[e.g.][]{Hosokawa2013}. 
%%(Hosokawa et al. 2013; Umeda et al. 2016; Woods et al. 2017; Haemmerlé et al. 2018a,b).

Recently, \cite{Chon2020} have shown that the formation of a heavy seed may be enabled even in halos where fragmentation takes place, provided that the environment is only slightly enriched ($Z<10^{-3} \, \rm Z_\odot$). The infalling, metal-poor, material preferentially feeds the first star that forms in the cloud (primary proto-star) that can still grow supermassive. They refer to this as the super competitive accretion (SCA) scenario. 

In order for the gas in atomic cooling halos to be pristine (metal free) or metal poor, external pollution (contamination with metals/dust produced in nearby star forming galaxies) must be limited and genetic (in situ) enrichment (i.e. stellar production of metals and dust within the halo) must be avoided, preventing star formation before %within its lower-mass progenitors (the minihalos) 
and after the halo reaches the atomic cooling stage.

As mentioned in the previous sections, photodissociating feedback (i.e. the suppression of $\rm H_2$ cooling and star formation due to LW photons) determines the efficiency of star formation in minihalos. Even a modest LW radiation field ($J_{\rm LW} \sim 1-100 J_{21}$) can rapidly suppress the cooling efficiency in minihalos, unless their gas is already enriched to $Z\gtrsim 0.1 \, Z_\odot$ (e.g., \cite{Valiante2016, deBennassuti2017}).
%
%Conversely, a strong LW flux ($>10^{2-3} J_{21}$) is required to photo-dissociate $\rm H_2$ molecules and inhibit star formation in atomic cooling halos, especially when the effect of photodissociating feedback is reduced by $\rm H_2$ self-shielding \cite[e.g.][]{Hartwig2015}, 
%%so that Pop~III star formation may still occur under moderate LW radiation, $J_{\rm LW}\sim 100-500 J_{21}$ (see e.g. \cite{LatifVolonteri2015} and the recent simulation by \cite{Latif2021b}, mentioned in Section~\ref{sec:mwSeeds})
%and/or by an external X-ray ionising field (that increases the fraction of free electrons promoting $\rm H_2$ formation; e.g. \cite[][]{Inayoshi2011, Johnson2014}.
Alternatively, star formation in minihalos may be efficiently suppressed via the dynamical heating of the gas, during the rapid mass growth of the halos (via mergers) in high-redshift overdense regions \cite{Wise2019}, or under the influence of baryonic streaming motions at high velocities (e.g., \cite{Schauer2017}).

Conversely, a strong LW flux ($>10^{2-3} J_{21}$) is required to photo-dissociate $\rm H_2$ molecules and inhibit star formation in atomic cooling halos. To date, the critical level of LW radiation, $J_{crit}$, above which $\rm H_2$ can be completely destroyed is still uncertain, %(see e.g. the discussion in the review by \cite[][]{valiante2017}), 
ranging from few tens to few $10^4-10^5 \, J_{21}$
%where $J_{21} = 10^{−21} \rm erg s^{−1} cm^{−2} Hz^{−1} sr^{−1}$$
%\cite[e.g.][]{Sugimura2014, Latif2015} 
(e.g., \cite{Latif2015}) depending on the fraction of $H_2$ present in the halo and on the spectral energy distribution of the emitting source %\cite[e.g. PopIII vs PopII stellar populations][]{Omukai2001, BrommLoeb2003, Omukai2008, Agarwal2012, Agarwal2015, Sugimura2014}
%\cite[e.g. PopIII vs PopII stellar populations][]{Omukai2001, Omukai2008, Sugimura2014, Agarwal2015}. 
(e.g., \cite{Omukai2001, Omukai2008}).
The required value of $J_{crit}$ is higher when $\rm H_2$ self-shielding (e.g., \cite{Hartwig2015}) and/or an external X-ray ionising field, that increases the fraction of free electrons promoting the formation of $\rm H_2$ molecules 
%\cite[e.g.][]{Inayoshi2011, Johnson2014}, 
(e.g., \cite{Inayoshi2011}), reduce the effect of photodissociation feedback.
%
%In addition, the value of $J_{crit}$ is higher when $\rm H_2$ self-shielding \cite[e.g.][]{Hartwig2015}%, so that Pop~III star formation may still occur under moderate flux levels, $J_{\rm LW}\sim 100-500 J_{21}$ (see e.g. \cite{LatifVolonteri2015} and the recent simulation by \cite{Latif2021b}, mentioned in Section~\ref{sec:mwSeeds}) 
%and/or an external X-ray ionising field, that increases the fraction of free electrons promoting the formation of $\rm H_2$ molecules \cite[e.g.][]{Inayoshi2011, Johnson2014}, reduce the effect of photodissociation feedback. %Under these conditions a higher intensity of the photo-dissociating flux is required to set the conditions for heavy seeds formation.

Several models suggest that atomic cooling halos are commonly illuminated by LW flux with intensities $>10^3 J_{21}$ if the source of radiation is a nearby galaxy, typically another atomic cooling halo located within $\geq 1$ kpc, that started to form stars only few Myr earlier (the so-called ``synchronised halo pair" scenario; e.g., %\cite{Dijkstra2008, Agarwal2012, Dijkstra2014, Visbal2014c, Chon2016, Lupi2021}).
\cite{Visbal2014c} and references therein).
%LW flux intensities $>10^3 J_{21}$ are commonly reached if an atomic cooling halo is illuminated by a nearby (within $\geq 1$ kpc) star forming galaxy, typically another atomic cooling halo, that started to form stars only few Myr earlier (the so-called "synchronised halo pair" scenario; e.g. \cite{Dijkstra2008, Agarwal2012, Dijkstra2014, Chon2016, Lupi2021}.

Alternatively, the formation of massive BHs may be aided by merger-driven direct collapse of gas, triggered by mergers of massive, gas-rich galaxies at $z\sim 8-10$, even in metal-enriched gas (up to to solar composition $Z=Z_\odot$). As a result, the gas rapidly infalls towards the nuclear regions (at rates $>10^3-10^4 \msun/ \rm yr$), leading to the formation of massive BHs of $10^5-10^6 \msun$ and up to $\sim 10^8 \msun$, with or without going through the SMS stage 
%\cite{Mayer2010, Mayer2015, MayerBonoli2019}.
(e.g., \cite{MayerBonoli2019}, and references therein).
%either from the direct collapse of a SMS or via the radial general-relativistic instability of a supermassive protostellar precursor.

As a consequence of the tight birth environmental conditions required for their formation, heavy seeds are predicted to be rare 
%\cite{Valiante2016, Dijkstra2014, Habouzit2016c}.
\cite{Valiante2016, Habouzit2016c}.
However, their formation mechanism is still poorly understood, thus, the number density of heavy seeds is still uncertain %\cite[see e.g.][and references therein]{valiante2017, Habouzit2016c}.
(e.g., \cite{Habouzit2016c}, and references therein).

\subsection{Primordial black holes and exotic candidates} \label{sec:pbh}
Along with the astrophysical BH seeds introduced above, more exotic formation channels have also been considered %as possible of the earliest SMBHs
such as ``dark stars", i.e. stellar objects composed of H and He, similar to Pop~III stars but powered by dark matter annihilation %(rather than by fusion) 
(see the review by \cite{Freese2016})
and BHs of $\sim 10^4 \msun$ (comparable to the mass of medium-weight seeds) formed from dissipative dark matter during the first structure formation epoch (e.g., \cite{Damico2018}).

In particular, primordial black holes (PBHs), originally proposed as dark matter particle candidates (e.g., \cite{Hawking1971}), %(Zel’dovich & Novikov 1967), 
are now becoming popular as possible early BH seeds %\cite[e.g.][]{Cappelluti2021}.
Dark matter PBHs are predicted to emerge from large primordial density fluctuations (the same overdensities that lead to the formation of galaxies) collapsing onto themselves soon after the Big Bang, probably before recombination, during the radiation dominated era (see, e.g., the review by \cite{Carr2005}) and are expected to cover a wide mass range, from $10^{-10} \msun$ up to $10^{7} \msun$ %\cite[][]{khlopov2005, GarciaBellido2019, Carr2020}. 
%\cite[][]{GarciaBellido2019, Carr2020}.
\cite[][]{GarciaBellido2019}. 

Finally, gas heating %via ambipolar diffusion,
induced by primordial magnetic fields (of the order of nano Gauss) has also been proposed as an alternative mechanism to suppress $\rm H_2$ cooling  (and thus fragmentation), maintaining the gas at temperatures of $\sim 10^4$ K and fostering the formation of massive seeds %\cite[][]{Schleicher2009, Sethi2010}.
\cite[][]{Schleicher2009}.

All these alternative/exotic scenarios still require further studies to asses their role as possible pathways to $z>6$ SMBHs.

\section{From seeds to SMBHs} \label{sec:seedGrowth} 
%\subsection{Growing the seeds}

Many attempts have been done so far to develop accurate theoretical models, based on different techniques, to study how the first SMBHs formed in the early Universe. %(see e.g. \cite{valiante2017} for a review).  
Usually, a semi-analytical modelling approach is employed to investigate the statistical properties of BH populations, whilst cosmological, hydrodynamical, simulations %of early galaxy formation. 
are best suited to describe the properties of large-scale structure formation.

Semi-analytical models (SAMs) describe the physical processes regulating galaxies' evolution (e.g., star formation, BH growth, chemical enrichment of the ISM, stellar/BH feedback) by means of a set of analytic equations (called prescriptions), linked to the hierarchical assembly history of structure formation.
In this approach, the hierarchical assembly, trough cosmic times, of massive halos hosting QSOs (the so-called merger trees, or merger history) is reconstructed either analytically (e.g., using Monte Carlo algorithms) on the basis of formalism describing the halo mass function at different times (e.g., \cite{Parkinson2008} and references therein) or by means 
of N$-$body, dark matter only, cosmological simulations of large volumes of the Universe ($>100 \, \rm Mpc^{3}$ comoving; e.g., \cite{Springel2005Nature}).

%SAMs combined with analytical merger trees, are best suited to perform statistical studies (exploring a large number of merger histories of the same halo) simultaneously resolving the low-mass end of the halo mass function (the mini-halos) over relatively short computational timescales. However, this technique does not include the information on the halo clustering properties. 

%Conversely, SAMs applied to dark-matter simulation based merger trees provide information on halo properties also tracking their spatial distributions. This, however comes at the expenses of the statistical information and computing costs.

Hydrodynamical simulations have the advantage of capturing the complex connection between the process of BH growth/feedback and its host galaxy evolution (see Chapter 3).
In this approach, dark matter, gas and stars are represented by particles, or grid cells, and their evolution is described by solving the relevant equations of gravity, hydrodynamics and thermodynamics. %to track the interplay between baryons and dark matter and to predict the resulting properties of galaxies (and of large scale structures in general).

To produce a statistically significant number of massive halos ($10^{12}-10^{13} \, \rm M_\odot$) hosting the rare $z>6$ QSOs ($1~\rm Gpc^{-3}$ \cite{deRosa2014}) via N$-$body or hydrodynamical techniques, it is necessary to simulate very large scales/volumes. %, that are computationally more expensive than the semi-analytical Monte Carlo method. 

However, due to the complexity of the approach, hydrodynamical simulation are computationally expensive and the physical processes occurring on different scales (from small to large) can not be tracked at the same time within a simulation (see, e.g., \cite{Habouzit2016c, Volonteri2020}). 
Small-scale simulations (box size of $\sim 1$ Mpc comoving) with high dark matter resolution (i.e. resolving the smallest halos, $\sim 10^3 \, \rm M_\odot$) are best suited to study local processes in detail and the environment in which seed BHs form (e.g., the conditions for heavy seed formation). 
Larger simulations (e.g., box size of $\sim 10$ Mpc comoving) resolving intermediate dark matter halos ($\sim 10^7 \, \rm M_\odot)$ instead allow to study the seed number density in a significant/representative, average, volume of the Universe and the effect of stellar/AGN-driven feedback on their occurrence. Finally, to track the cosmological growth of the seeds, as possible progenitors of high-redshift QSOs, across cosmic epochs we need to perform simulation with a very larger box size ($>$100 Mpc comoving), at the cost of the mass resolution that is lower in this case ($\sim 10^8 \, \rm M_\odot$ must be employed).

%Simulations tracking galaxy collisions and the formation of large-scales structures across cosmic epochs are usually referred as cosmological simulations.

With respect to SAMs, cosmological simulations of early galaxy formation provide a much more detailed description of the properties of large-scale structures. However, the dynamical range described is not adequate to allow global statistical predictions, simultaneously resolving lower-mass structures (minihalos), at the same level as SAMs. 
%In addition, due to their high computational costs, and depending on the box size, they must employ sub-grid modelling to describe the physical processes occurring on scales smaller than the their resolved physical scale (e.g. for gas cooling, star formation, BH formation, accretion and dynamics).

The main issue in the assembly of early QSOs is the timescale over which the seed BHs are expected to form and grow at early cosmic epochs. At $z\sim 6$ the available cosmic time is relatively short, less than $\sim 1$ Gyr (about the age of the Universe at $z=6$), and even shorter in the case of $z>7$ systems ($\lesssim 800$ Myr). 
In general, a BH can increase its mass via accretion of material infalling %towards it 
%its surface 
from a surrounding disc 
%\cite[e.g.][]{ShakuraSunyaev1976, Jiang2014}
(e.g., \cite{ShakuraSunyaev1976}) or from random directions (chaotic cold accretion, %e.g. \cite{Gaspari2013, Gaspari2015}).
e.g., \cite{Gaspari2013}).
A light seed origin was the first, natural, hypothesis for seeding SMBHs at high redshifts, as they are the remnants of rapidly evolving, massive stars expected to form at cosmic dawn (as discussed in Section~\ref{sec:lightSeeds}).

One possibility to grow SMBHs by $z\sim 6$ is that at least a small fraction of light seeds were able to grow %continuously 
almost uninterruptedly near or at the Eddington rate for a relatively long period of time %(e.g. Shapiro 2005 \cite{Pelupessy2007, TanakaHaiman2009}
(e.g., \cite{TanakaHaiman2009}).
%(\cite[e.g.][]{Alvarez2009, Tanaka2014}). 
Such a rate is defined as: 
\begin{equation}\label{eq:EddRate}
\dot{M}_{\rm Edd}=\frac{L_{\rm Edd}}{\epsilon c^2}%=\frac{4\pi G M_{\rm BH} m_p}{\sigma_{\rm T} \epsilon c} \label{eq:EddRate}
\end{equation}
%$\dot{M}_{\rm Edd}=L_{\rm Edd}/ \epsilon c^2$, where $c$ is the speed of light and $\epsilon$ is the radiative efficiency, i.e. the fraction of the rest-mass energy that is radiated away as light during the process.
where $L_{\rm Edd}$ is the Eddington luminosity defined in Eq.~\ref{eq:EddingtonLum} and $\epsilon$ is the radiative efficiency, i.e. the fraction of the rest-mass energy that is radiated away as light during the accretion process.

Assuming that the BH can accrete a fraction $1-\epsilon$ of the infalling material, %infalling towards its surface, 
the BH mass evolution with time ($t$) can be described as an exponential growth: 
\begin{equation}
    M_{\rm BH}=M_{seed} e^{f_{duty}(\frac{1-\epsilon}{\epsilon})(\frac{t}{t_{\rm Edd}})}
    \label{eq:exponentialgrowth}
\end{equation}
where $f_{duty}$ is the accretion duty cycle, i.e. the fraction of cosmic time over which the seed is actively accreting and $t_{\rm Edd}=M_{\rm BH}c^2/L_{\rm Edd}=0.45$ Gyr is the Eddington time.
Note that the timescale in Eq.~\ref{eq:exponentialgrowth} can be also expressed via the $e-$folding (or Salpeter) time,  $t_{e}=\frac{\epsilon}{(1-\epsilon)}\frac{t_{Edd}}{f_{duty}}$, so that $M_{\rm BH}=M_{seed}e^{(t/t_e)}$. 

\begin{figure}
    \includegraphics[width=\columnwidth]{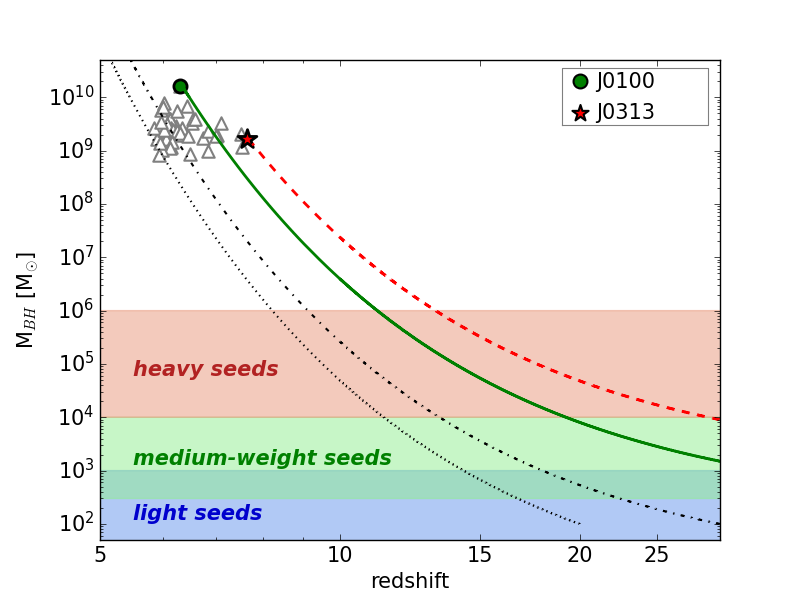}
    \caption{Black hole growth tracks for a sample of $z>6$ QSOs. %{\bf (add ref. to the sample)}. 
    To prepare this figure, BH masses have been consistently re-calculated using the same virial estimator, based the observed MgII 2798 $\AA$ emission line (mass calibration from \cite{ShenLiu2012}). The dotted and dot-dashed lines show the time evolution of a light seed of $100 \, \rm M_\odot$ forming at $z=20$ and $z=30$, respectively. The remaining curves instead give the mass of the seed required to grow the final observed SMBH in J0100 (green solid) and J0313 (red dashed), in less than $\sim 700-800$ Myr. In this figure each SMBH grows from a single seed that accretes gas at the Eddington rate ($\dot{M}=\dot{M}_{\rm Edd}$) at all times.%: the BH evolution with time, $t$, is described by the exponential growth $M_{\rm BH}=M_{seed} e^{(\frac{1-\epsilon}{\epsilon})(\frac{t}{t_{\rm Edd}})}$ with an e-folding time $t_{\rm Edd}=50$ Myr and assuming a radiative efficiency $\epsilon=0.1$ (i.e. the BH accretes is a fraction $1-\epsilon$ of the infalling material). 
    The three shaded areas approximately indicate the mass range of the different seed populations: \textit{light seeds} (blue),  \textit{medium-weight seeds} (green) and \textit{heavy seeds} (red).}
    \label{fig:seedgrowth}
\end{figure}

However, according to Eq.~\ref{eq:exponentialgrowth}, assuming an uninterrupted accretion ($f_{duty}=1$) at the Eddington rate and a radiative efficiency $\epsilon=0.1$ ($10\%$ of the material falling towards the BH is converted into radiation,  as described by the standard, radiatively efficient, thin-disc accretion model; e.g., \cite{ShakuraSunyaev1973}), a light seed of $100 \, \rm M_\odot$ barely reaches a mass of $10^9 \msun$ by $z=6 (5)$ if it forms at $z=30 (20)$ and can not explain the formation of the most massive QSOs at $z>6$ (see Figure~\ref{fig:seedgrowth}). 

A growth timescale, $t_{growth}=\frac{t_{\rm Edd \epsilon}}{(1-\epsilon) f_{duty}} ln(\frac{M_{\rm BH}}{M_{seed}})$ Gyr, of almost 1 Gyr would be required to grow a $> 10^9 \msun$ SMBH by $z=6-7$ from a $100 (10) \, \rm M_\odot$ seed, a timescale comparable to (longer than) the age of the Universe at $z\sim 6-7$ %(e.g. \cite{HaimanLoeb2001, Volonteri2003}).
(e.g., \cite{HaimanLoeb2001}).

At $z\geq 7$ the mass/timescale constraint is even more stringent. Provided they accrete gas at the Eddington limit, light seeds ($<10^3 \msun$) have not enough time to grow supermassive (see Figure \ref{fig:seedgrowth}). The seed should be already massive, $\sim 10^4 (10^5) \, \rm M_\odot$ at $z=30(20)$ in order to grow, via continuous gas accretion at the Eddington rate, $\dot{M}_{\rm Edd}$, up to billion solar masses by $z>7$ (i.e. in few hundreds Myr), as shown  Figure \ref{fig:seedgrowth}. However, even in the case of massive seeds, a long-term accretion of gas at the Eddington rate may be limited by the available gas reservoir and by the strong radiative feedback from the BH accretion process itself, that may halt the growth (e.g., \cite{Chon2021} and references therein).

Different semi-analytical and numerical studies show that it is very difficult to sustain Eddington accretion uninterruptedly over the entire seed BH evolution history as the process is limited by the available steady, low-angular momentum, gas around the BH and by the effects of radiative (ionising) and mechanical feedback %(e.g. \cite[][]{Whalen2008, Alvarez2009, Milosavljevic2009a, Pacucci2015, Park2016, Pacucci2017}
%\cite[e.g.][]{Alvarez2009, Pacucci2015}
(e.g., \cite{Pacucci2015}, and references therein)
%% add Tanaka et al. 2012, Jeon et al. 2012 and Smith et al. 2018
 so that there might be a significant time delay ($\sim 10^8$ yr) between the formation of the seed and the onset of the condition for its efficient growth \cite[][]{JohnsonBromm2007}.

Moreover, several (high-resolution) simulations suited to study the process of the first stars formation (i.e. resolving the physical scales on which the process occurs),  suggests that Pop III stars may be less massive than expected %\cite[$<100 \, \rm M_\odot$; e.g.][]{OsheaNorman2008, Hosokawa2011, Greif2011, Stacy2012, Hirano2015, Latif2021a} 
%\cite[$<100 \, \rm M_\odot$; e.g.][]{Hirano2015, Greif2011}
($<100 \, \rm M_\odot$; e.g., \cite{Hirano2015})
% e.g. as a consequence of ionizing radiation feedback from massive stars and/or ejections due to 3-body interactions
and their resulting remnant BHs ($\sim 20-30 \, \rm M_\odot$) may be kicked out from the host halos soon after their formation, via dynamical 3-body interactions, and thus their Eddington-limit growth may be prevented 
%\cite[e.g.][]{WhalenFryer2012, JohnsonBromm2007}.
(e.g., \cite{JohnsonBromm2007}). Nonetheless, such low-mass seeds unlikely settle in the centre of their host galaxies but they rather wander in the halo, without accreting gas (see, e.g., the discussion presented by \cite{volonteri2010}).

However, a recent study by \cite{ZubovasKing2021} suggests that even the SMBHs at $z>7$ can grow from a stellar-mass seed of $10 \msun$ if the seed accretes almost continuously via a chaotic accretion mode (i.e. through a large number of accretion episodes of mass transfer, initially at the Eddington rate, flowing from uncorrelated directions).\\
\newline   %% PER ACCORCIARE TAGLIARE I DUE PARAGRFAFI SOTTOSTANTI
\noindent All the considerations above motivated the exploration of alternative scenarios, invoked to decrease the BH growth timescale (the $e-$folding time, $t_e$ defined above) and grow SMBHs by $z\sim 6-7$,
%% TOGLIERE LA FRASE SOTTO?
such as the formation of populations of more massive seeds (the heavy channel discussed in Section~\ref{sec:seeds}), mergers and/or a very rapid/efficient mass growth mechanisms (see, e.g., the recent review by \cite{inayoshi2020}). 
%\cite[see e.g. the recent reviews by][]{Woods2019, inayoshi2020}. 
%We refer the reader to \cite[e.g.][]{Haiman2013, Natarajan2014, valiante2017,  Woods2019, inayoshi2020} for reviews.
%We refer the reader to \cite[e.g.][]{valiante2017,  Woods2019, inayoshi2020, Natarajan2014} for reviews.
%\cite[e.g.][]{VolonteriBellovary12, Haiman2013, Natarajan2014, valiante2017,  Woods2019, inayoshi2020} for reviews.
%
%In the literature, a large number of theoretical models offer different viable explanations for the early formation of SMBHs, on the basis of different seed BHs formation channels (see e.g. \cite{LatifFerrara2016} for a recent review) and/or different accretion modes/models (see e.g. the recent review by \cite{JohnsonHaardt2016}).
%Detailed observations of a large sample of the highest redshift quasars are needed to test these models and to improve our understanding of SMBH formation and evolution. 

\subsection{SMBH assembly in a cosmological context}\label{sec:CosmoContext}
If light, medium-weight and heavy seeds actually form at early cosmic epochs, we may now ask where, when and how often they form, alongside the cosmological evolution of high-redshift galaxies, %the cosmic history of structure formation 
and what is their relative role in the origin of SMBHs.
To answer these questions, SAMs and simulations (large-volumes and zoom-in) must describe the formation and evolution of galaxies/QSOs in a cosmological context, i.e. within the process of structures formation through cosmic times (we refer to these kind of tools as cosmological models). %In the framework of structures formation seeds are expected to grow accreting material at a rate that is determined by the reservoir of dense gas in their vicinity. In turn, the available gas budget is set by the processes regulating the so-called baryon cycle, i.e. the star formation (that consumes gas), inflows of material from the external intergalactic medium (replenishing the gas reservoir) and fast galaxy-scales winds (gas outflows) driven by supernova explosions and accreting BH feedback.

%shortcomings of cosmological models (both SAMs and simulations)
However, since they target the evolution of the rarest, most massive systems in the early Universe, cosmological models can not resolve the physical scales on which seed BH formation occurs nor the properties of the gas in the vicinity of a BH (i.e. on scales of the so-called Bondi radius), thus, they do not properly trace the accretion and feedback process in the nuclear regions of their host galaxies (see, e.g., \cite{CurtisSijacki2015} for a discussion). Thus, they rely on approximated analytical prescriptions/sub-grid models. 

\subsubsection{Seeding galaxies with the earliest BHs}
%In general, SAMs are efficient tools to study the early formation of SMBHs at $z>6$ (e.g. \cite{HaimanLoeb2001, Volonteri2003, Haiman2004, Yoo2004, VolonteriRees2005, TanakaHaiman2009, Volonteri2015a}) enabling to perform statistical studies in a cosmological context on the relative role  \cite[e.g.][]{Valiante2016, Dayal2019, Sassano2021}, host galaxy properties \cite[e.g.][]{Valiante2018a},  observational signatures \cite[e.g.][]{Dayal2019, Valiante2018b, Ricarte2018} and/or accretion modes \cite[e.g.][]{Pezzulli2016} of seed populations as they run over relatively short timescales, thus, enabling an extensive exploration of the parameters spaces and of different prescriptions for the formation and growth of BHs (i.e. to test different model variations). Recent models have been developed with the aim of investigating the formation of $z>6$ QSOs simultaneously exploring the contribution of progenitor seed BHs of different origin. We discuss these studies in Section~\ref{sec:seedsRelativeRole}. 
SAMs are efficient tools to study the early formation of SMBHs at $z>6$ enabling to perform statistical studies, in a cosmological context, on the relative role, the host galaxy properties, the observational signatures and/or of different accretion modes of seed populations over relatively short timescales (wide and fast exploration of the parameter space). Recent models have been developed with the aim of investigating the formation of $z>6$ QSOs simultaneously exploring the contribution of progenitor seed BHs of different origin. We discuss these studies in Section~\ref{sec:seedsRelativeRole}. 

\begin{figure*}[!]
    \includegraphics[width=\textwidth]{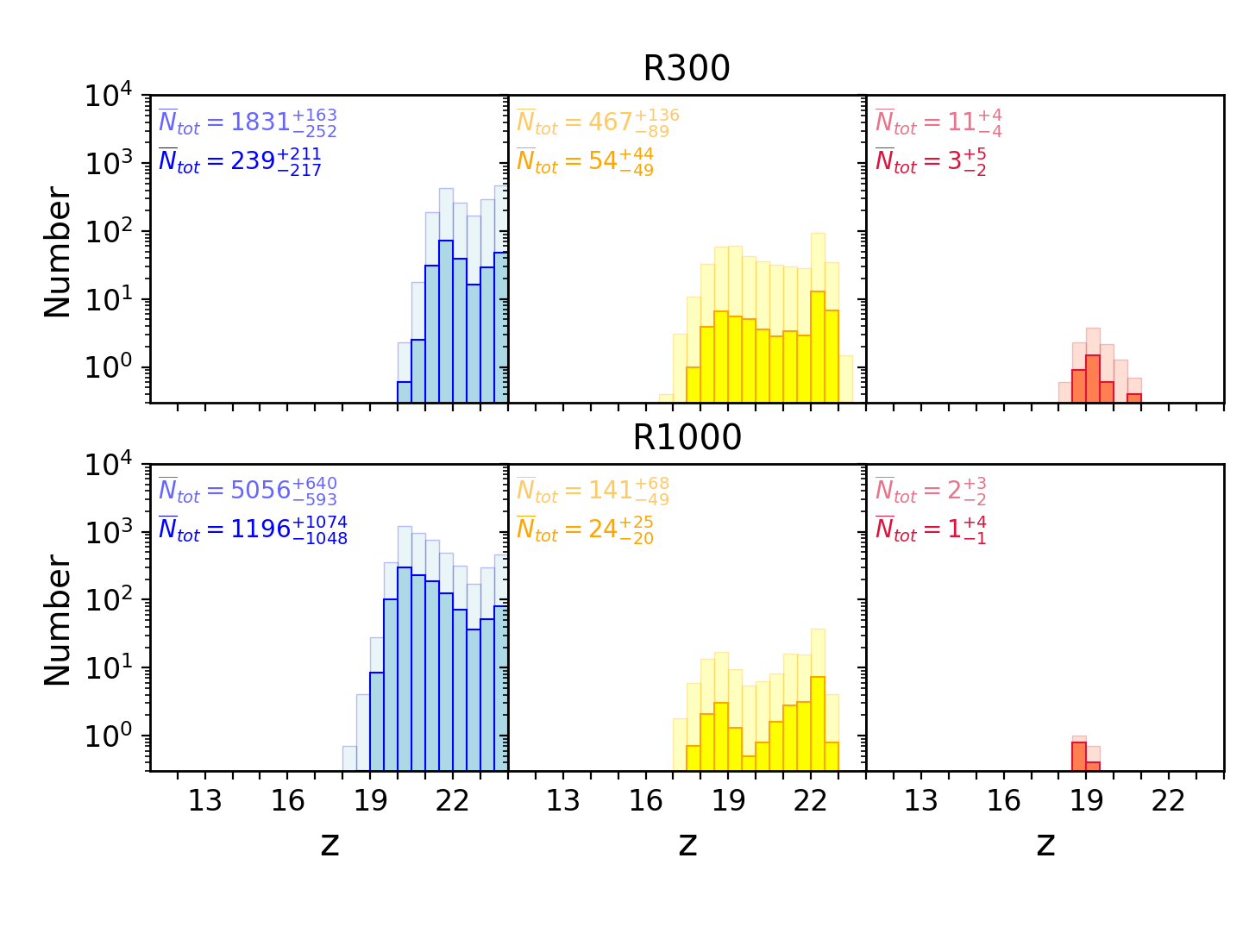}
    \caption{
     Distribution of formation redshifts of light (left), medium-weight (middle), and heavy (right) BH seeds from \cite{Sassano2021}. The histograms show
     the mean values averaged over 10 different simulations of the host dark matter halo merger trees. Lighter colours show the global BH seed population, whereas histograms in heavier colours count only the progenitors of the final SMBH at $z=6.4$ (i.e. those that are not ejected as satellite in minor halo mergers along the host galaxy cosmic assembly). Labels in each panel indicate the total number of seeds formed, on average, in the corresponding population. Corresponding upper and lower values indicate the difference with the maximum and minimum values for each family over 10 simulations. 
     The top panels show the distribution of the reference model (R300) in which a critical LW flux level of $J_{\rm cr} = 300$ is adopted. For comparison, the results obtained assuming a higher threshold, $J_{\rm cr} = 1000$ are reported in the bottom panels (model R1000).
    %Distribution of light (left), medium-weight (middle) and heavy (right) seeds as a function of their formation redshift obtained adopting two different metallicity/dust-to-gas ratio mass criteria for the seeding prescription within the data-constrained SAM presented by \cite{Sassano2021}. In the Super Competitive Accretion scenario (SCA model, lower panel) medium-weight and heavy seeds are allowed to form in higher metallicity galaxies with respect to the reference model (R300, upper panel). A critical LW threshold of $300\times 10^{-21}$ erg/s/Hz/cm$^2$/sr has been adopted in both scenarios. The histograms are obtained averaging over 10 different semi-analytical reconstruction of the merger three of a $10^{13} \msun$ QSO host dark matter halo. Lighter colours show the entire population of BHs formed while histograms with heavier colours indicate the actual progenitors of the $z\sim 6$ SMBH (see e.g. \cite{Valiante2016} for details), representing only a small fraction of ($< 10 - 20\%$) the whole population of BHs that form in galaxies at $z>15$. The total number (averaged over the 10 merger tree realisations) of seeds with upper and lower values indicating the difference with the maximum and minimum value for each seed family reached over 10 realisations. Figure adapted from \cite{Sassano2021}.
    }
    \label{fig:sassanoSeeds}
\end{figure*}

The flexibility of SAMs allow to implement and test increasingly more accurate, physically-motivated, seeding prescriptions (e.g., \cite{Valiante2016, Sassano2021}) based on the specific environmental conditions required for their formation (in particular, the metals and dust content of the ISM and the level of external LW photo-dissociating radiation illuminating the halo), as discussed in Section~\ref{sec:seeds}.
%based on the interplay of the processes regulating the evolutionary properties of galaxies, such as the chemical enrichment of the interstellar medium (i.e. the abundance of gas, metals and dust produced/injected by stars), the radiative feedback (i.e. the effect of ionising/photo-dissociating ultraviolet radiation emitted by stars and accreting BHs) and mechanical feedback (i.e. galaxy-scale winds, gas outflows, driven by supernova explosions and the BH accretion process).
As an example, Figure \ref{fig:sassanoSeeds} shows the results of the statistical analysis performed by \cite{Sassano2021}. They explore the relative birth rate of light, medium-weight and heavy seeds in a highly biased region of the Universe at $z>6$ by modelling the formation and evolution of a $z=6.4$ QSO along different analytical merger trees of the $10^{13} \msun$ host halo. In order to set the physical conditions for BH seeds formation, they implement a statistical description of the spatial fluctuations of LW photo-dissociating radiation and of metal/dust enrichment. 
%The free parameters of the model are chosen to reproduce the observed SMBH mass and host galaxy physical properties of a typical QSO at $z\sim 6$, SDSS J1148+5251 (see \cite{Valiante2011, Valiante2014, Valiante2016} for details). 
%In this model, light, medium-weight and heavy seeds form consistently with the environmental properties determined by the in-homogeneous chemical enrichment process (i.e. the non instantaneous mixing in the ISM of metals and dust produced by stars) and the fluctuations of the illuminating LW flux, with respect to the mean local background (i.e. the non-linear spatial distribution of the UV emitting sources around a given halo). 
In this model light, medium-weight and heavy seeds form consistently with the environmental properties, i.e. the metallicity ($Z$, the ratio between the mass of metals and gas) and dust-to-gas mass ratio ($\mathcal{D}$) of the ISM and the intensity of the illuminating LW flux ($J_{LW}$).

In their reference model, seed BH formation is activated in metal-poor halos with metallicity $Z<Z_{cr}=10^{-3.8} \, Z_\odot$, independently of the seed mass. Then, the emergence of a specific seed flavour depends on the LW flux level and on the ISM dust-to-gas mass ratio (see \cite{Valiante2016} and \cite{Sassano2021} for details). Light seeds of $[40-140] \msun$ and $>260 \msun$ form if both the dust-to-gas mass ratio of the ISM and the LW flux are relatively low, i.e. if $\mathcal{D} \leq 4.4 \times 10^{-9}$ and $J_{LW}< J_{cr}$. More massive seeds are instead assumed to form in atomic cooling halos illuminated by a super-critical LW radiation ($J_{LW}\geq J_{cr}$): medium-weight seeds of $10^3 \msun$ form in the presence of dust ($\mathcal{D} \geq 4.4 \times 10^{-9}$) whereas $10^5 \msun$ heavy seeds form in dust-poor/free halos ($\mathcal{D} < 4.4 \times 10^{-9}$). The results of this reference model, called R300 (to indicate the adopted LW critical threshold $J_{cr}=300 \times 10^{-21}\rm erg/s/Hz/cm^2/sr$), are shown in the upper panels of Figure~\ref{fig:sassanoSeeds}.

In this scenario, $\sim 1830$ light seeds (average number over 10 merger tree simulations) form at $z>20$, while the formation of medium-weight seeds is extended towards lower redshifts, as shown in the upper left and central panels of Figure~\ref{fig:sassanoSeeds}. The heavy seeds are the rarest ones, $< 1\%$, forming only during a relatively short period of time, as a consequence of their tight birth environmental condition requirements (see Figure~\ref{fig:sassanoSeeds}). The number of medium-weight and heavy seeds is significantly reduced if a higher intensity of the LW flux is required (i.e. a higher critical LW flux threshold, $J_{cr}=1000 \times 10^{-21}\rm erg/s/Hz/cm^2/sr$ is assumed; see lower panels of Figure~\ref{fig:sassanoSeeds}, model R1000).
In addition, \cite{Sassano2021} also discuss the dependence of their results on the adopted seeding prescription, showing how the number and redshift distribution of the seeds vary with the adopted critical metallicity threshold ($Z_{cr}$; see \cite{Sassano2021} for details). %being higher and extended towards lower redshift when medium-weight seeds are allowed to form in higher metallicity galaxies, via the so-called super-competitive accretion scenario
Note that, we expect that only a small fraction ($<15-20\%$) of all the seed BHs forming along the hierarchical assembly of the QSO eventually contribute to the growth of the final SMBH (e.g., \cite{Valiante2016, Sassano2021}).
%underlying the effect of different seeding criteria on the relative number and redshift distribution of newly formed seeds (light, medium-weight and heavy), tracked in a cosmological context.
\newline
\newline
Hydrodynamical techniques are also widely adopted to study the formation of high redshift QSOs. As mentioned above %in Section~\ref{sec:simulations}, 
simulations of large cosmological volumes are required to account for the low number density of $z\sim 6$ QSOs.
%Simulations of high redshift quasar formation must model a large cosmological volume to accommodate the low abundance of this population, have a large dynamic range to follow the hierarchical buildup of the quasar hosts, and include the hydro-dynamics of the gas flows in galaxy mergers. 
%
%Some examples of large volume hydrodynamical simulations (box size $>100$ Mpc) including BHs are: the Horizon-AGN \cite{Dubois2014, Volonteri2016}; Illustris \cite{Vogelsberger2014, DeGraf2020}; IllustrisTNG \cite{Weinberger2018};  EAGLE \cite{Schaye2015};  MassiveBlack \cite{Khandai2015}; BlueTides \cite{Feng2016}; Magneticum \cite{Dolag2016} and Simba \cite{Dave2019}. 
Some examples of large volume hydrodynamical simulations (box size $>100$ Mpc) including BHs are: the Horizon-AGN \cite{Dubois2014}; Illustris \cite{Vogelsberger2014}; IllustrisTNG \cite{Weinberger2018};  EAGLE \cite{Schaye2015};  MassiveBlack %\cite{Khandai2015};
\cite{DiMatteo2012}; BlueTides \cite{Feng2016}; Magneticum \cite{Dolag2016} and Simba \cite{Dave2019}. 

These simulations can describe the hydro-dynamics of gas flows, on galaxy-scales, more accurately than SAMs.
%These simulations can describe %resolve
%the physical processes regulating the evolution of QSOs (the hierarchical build up of their hosts and the hydro-dynamics of gas flows) on galaxy-scales more accurately than SAMs.
%However, due to %high computational costs, and 
%their low mass resolution, these simulations can not directly trace the formation of seed BHs (or the BH accretion process). They instead rely on sub-grid models seeding specific, well-resolved, massive dark matter halos (usually $10^9-10^10 \msun$) with BHs of $10^5.10^6 \msun$) and thus can not distinguish between different seed formation channels or test their relative contribution/occurrence within the same simulation run.
%However, they do not have the necessary resolution to resolve the physical scales on which seed BH formation (and accretion) process occurs. They instead rely on sub-grid models seeding specific, well-resolved, massive dark matter halos (usually $10^9-10^10 \msun$) with BHs of $10^5.10^6 \msun$) and thus can not distinguish between different seed formation channels or test their relative contribution/occurrence within the same simulation run. 
However, they often employ simple sub-grid prescriptions to describe the formation of seed BHs, usually seeding specific, well-resolved, massive dark matter halos above a given mass threshold (usually $10^9-10^{10} \msun$) with a fixed seed BH mass $10^5-10^6 \msun$; ``sink particles"). 
Both the seed mass and the halo mass threshold to host a seed are free parameters of the simulations. Such a seeding prescription is less accurate than those adopted in SAMs, which are instead based on environmental properties. Thus, different seed formation channels or their relative contribution/occurrence can not be tested within the same simulation run.

%Both, the seed mass and the minimum halo mass (in which the seeds are planted) are free parameters of the simulations, thus, this assumption is less accurate than the physically motivated prescriptions/criteria that is possible to implement and test within SAMs and does not allow to distinguish between different seed formation channels or test their relative contribution/occurrence within the same simulation run. 

Using this simple halo-based seeding prescription (and assuming Eddington-limited Bondi accretion, see next section), this kind of simulations successfully reproduce the properties of SMBHs observed in the local Universe %\cite[e.g.][]{Weinberger2018, Khandai2015, Volonteri2016}, 
%\cite[e.g.][]{Weinberger2018, Khandai2015}, 
(e.g., \cite{Weinberger2018}), but they do not produce the rarest $>10^9 \msun$ SMBHs at $z>6$, as these large-size (low-mass resolution) simulations primarily target average, uniform, volumes of the Universe. Only in few cases, large volume hydrodynamical simulations, such as the MassiveBlack ($0.75 Gpc^3$; e.g., \cite{DiMatteo2012}) %, its high-resolution zoom-in runs (\cite{Khandai2012, Feng2014}) 
and the largest one, BlueTides ($\sim 400$ Mpc comoving box size, e.g., \cite{Feng2016}) were able to reproduce the formation of $z>7$ QSOs. %(e.g. \cite{Tenneti2019, Marshall2020}).

\begin{figure*}[!]
\centering
    \includegraphics[width=3.85cm]{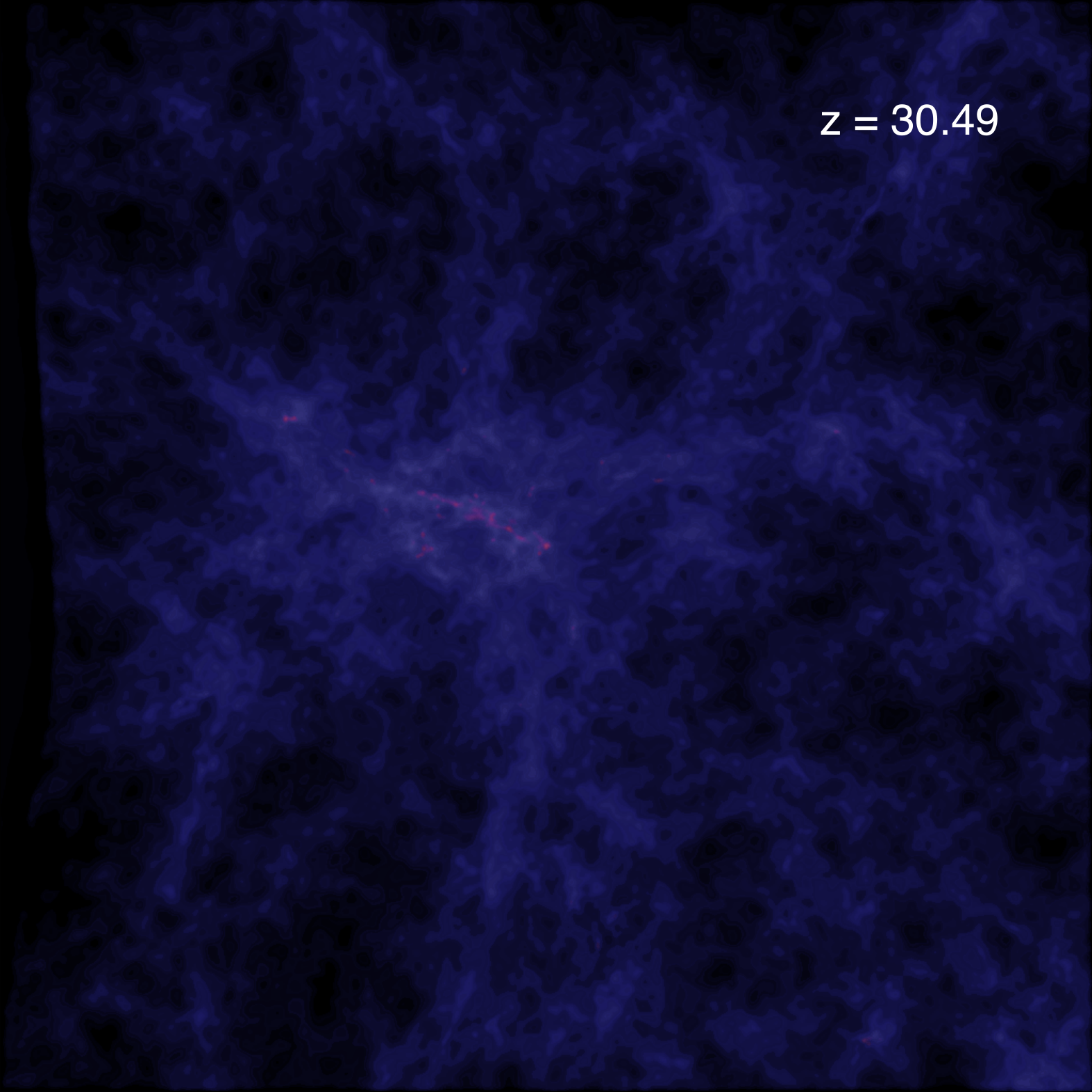}
    \includegraphics[width=3.85cm]{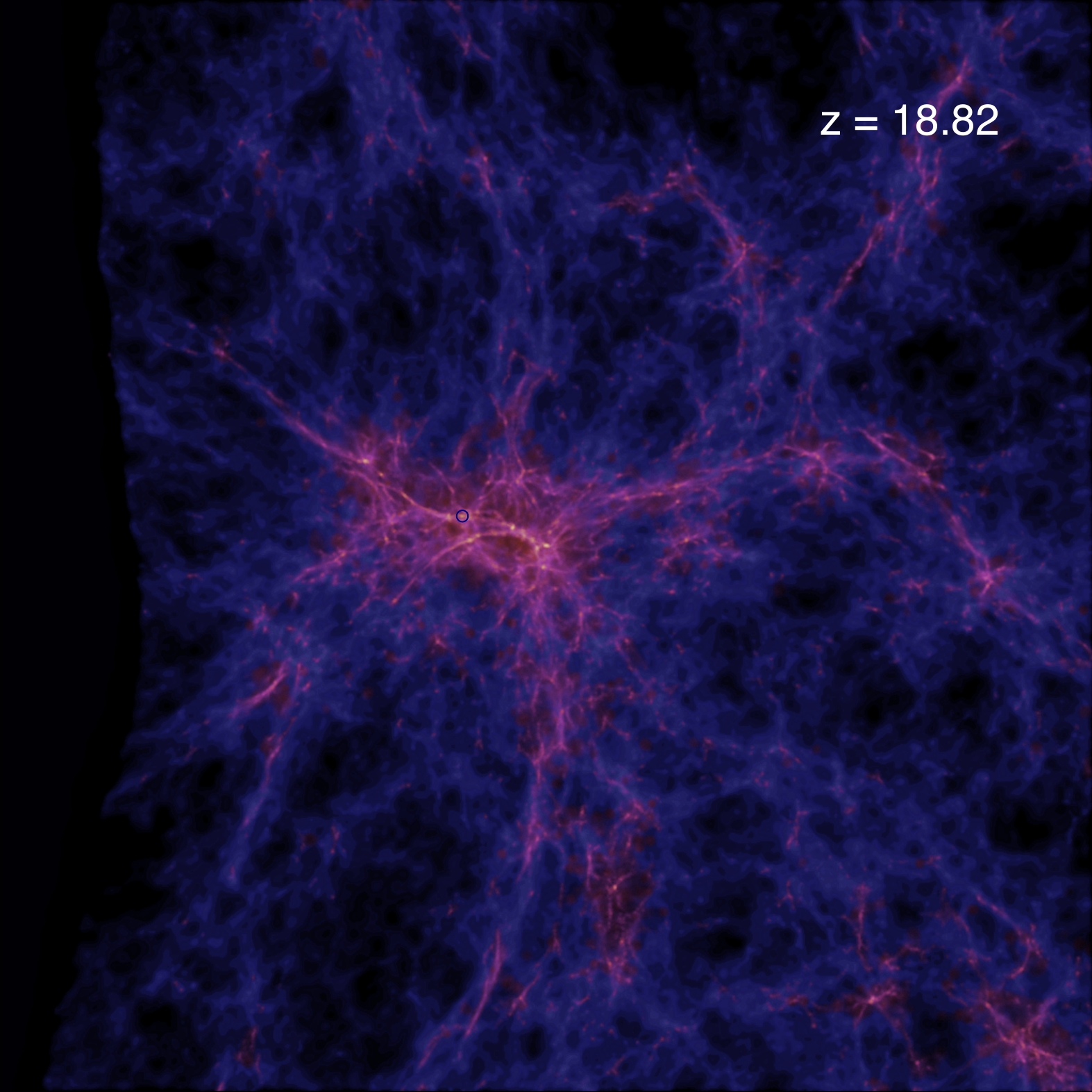}
    \includegraphics[width=3.85cm]{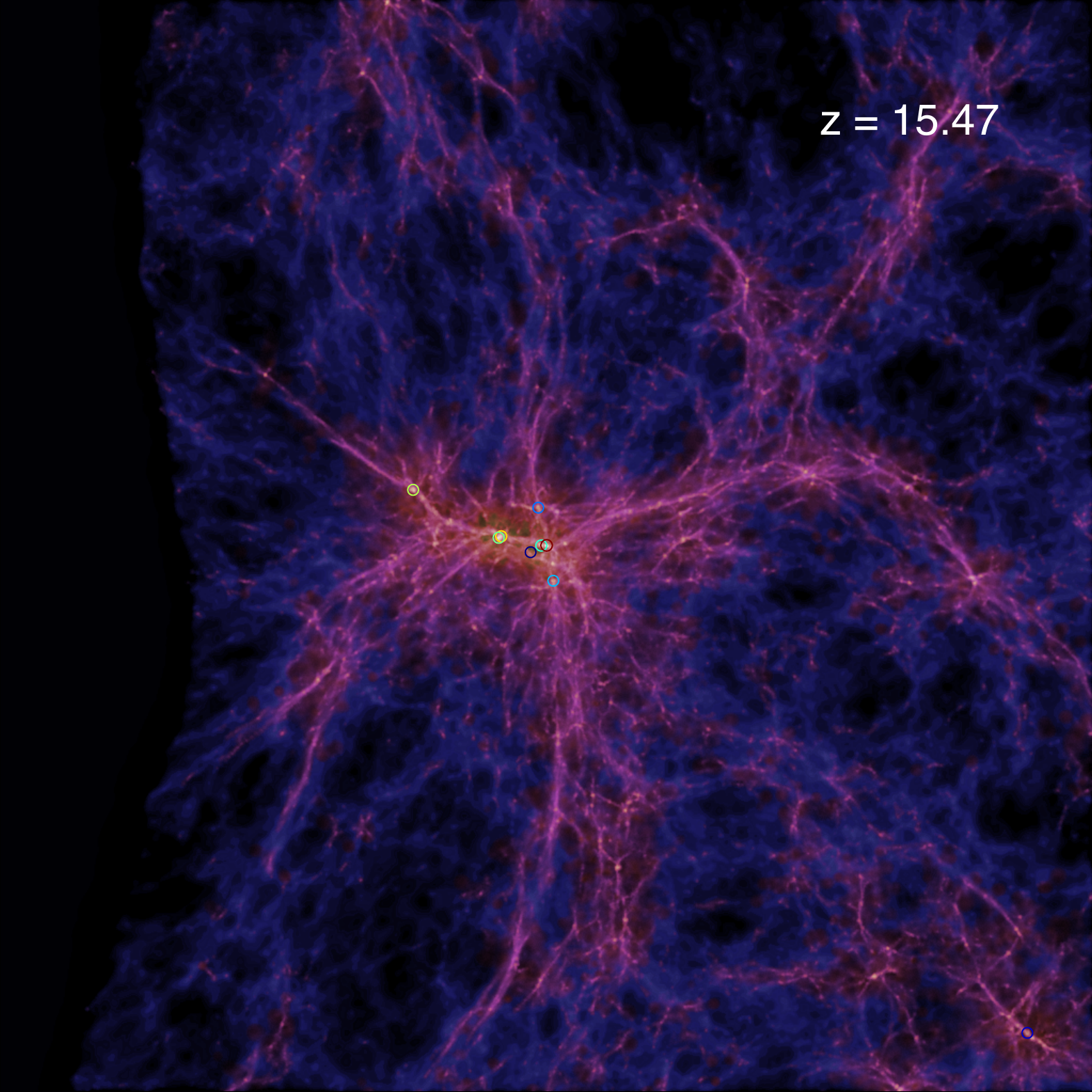}
    \includegraphics[width=3.85cm]{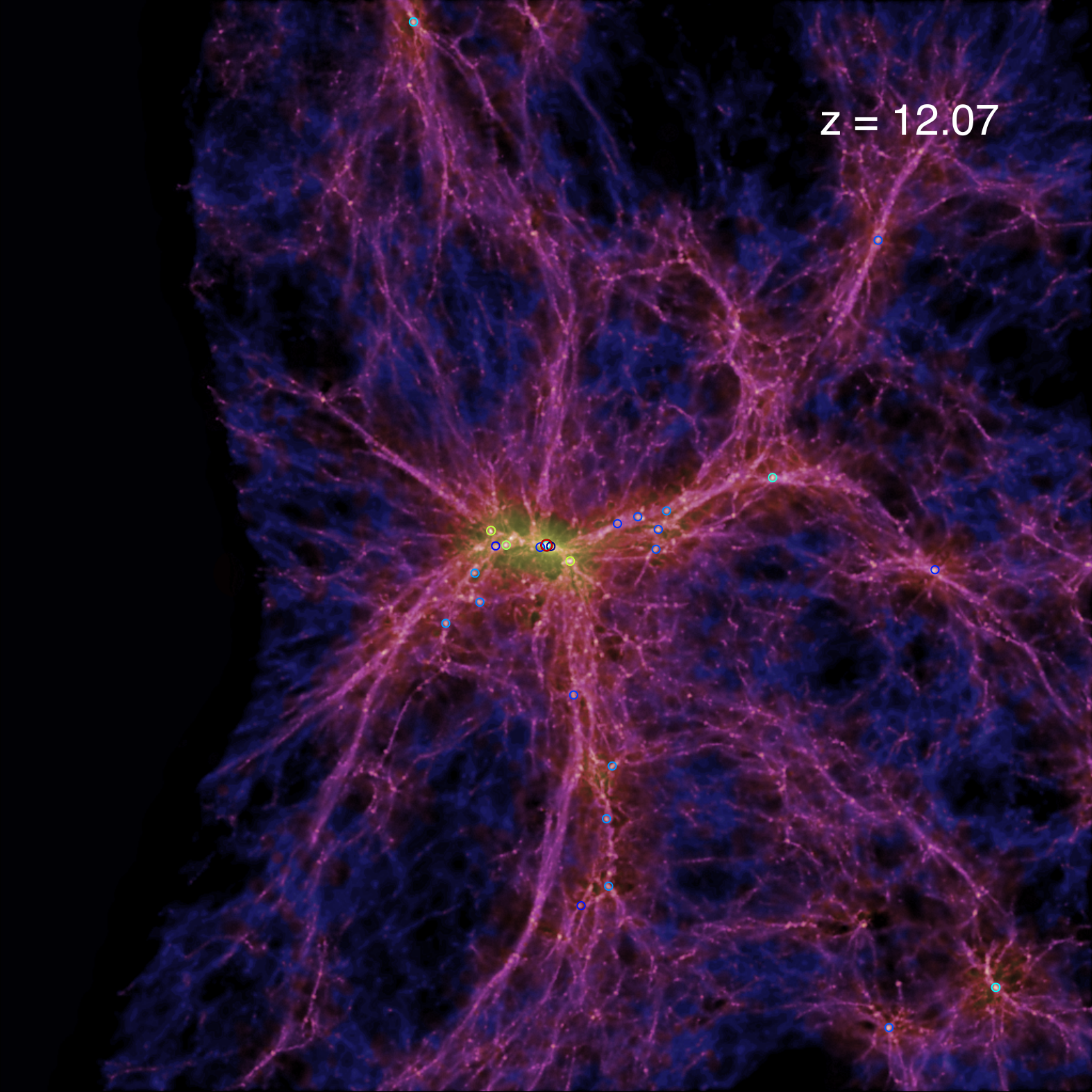}
    \includegraphics[width=3.85cm]{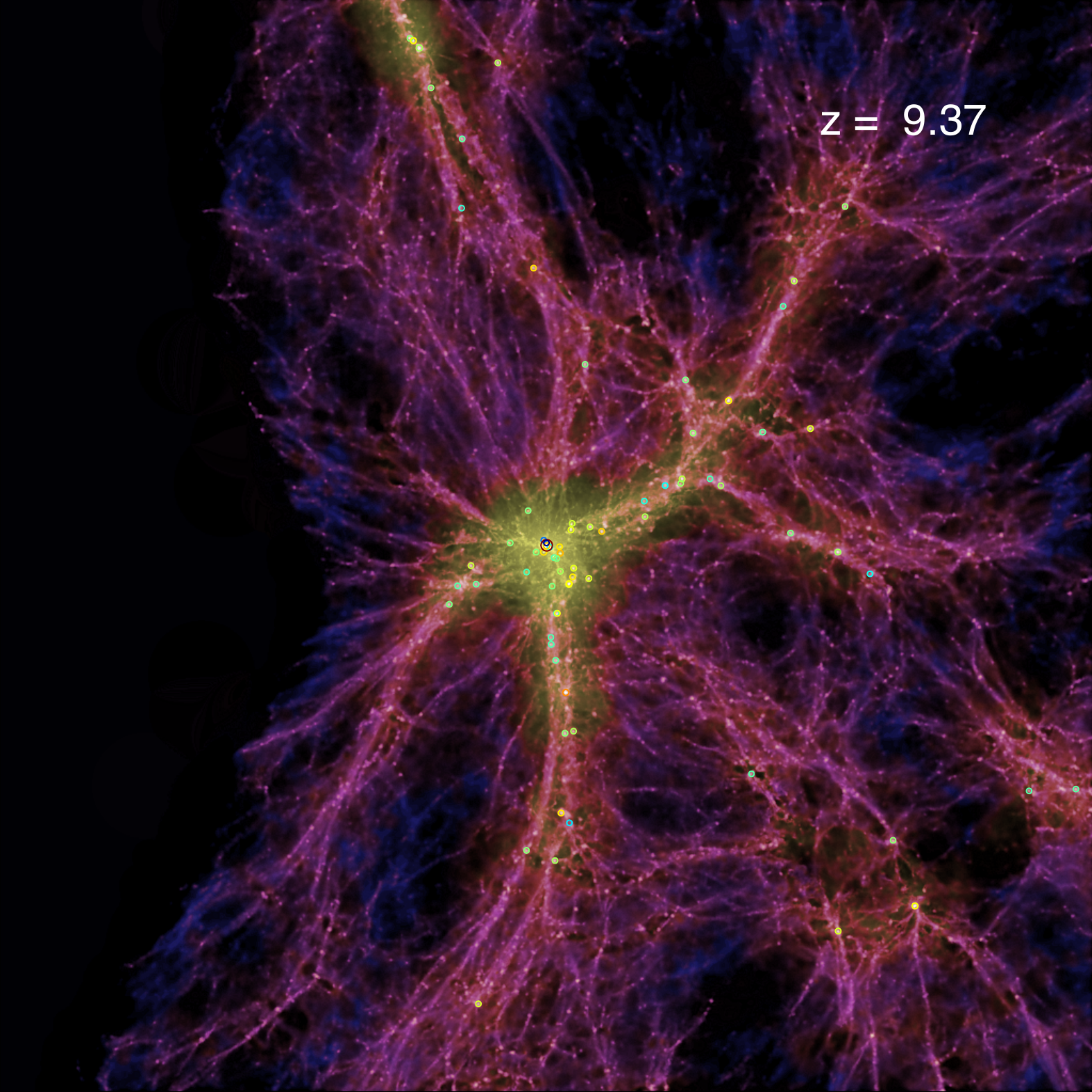}
    \includegraphics[width=3.85cm]{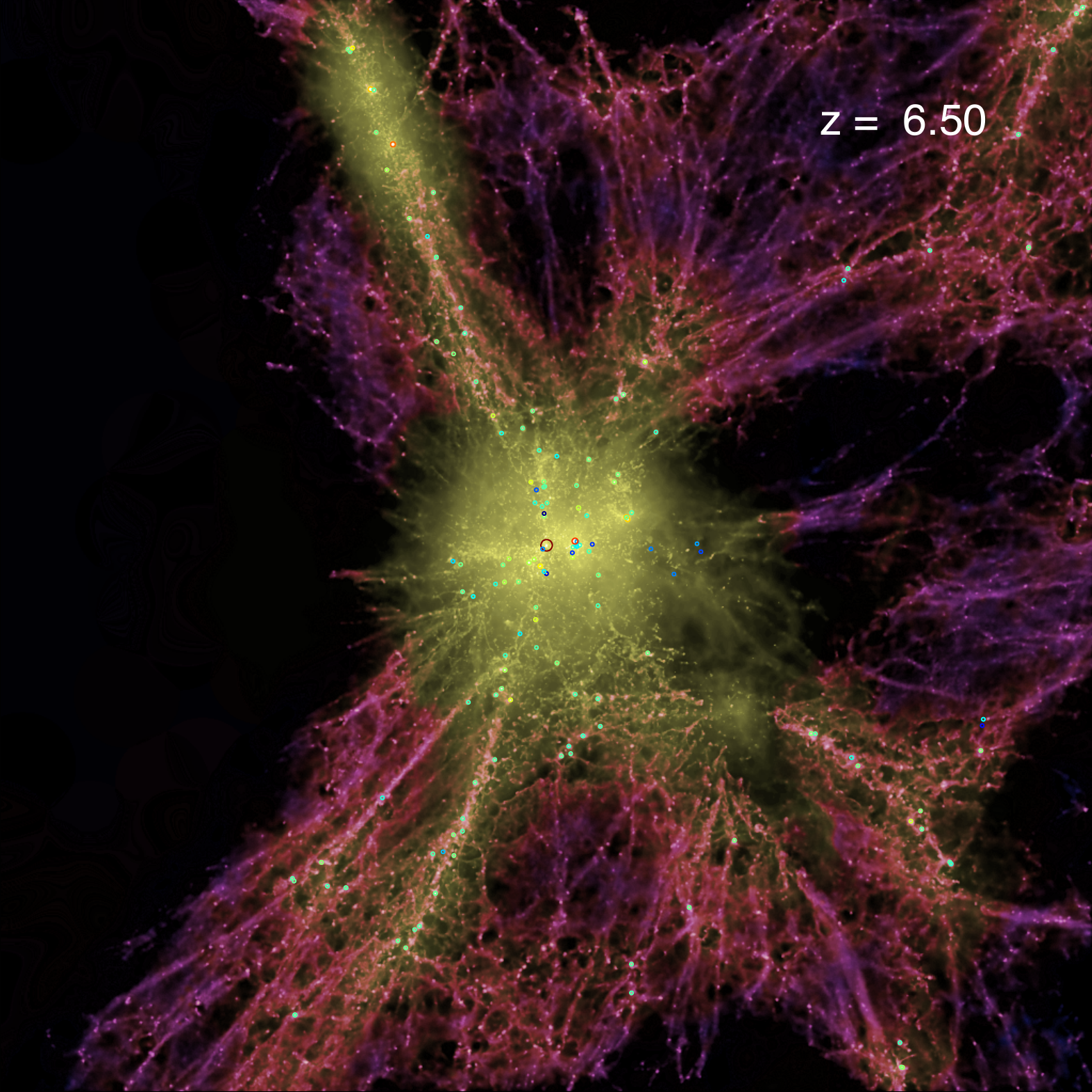}
    \caption{Cosmological zoom-in simulation from \cite{Zhu2020}. The different images (snapshots) show the evolution of the first galaxies and BHs from $z\sim 30$ to $z\sim 6.5$, simulated within a zoom-in region of $10$ comoving Mpc size centred on a $10^{13} \msun$ dark matter halo. Each snapshot shows the projected gas density colour-coded with temperature: cold gas is in blue while the hot gas (heated and ionised by SN and accreting BH feedback) is represented in yellow. In this model the author assume that a seed BH of $\sim 10^5 \msun$ is planted with halos of $\sim 10^{10} \msun$ and subsequently grows via Eddington-limited chaotic accretion assuming the standard, thin-disc accretion model. Coloured circles indicate the locations of BHs.
    }
    \label{fig:zhu2020}
\end{figure*}

Zoom-in simulations, i.e. higher mass resolution re-simulations of individual halos (smaller volumes) targeting the high dark matter density peaks of $10^{12}-10^{13} \msun$ (highly biased regions of the Universe) at $z\sim 6$ are instead best suited to study the formation of high redshift QSOs, producing SMBHs of $\sim 10^9 \msun$ at $z\geq 6$ %\cite[e.g.][]{Li2007, Sijaki2009, khandai2012, Feng2014, Costa2014, Smidth2018} 
%\cite[e.g.][]{Li2007, Sijaki2009, khandai2012, Costa2014}
(e.g., \cite{Li2007, Sijaki2009}) and allowing to explore the impact (still with sub-grid models) of different seed masses 
%(e.g. \cite{Lupi2019, Zhu2020, Huang2020}) 
(e.g., \cite{Zhu2020, Huang2020}) %seeding criteria
and/or BH accretion modes (e.g., \cite{Zhu2020}).
An example of a zoom-in simulation is reported in Figure~\ref{fig:zhu2020}. %{\bf Add figure 1 from Zhu et al. 2020}.

Some advancements have been done to include more physically motivated BH seeding prescriptions (in smaller-scale/zoom-in simulations) based on the properties of the gas (or stars), within cosmological hydrodynamical simulations 
%(see e.g. \cite{Bellovary2011, Habouzit2017, Dave2019, Lee2021, Bhowmick2021b} and references therein), 
%\cite[see e.g.][and references therein]{Dave2019, Bhowmick2021b}
(e.g., \cite{Dave2019, Bhowmick2021b}, and references therein)
usually linking the formation either of low-mass (light and medium-weight) seeds on local gas/stellar properties \cite{Habouzit2017}, or of heavy seeds to low metallicity, high-density gas cells (e.g., the \textsc{romulus} simulation presented by \cite{Tremmel2017}) or to regions with high gas and stellar density (e.g., \cite{Kaviraj2017}). 

%%% PER ACCORCIARE ELIMINARE IL PARAGRAFO SOTTOSTANTE
%Despite the recent attempts made to test the impact of different halo- and gas-based seed models %, in a wider range of masses ($10-10^6 \, \rm M_\odot$), on the formation of high-z QSOs (within zoom-in cosmological hydrodynamical simulations; see e.g. \cite{Zhu2020, Huang2020, Bhowmick2021} and references therein), exploring different physical parameters regulating the seed formation (or growth history) is still computationally prohibitive for hydrodynamical simulations. The only way to overcome this limitation is through post-processing methods \cite{DeGraf2020}.
%
%Note that although the seeding prescriptions adopted in cosmological models usually assume a fixed mass for the heavy seed channel (but see \cite{PacucciLoeb2020}), the initial mass fucntion of heavy seeds may be described by a Gaussian distribution \cite{Ferrara2014}.

\subsubsection{Eddington-limited growth}
Theoretical models and observations suggests that BH growth across cosmic times occurs mainly via gas accretion %(e.g. \cite{Soltan1982, Marconi2004, Shankar2009}).
%\cite[e.g.][]{Marconi2004, Shankar2009}).
(e.g., \cite{Marconi2004}).
A commonly adopted approach in cosmological models (SAMs and simulations) to describe the BH accretion process is to implement an approximated/sub-grid prescription, usually based on the Bondi–Hoyle–Lyttleton, spherically symmetric, approach \cite{Hoyle1941, Bondi1952}:
\begin{equation}
\dot{M}_{\rm BHL} =  \alpha \frac{4 \pi G^2  M_{\rm BH}^2 \rho_{\rm gas}(r_B)}{(c_s^2 + v^2)^{3/2}}
\label{eq:BHL}
\end{equation}
where $c_s$ is the sound speed of the gas, $v$ is the relative velocity between the BH and the gas, and $\rho_{\rm gas}(r_B)$ is the gas density evaluated at the Bondi-Hoyle radius, namely the radius of gravitational influence of the BH, $r_B = 2G M_{\rm BH}/(c_s^2+v^2)$ \footnote{Note that sometimes Eq.~\ref{eq:BHL} and the expression of the Bondi-Hoyle radius given above are simplified by setting the velocity term, $v$, to zero.}.
For the (original) Bondi-Hoyle-Lyttleton rate to be accurate, the models should resolve the gas properties down to the scale of the Bondi-Hoyle radius, $r_{B}$. This is currently not feasible for the vast majority of simulations and SAMs. For this reason, the free parameter $\alpha$ is introduced in the original formulation of the rate to account for the enhanced (non-resolved) gas density in the inner regions around the BH. This parameter is calibrated using simulations of galaxy mergers (e.g., \cite{DiMatteo2005}) against the local scaling relations and is usually set to a constant value (around $\sim 100$), but it should depend on the gas density (e.g., \cite{BoothSchaye09}).
In any case, without this contribution, the actual accretion rate would be strongly underestimated due to the lack of resolution of the models %\cite{[e.g.][]DiMatteo2012, Schaye2015}).
(e.g., \cite{DiMatteo2012}).

Often theoretical models describe the growth of BHs within the so-called Eddington-limited gas accretion regime. In other words they assume that the BH accretion can not exceeds the limit set by the Eddington rate for a given BH mass (Eq.~\ref{eq:EddRate}), i.e. thus $\dot{M}_{\rm BH}=min(\dot{M}_{BHL}, \dot{M}_{\rm Edd}$).

It is worth mentioning that a new numerical technique, suited to resolve small-scales flows of gas around BHs (at the level of the Bondi radius), has been presented by \cite{CurtisSijacki2015}. Their method enables to increase the mass/spatial resolution in selected regions around accreting BHs without increasing too much the computational costs and enabling to effectively evaluate the Bondi-Hoyle-Lyttleton accretion rate within cosmological simulations of galaxy formation.

%An alternative SMBH growth mechanism is represented by the chaotic accretion of cold gas, proposed in several studies (see e.g. \cite{Zhu2020, ZubovasKing2021} and references therein). In this scenario the BH growth proceeds via individual episodes of accretion characterised by uncorrelated initial directions of angular momentum. As a result the spin of the BH is low and allows to achieve low radiative efficiencies ($\sim 0.06$; e.g. \cite{King2008}).
An alternative SMBH growth mechanism is represented by the chaotic accretion of turbulent, cold gas, proposed in several studies (see \cite{ZubovasKing2021}, and references therein). In this scenario, the gas flows towards the BH from uncorrelated directions and the BH spin is kept low, allowing to achieve low radiative efficiencies %($\sim 0.06$; e.g. \cite{King2008})
($\sim 0.06$), thus providing more efficient growth with respect to radiatively efficient ($\epsilon \geq 0.1$) accretion modes.

\subsubsection{SMBH growth boosted by heavy seeds}
The abundance of seed BHs, as well as their role in the growth of SMBHs, strongly depends on the fraction/number of halos in which the conditions for the formation of the different channels are met (e.g., \cite{LodatoNatarajan2007, Dijkstra2014, Habouzit2016c}).

Heavy seeds are considered one of the most promising solutions to the early SMBH growth problem as they form already massive, thus providing a significant (initial) mass boost to the subsequent efficient growth (as it can be inferred from Eq.~\ref{eq:exponentialgrowth}, following an exponential growth, the larger the initial seed mass the faster is the SMBH growth) and within an environment characterised by higher gas densities \cite{Latif2013b}).

Both SAMs and numerical simulations pointed out that a SMBHs at $z>6$ can grow from (at least) few heavy seeds of $10^5-10^6 \, \rm M_\odot$ forming at early times ($z>10-15$), and increasing their mass mainly via Eddington-limited gas accretion 
%(e.g. \cite{Valiante2016, Sassano2021, DiMatteo2012, Sijaki2009, Costa2014, DiMatteo2017, Smidth2018, Lupi2019, Zhu2020, Valentini2021}. 
(e.g., \cite{Sassano2021, Zhu2020}, and references therein).
In Figure \ref{fig:SassanoBHevo} we show an example of SMBH growth boosted by the formation of heavy seeds (already at $z\lesssim 20$).
%as they have the advantage of providing a significant (initial) mass boost to the subsequent efficient growth (as it can be inferred from Eq.~\ref{eq:exponentialgrowth}, following an exponential growth, the larger the initial seed mass the faster is the SMBH growth) and of forming in a environment rich gas at higher densities (e.g. \cite{Latif2013b, Pacucci2015, Volonteri2016}).
%with mergers with other BHs providing a contribution mainly at early epochs  
%
%On the one hand heavy seeds are promising pathways but they are expected to for under extremely tight conditions (i.e. in a cosmological context they may be rare with respect to other seed formation channels e.g. Valiante et al. 2016, Sassano et al. 2021). On the other hand the more common seeds (light and medium-weight) may struggle to sustain efficient/fast enough gas accretion (at the Eddigton or super-Eddington rates)
%
%For example, heavy seeds are expected to be rare (e.g. Lodato & natarajan 2006; Dijkstra et al. 2008, 2014; (Habouzit et al. 2016; Tremmel et al. 2017; Dunn et al. 2018; Luo et al. 2020; Chon et al. 2021; Bhowmick et al. 2021b), especially compared to the other seed populations in a cosmological context (e.g. Valiante et al. 2016, Sassano et al. 2021) as they require quite tight conditions to form (see Section \ref{sec:heavySeeds}). Thus, in environmentin in which theyr formation may be prevented

The first hydrodynamical simulation of the formation of a high-redshift QSO has been presented by \cite{Li2007} who follow the hierarchical evolution of a $\sim 10^{13} \, \rm M_\odot$ dark matter halo at $z=6.4$, the most massive halo found in a simulated volume of $\sim 2.5 \, \rm Gpc^3$. The authors find that a $10^9 \, \rm M_\odot$ SMBH form via efficient accretion of gas onto massive seeds of $\sim 10^5 \, \rm M_\odot$, facilitated (triggered) by gas-rich mergers. 
By means of a semi-analytical approach, \cite{TanakaHaiman2009} found similar results, confirming that the hierarchical history of $z\sim 6$ QSO hosts (plausibly $10^{12}-10^{13} \, \rm M_\odot$ dark matter halos), and in particular the frequency (number) of major galaxy mergers (i.e. mergers between halos with comparable mass, or at above a mass ratio of $1:10$), plays a crucial role in driving efficient near-Eddington seed growth. In addition, \cite{TanakaHaiman2009} show that the formation of a $z\sim 6$ QSO does not require super-Eddington rates, for either light or heavy seeds, if the QSO hosts halo become atomic-cooling at $z>30-40)$, at earlier epochs with respect to what found in cosmological simulations. 

Zooming-in on a region of $15 \, h^{-1}$ Mpc$^3$ selected within the BlueTides cosmological hydrodynamic simulation, \cite[][]{Huang2020} explored the early growth of $z>6$ SMBHs, concluding that the formation of a $10^9 \msun$ BH does not depend on the seed mass (for seeds of $5\times 10^3$, $5\times 10^4$ and $5\times 10^5 \, h^{-1} \msun$),
provided that the ratio between the dark matter halo mass threshold (i.e. the minimum mass for halos to be seeded with BHs) and the BH seed mass ratio is the same (constant) for all the seeding scenarios.
%the dark matter halos are seeded according to a constant ratio between the seed mass and the halo mass threshold for planting the seeds. %As a result of this assumption, the number of seeds planted in the simulated volume is larger when assuming less massive seeds.
%
%Some advancements have been done towards a more realistic/accurate description of the seeding prescriptions, based on the properties of the gas (or stars), within cosmological hydrodynamical simulations (see e.g. \cite{Habouzit2017, Dave2019, Lee2021, Bhowmick2021b} and references therein), usually linking the formation either of low-mass (light and medium-weight) seeds on local gas/stellar properties (\cite{Habouzit2017}) or of heavy seeds to low metallicity, high-density gas cells (e.g. the \textsc{romulus} simulation presented by \cite{Tremmel2017}) or regions with high gas and stellar density (e.g. \cite{Kaviraj2017}). 
%Despite the recent efforts to test the impact of different halo- and gas-based seed models, in a wider range of masses ($10-10^6 \, \rm M_\odot$), on the formation of high-z QSOs (within zoom-in cosmological hydrodynamical simulations; see e.g. \cite{Zhu2020, Huang2020, Bhowmick2021} and references therein), exploring the parameters space of light, medium-weight and heavy seed models together and self-consistently (i.e. according to environmental properties), at the same level as in SAMs, is still computationally prohibitive for hydrodynamical simulations.
%based on gas properties

In a recent work, \cite{DeGraf2020} proposed an alternative approach, in which the growth of BHs is re-calculated by means of a post-processing method applied to the results of the Illustris simulation (on BHs and their surrounding gas properties). This approach enables the authors to test the impact of different heavy seed BHs formation models (i.e. stochastic vs physically-
motivated prescriptions), simultaneously saving computational time, with respect to full hydrodynamical simulations. 
%They find
In agreement with the previous works mentioned above, they find that the assembly of massive BHs as well as their merger history are strongly affected by the seeding model and suggest that the current observational constraint implies a more efficient accretion onto heavy seeds (possibly through super-Eddington phases) or a non-negligible contribution from alternative seed formation channels.

In addition, heavy seeds are expected to be rare, i.e. only a small fraction of halos/galaxies satisfies the tight environmental conditions for their formation %\cite[e.g.][]{Dijkstra2014, Habouzit2016c, Valiante2016, Chon2021, Sassano2021, Lodato2006}).
(e.g., \cite{Dijkstra2014, Habouzit2016c, Valiante2016, Chon2021, Sassano2021}).
Thus, in environments hostile to their formation, additional/alternative scenario should be invoked to grow a SMBH (see Section~\ref{sec:superEddingtonGrowth}). %(e.g. Lodato & natarajan 2006; Dijkstra et al. 2008, 2014; (Habouzit et al. 2016; Tremmel et al. 2017; Dunn et al. 2018; Luo et al. 2020; Chon et al. 2021; Bhowmick et al. 2021b), especially compared to the other seed populations in a cosmological context (e.g. Valiante et al. 2016, Sassano et al. 2021) as they require quite tight conditions to form (see Section \ref{sec:heavySeeds}). Thus, in environmentin in which theyr formation may be prevented

\subsubsection{The relative role of seed BH populations}\label{sec:seedsRelativeRole}

SAMs devoted to study the formation of SMBHs usually explore the role of either light or heavy seeds  %\cite[e.g.][]{VolonteriRees2005, TanakaHaiman2009, Volonteri2003, Volonteri2015a}.
%\cite[e.g.][]{TanakaHaiman2009, Volonteri2003, Volonteri2015a}).
(e.g., \cite{TanakaHaiman2009, Volonteri2003}).
The first attempt made to explore the combined action of light and heavy seeds, with a SAM, has been presented by \cite{Petri2012}. They show that the SMBH assembly strongly depends on the fraction of halos that are seeded with heavy seeds, along the merger tree, that must be close to $100\%$ to form of a $\sim 10^{10} \msun$ BH.
However, in previous studies the seeds were not assigned to halos according to the environmental conditions and taking into account the chemical evolution of the ISM (in particular, its metallicity and dust-to-gas mass ratio).

\begin{figure*}[!]
\centering
    \includegraphics[width=\textwidth]{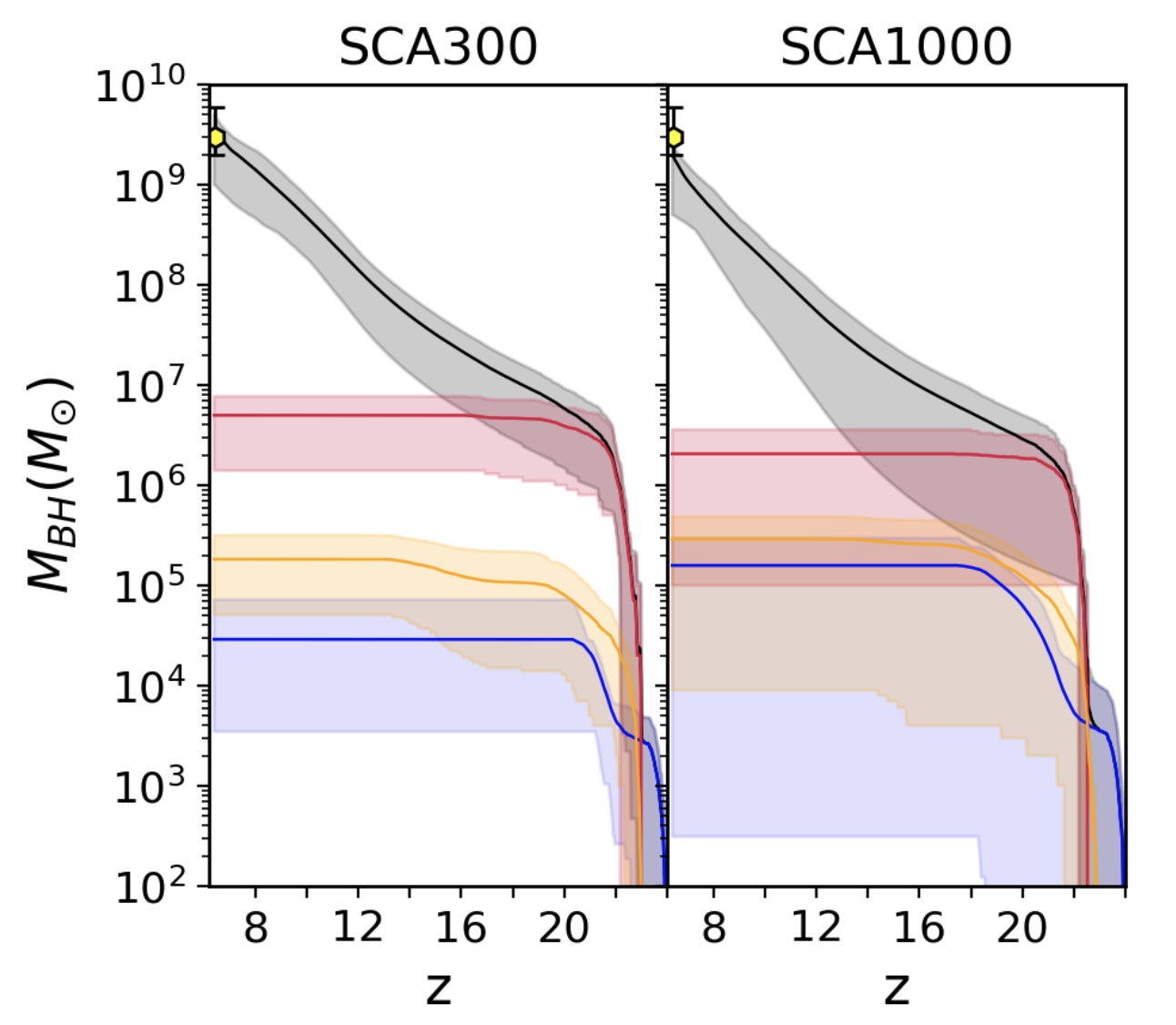}
    \caption{Redshift evolution of the SMBH powering a $z>6$ QSO via Eddington-limited gas accretion and mergers with other BHs (black line), in the SCA scenario (see Section~\ref{sec:mwSeeds} and \cite{Sassano2021} for details). The blue, yellow and red lines instead show the relative contribution, without gas accretion, of light, medium-weight, and heavy BH seed progenitors, respectively. Each curve is obtained averaging on 10 different merger tree realisations of the $10^{13} \msun$ QSO host halo, and the shaded areas bracket the minimum and maximum mass reached at each redshift. The two panels are for two different critical LW radiation threshold $J_{\rm cr} = 300$ (left panel) and $J_{\rm cr} = 1000$  (right panel). The data point at $z = 6.4$ indicates the estimated SMBH mass of the QSO SDSS $J1148+5251$ assumed as a prototypical system and against which the model parameters have calibrated. %(see \cite{Valiante2011, Valiante2014, Valiante2016} for details). 
    The figure is taken from \cite{Sassano2021}.
    }
    \label{fig:SassanoBHevo}
\end{figure*}

More recently, the relative, combined role of different seed formation channels has been extensively investigated in the cosmological pure SAMs presented by \cite{Valiante2016, Sassano2021}, showing that the relative contribution of light, medium-weight and heavy seeds to the assembly of $z\sim 6$ SMBHs is determined by a complex interplay between chemical, mechanical and radiative feedback processes in which the mass and redshift distribution of the seeds have a fundamental impact, as they strongly affect the evolution history of the cold gas inside the QSO progenitor galaxies. 

The study presented by \cite{Valiante2016} suggests that, although both light and heavy seeds are able to form along the hierarchical assembly of a $z>6$ QSO, the growth of the SMBH in an Eddington-limited gas accretion regime relies on the formation of few heavy seeds at $z\sim 16-18$ (from 3 to 30, according to the specific halo merger tree).

In their study, \cite{Sassano2021} show that, %if the spatial fluctuations of the LW radiation and of metals and dust enrichment of the ISM are taken into account (e.g. \cite{Dijkstra2014, Salvadori2014}) 
although light and medium-weight seeds are able to form at higher rates (i.e. are more common), in particular in the SCA scenario \cite{Chon2020}, heavy seed BHs provide the necessary ``head-start" to grow a SMBH by $z>6$ if BHs grow via Eddington-limited gas accretion, independently of the critical level of LW radiation adopted to set favourable conditions for massive seed formation channels (see Figure~\ref{fig:SassanoBHevo}).

\subsubsection{Super-Eddington driven growth}\label{sec:superEddingtonGrowth}
%\subsection{Exceeding the Eddington rate limit}
As an alternative scenario for SMBH growth, several studies (both SAMs and simulations) suggest that the mass growth of seed BH (especially the light ones) may be accelerated by short (intermittent) periods of super/hyper-Eddington accretion 
%(i.e. exceeding the Eddington limit; \cite{Natarajan2014, VolonteriRees2005, Volonteri2015a, Pezzulli2016, Alexander2014, Madau2014, VolonteriSilkDubus2015, BegelmanVolonteri2017}),
(i.e. exceeding the Eddington limit; \cite{Natarajan2014, Pezzulli2016, Madau2014} and references therein), often promoted by major (gas-rich) galaxy mergers (e.g., \cite{Pacucci2015, Regan2019}), provided that the angular momentum of the gas settled in the disc around the BH is efficiently suppressed  %\cite[e.g.][]{Lupi2016, SadowskiGaspari2017, Sugimura2018}
(e.g., \cite{Pezzulli2017b, SadowskiGaspari2017}).
%Episodes of super-Eddington accretion are often promoted by major (gas-rich) galaxy mergers (e.g. \cite{Regan2019}) and enable to accrete up to $80-100 \%$ of the available gas onto the BH (e.g. \cite{Pacucci2015}).
%%%%(Haiman 2004; Yoo & Miralda-Escud ́e 2004; Shapiro 2005; Volonteri & Rees 2005, 2006; Pelupessy, Di Matteo & Ciardi 2007; Tanaka & Haiman 2009; Madau, Haardt & Dotti 2014; Volonteri, Silk & Dubus 2015).

%{\bf The following paragraph could be reduced as this argument will be probably covered in the BH accretion chapter}\\
Different super-Eddington accretion models have been proposed in the literature so far 
%(see e.g. \cite{Jiang2014} and the comprehensive review by \cite{Mayer2019}). 
(see, e.g., the comprehensive review by \cite{Mayer2019}).
One of the most widely adopted is the slim disc solution, a generalisation of the Shakura-Sunyaev thin disc solution, derived numerically integrating the Navier-Stokes equations \cite{Abramowicz1988, Sadowski2009}, that better describes super-critical regimes ($\dot{M}>0.5 \dot{M_{\rm Edd}}$). 
%Note that, following \cite{volonteri2010}, we distinguish super-Eddington and super-critical accretion regimes, with the latter not implying super-Eddington luminosity, according to the accretion disc properties). In addition, in the slim disc solution super-Eddington rates $\dot{M}\geq 100 \, \rm M_\odot/yr$ translate in a only mildly super-Eddington disc luminosity $L\sim (2-4) L_{\rm Edd}$ as in this regime, as an effect of photon trapping.
In the slim disc model, the super-Eddington flow of gas towards the BH proceeds via an optically thick accretion disc. The photons emitted by the accreting gas are trapped (i.e. can not escape the accretion flow) and are advected (dragged) inward, towards the central BH, by the accretion flow itself (see however \cite{SadowskiNarayan2016}), thus resulting in a radiative efficiency ($\epsilon$) %which indeed depends on the accretion rate
lower than the typical $10\%$ adopted in the standard optically thin disc model, $0.002\lesssim \epsilon \lesssim 0.05$ (independently of the BH spin), as shown by \cite[][]{Madau2014}.  
In other words, in this radiatively inefficient accretion regime, super-Eddington rates $\dot{M}\leq 100 \, \rm M_\odot/yr$ translate in a only mildly super-Eddington disc luminosity $L\sim (2-4) L_{\rm Edd}$ and, consequently, the quantity $1-\epsilon$ is higher than in the Eddington-limited accretion scenario.
%Black holes accreting at rates above the Eddington limit have been shown to generate powerful (collimated) jets (McKinney, Tchekhovskoy, & Blandford 2013; Sa  ̧dowski et al. 2014, 2016; Jiang et al. 2014) which may significantly disrupt the accretion flow.
A lower radiative efficiency is also supported by observational evidences up to $z>7$ QSOs %\cite[e.g.][and references therein]{KellyShen2013, Page2014, Davies2019}.
(e.g., \cite{Davies2019}, and references therein).
%Cross-ref to BH accretion chapter?

In the super-Eddington growth scenario, light seeds embedded in dense gas clouds (in atomic-cooling halos) may be able to grow rapidly, reaching masses of $\sim 10^4-10^5 \, \rm M_\odot$, comparable to that of more massive seeds, within few Myr %\cite[e.g.][]{Pacucci2015, Lupi2016}.
(e.g., \cite{Pacucci2015}).
%%\cite{Pacucci2015, Inayoshi2016, Lupi2016, Ryu2016}.
%
%% SAMs
%Semi-analytical models 
%In cosmological frameworks, SAMs show that if short, intermittent, episodes of super-Eddington growth onto light seeds can be sustained down to $z\sim 10$, mildly super-Eddington accretion rates (from $3-4$ up to $\sim 20 \dot{M}_{\rm Edd}$) is enough to trigger the formation of a SMBH by $z\sim 6$ \cite{Madau2014, Pezzulli2016}, even if the subsequent BH growth proceeded at sub-Eddington rates \cite[see e.g.][]{Pezzulli2016, Pezzulli2017b, Madau2014}

Several simulations (see, e.g., \cite{Dubois2014, Habouzit2017, Lupi2019} and references therein) show that also massive seeds of $10^5 -10^6 \, \rm M_\odot$ experience an efficient, but short (few Myr) BH accretion phase, triggered by the high gas density characterising the centre of massive galaxies in which the seeds are assumed to be located. Subsequently, the BH growth is suppressed by SN- and AGN-driven outflows for a long period of time, until the host galaxy reaches a mass of $\sim10^9-10^{10} \, \rm M_\odot$, fuelling again the BH. 

%% \cite{Zhu2020} perform a suite of 15 simulations of a £z\sim 6$ dark matter halo of 10^{13 \msun}$ exploring the impact of different masses for the seed BH %(from 10 to $10^6 \msun$) 
%(assuming either $10, 100, 10^3 10^4 10^5 $or $10^6 \msun$) different accretion modes (from near- to super/hyper Eddington) and BH feedback models (thin vs slim accretion/radiation).
Contrary to previous studies, the recent suite of zoom-in simulations of a $z\sim 6$ dark matter halos of $10^{13} \msun$ presented by \cite{Zhu2020} suggests that seeds with masses $\lesssim 10^4 \msun$ fail to grow to SMBHs by $z\sim 6$ even in the super-Eddington gas accretion scenario as a consequence of strong negative feedback (drastically reducing the accretion rate). In their simulations, the growth of SMBHs of $10^8-10^9 \msun$ is favoured if halos of $10^{10} \msun$ are seeded with $\gtrsim 10^5 \msun$ seeds that increase in mass via Eddington-limited or moderately super-Eddington accretion promoted by gas-rich galaxy mergers and self-regulated feedback.
%(in agreement with previous SAMs and numerical studies e.g. \cite{Valiante2016}).
Their results show that, although the presence of a large reservoir of gas and efficient angular momentum transport are fundamental to grow a SMBH, the final BH mass strongly depend on the adopted seed mass and radiative efficiency.

%%%% THE ROLE OF MERGERS
\subsubsection{The role of BH mergers}
Interactions and mergers with other BHs (e.g., following their host galaxies merger event) may also contribute to the growth of SMBHs across cosmic epochs, if BHs efficiently coalesce over timescales shorter than the Hubble time (e.g., \cite{Volonteri2003, %Volonteri2005, 
Sesana2007, Bonetti2019, Valiante2021}). During the merger of two galaxies, their BHs can form a binary system if they are efficiently dragged towards the centre of the newly formed system by dynamical friction against dark matter, stars and/or gas, bringing the BHs from kpc separations to pc-scale distances. At smaller scales, three-body scattering on stars and gas and/or multiple interactions with other incoming BHs (e.g., in a subsequent galaxy encounter) control the sinking of the binary down to milliparsec separation and its eventual coalescence, with the consequent emission of GWs (see, e.g., \cite{Colpi2014}).
%At lower scales, dynamical friction, three-body scattering on stars and gas particles and/or multiple interactions with other incoming BHs (e.g. in a subsequent galaxy encounter) may lead to the hardening (to sub-pc scale relative distances) and to the final coalescence of the binary (if they reach milliparsec relative separations) with the consequent emission of gravitational waves (see e.g. \cite{Colpi2014}) %(radiating away a small fraction, less than $\sim 10\%$, of the total BH mass, e.g. Healy2014). 
Thus, the formation of binary BHs, %in halo mergers 
and the evolution of the system to the coalescence phase are %on timescales shorter than the cosmic time 
multi-scale processes, challenging to be described within cosmological theoretical models. 
Different prescriptions to describe BH dynamics in cosmological frameworks have been implemented in several semi-analytical models 
%(e.g. \cite[][]{Volonteri2003, Barausse2012, Klein2016, Bonetti2019, Katz2020}, 
%\cite[e.g.][]{Volonteri2003, Bonetti2019, Barausse2012},
(e.g., \cite{Volonteri2003, Bonetti2019})
and, recently, in few large-scales simulations %\cite[e.g.][]{Volonteri2020, Kelley2017}.
(e.g., \cite{Volonteri2020}).
%\citep[associating time delays to BBHs in post-processing; see e.g.][]{2017MNRAS.464.3131K, Volonteri2020}.

BH growth via mergers is expected to have a dominant role in the formation of SMBHs (with respect to gas accretion), especially for $<10^{4-5} \msun$ BHs at high redshift 
%(e.g. \cite{Valiante2011, Valiante2016, Dayal2019, PacucciLoeb2020, Piana2021})
(e.g., \cite{Valiante2016, Dayal2019, Piana2021}). 
%and of $>10^8 \msun$ BHs at z<2 (e.g. \cite{PacucciLoeb2020}.
%For example, merger events are expected to drive the growth of $10^5 \, \rm M_\odot$ BHs at high redshift ($z>5.5$) and of $>10^8 \msun$ BHs at low redshift ($z<2$), while gas accretion dominates the evolution of SMBHs ($>10^8 \msun$) at $z>6$ \citep[e.g.][]{Pacucci2020}. 
%Although they may have an important role in the early phases of BH evolution at high redshift, BH growth via mergers only can not explain the formation of $z>7$ SMBHs and fast accretion at high rates are crucial in the growth of high-z SMBHs, almost independently of the seed mass (e.g. \cite{Sesana2007, TanakaHaiman2009, Natarajan2011}.
However, BH growth via mergers only can not explain the formation of $z>7$ SMBHs, and high accretion rates are crucial in the growth of high-redshift SMBHs, almost independently of the seed mass (e.g., \cite{Natarajan2011, TanakaHaiman2009, Sesana2007}).
In addition, the asymmetric emission of GWs in the merger process may lead to the ejection/displacement of the newly formed BH from the galaxy centre (gravitational recoil). Thus the BH may be removed from the reservoirs of dense gas from which it accretes (e.g., \cite{Haiman2004}). %Although kicked out BHs may be able to return back to the centre of the galaxy (where they can more easily accrete gas), within a Hubble time %(trigger additional coalescence in only a small fraction of cases \cite[$<20\%$][]{HoffmanLoeb07}), the high velocities imparted to the recoiling BH may have an impact on the growth of SMBHs mainly at high redshift \cite[$z>15$; ][]{TanakaHaiman2009, Pezzulli2016}.
The high velocities imparted to the recoiling BH (often called kick velocities) may have an impact on the growth of SMBHs, especially at high redshift ($z>15$; e.g., \cite[][]{TanakaHaiman2009, Pezzulli2016}).

If seed BHs pair and merge shortly after 
the formation of the binary system, %in the earliest halo-halo merger events, 
they may become loud sources of GW signals in the frequency domain of next-generation GW detectors that may reveal them back to very high redshift ($z\sim 20$; see Section~\ref{sec:futureProspects}).\\
%NB the role of feedback in the growth of heavy seeds is mentioned within the text, were necessary (e.g. as a caveat for super-critical and Eddington-limit growth)
\newline
%All together, the results of the studies mentioned above, from both semi-analytic and hydrodynamic approaches, suggests that in reality a combination of different seeding channels and growth mechanisms may be in action at high redshift, determining the formation of SMBHs. A process/channel may dominate with respect to another according to local environmental conditions.
In nature, a combination of different seeding channels and growth mechanisms may be in action at high redshift, determining the formation of SMBHs. A process/channel may dominate with respect to another according to local environmental conditions.

\subsubsection{The role of feedback}
The efficiency of BH growth is related to the available gas budget in the vicinity of the BH. However, the accretion process can be contrasted (delayed or even halted) by the negative effect of the energy released by SN explosions and the BH accretion process itself. SN- and AGN-driven feedback may affect the gas from subpc to kpc (or even Mpc) scales.
AGN feedback is expected to have a role in the (co-)evolution of BHs and their host galaxies, heating and/or blowing away the gas, preventing gas infalls and quenching star formation within the host galaxy (e.g., \cite{Dubois2013} and see Chapter 3).
%Finally it is important to mention that AGN- and/or SN-driven feedback (mechanical and radiative) may play a fundamental role in the growth of $z>6$ SMBHs.
Similarly, energetic SN feedback is expected to efficiently expel the gas from the host galaxy, limiting, or even quenching, the mass growth of seed BHs at high redshift 
%\cite[e.g.][]{Habouzit2017, Smidth2018, Dubois2015},
(e.g., \cite{Habouzit2017, Dubois2015}), that may not grow supermassive, by $z\sim 6$  remaining below $\sim 10^7 \msun$ in the most optimistic scenario (e.g., \cite{Habouzit2017}). In general, SN feedback delays the seed growth during the early phases of galaxy evolution, but efficient growth, enabling BHs to reach $\sim 10^9 \msun$, can be triggered as soon as the host galaxy becomes sufficiently massive \cite[][]{Li2007, Sijaki2009, Lupi2019}.

However, in massive dark matter halos of $10^{12}-10^{13} \msun$, presumably hosting $z>6$ QSOs, cold streams (flows of cold gas from large-scale structures) may rapidly replenish the gas reservoir, feeding the BHs, without SN/AGN feedback, as suggested, e.g., in the large-scale cosmological simulations by (e.g., \cite{DiMatteo2012}).

\section{Observational results on high-redshift QSOs}
\label{Sec_selection}

%From an observational point of view, the study of high-redshift QSOs provides unique information on the physics of SMBH growth in epochs as close as possible to the time of their formation. 

\begin{figure}
    \centering
    \includegraphics[width=\textwidth]{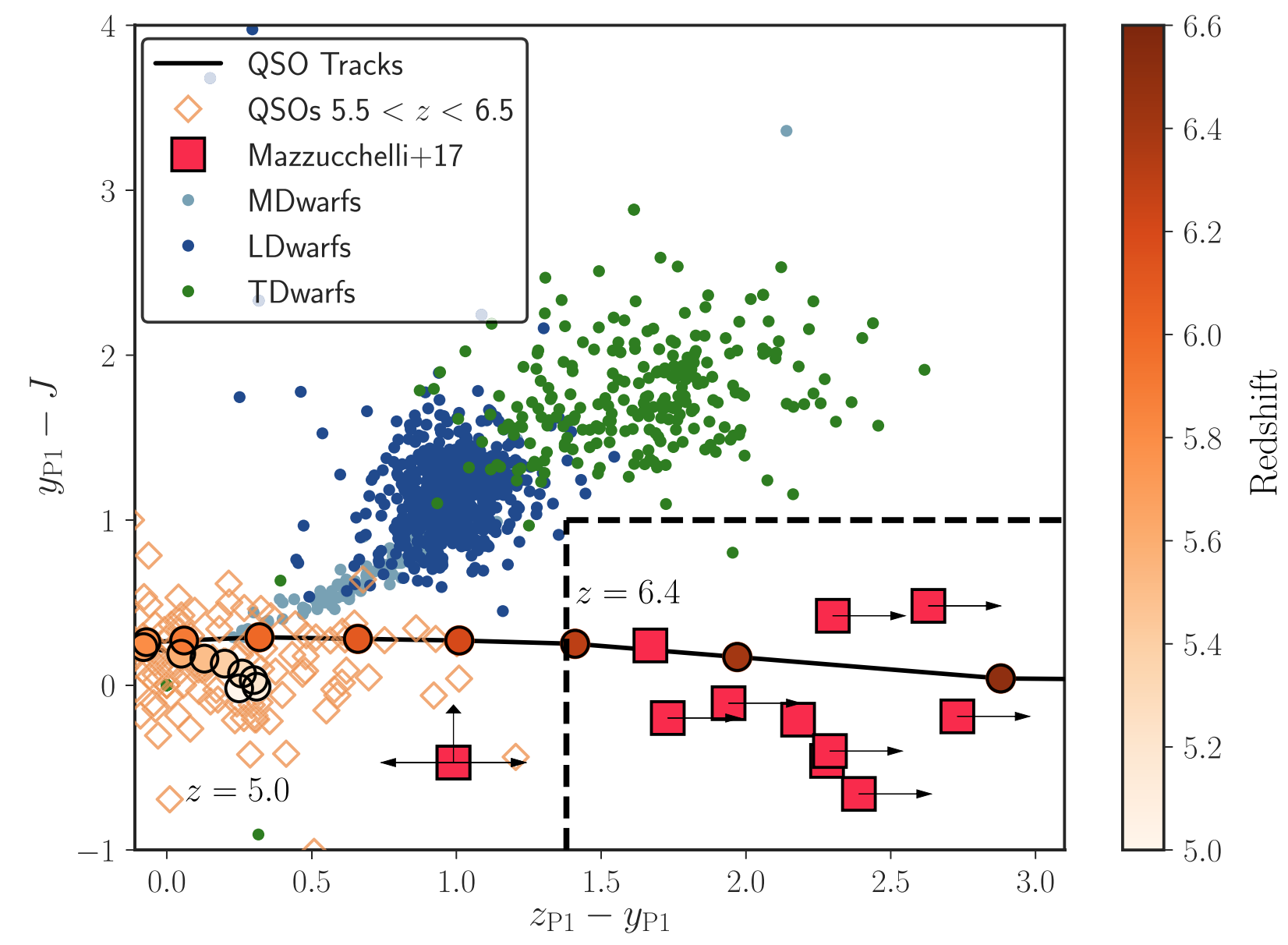}
    \caption{Example of colour-colour diagram used to select high-redshift QSOs. The location of known $5.5<z<6.5$ QSOs (open diamonds) and the expected colour track of a high-redshift QSO (black line and circles) are shown and colour-coded according to the redshift. Cyan, blue, and green circles show the location of cold stars. The dashed lines mark the colour cuts used to select a sample of $z>6.5$ QSOs (red rectangles), exploiting the flux drop between the $z$ and $y$ bands. From \cite{Mazzucchelli2017}. $\copyright$ AAS. Reproduced with permission.}
    \label{fig_selection}
\end{figure}

The identification of high-redshift QSOs relies typically on a rest-frame UV colour pre-selection of candidate samples, which aims at detecting a strong drop of the UV continuum at rest-frame wavelengths shorter than the $Ly\alpha$ emission line, due to absorption by intervening neutral hydrogen along the line of sight (i.e., the $Ly\alpha$ forest and the Lyman break). This spectral feature falls between the $i$ and $z$ photometric bands at $5.7\lesssim z \lesssim 6.5$. and shifts to redder wavelengths at $z\gtrsim6.5$ (e.g, Fig.~\ref{fig_selection}). 
Detections in the wide-field radio surveys have been used to ease the selection of a few $z\gtrsim6$ QSOs by discarding low-redshift interlopers that are not expected to be bright radio sources, allowing for the relaxation of commonly used rest-frame UV colour requirements (e.g., \cite{McGreer06, Belladitta20}). Spectroscopic follow-up campaigns are then required to discard low-redshift contaminants, mainly cold stars, and confirm the redshift of the QSOs. Multi-wavelength observations from radio to X-ray bands of high-redshift QSOs concluded that their typical physical properties (e.g., spectral energy distribution, emission-line ratios, radio-loud fraction) are similar to those of QSOs at much later cosmic times (e.g., \cite{deRosa2014, Shen2019, Vito19a, Schindler2020}), implying a general lack of strong evolution of the SMBH accretion physics. In this section, we focus on the observed X-ray properties of $z>6$ QSOs.

\subsection{X-ray observations of high-redshift QSOs}\label{Sec_observations}

X-ray emission from QSOs is thought to be produced by Comptonization of the thermal photons emitted by the accretion disc which interact with a population of hot electrons ($T\approx10^8-10^{9}$ K), generally referred to as ``hot corona" (see Chapter 2). While the geometry and physical details of this process are still highly unknown, it is widely accepted that the involved scales are of the order of a few gravitational radii, as implied by, e.g., X-ray variability studies. Therefore, parameters such as the X-ray luminosity and its relation with the UV luminosity, and the slope of the X-ray power-law emission can be used as proxies for the accretion physics in QSOs and the interplay between the accretion disc and the hot corona.

\begin{figure}
    \centering
    \includegraphics[width=\textwidth]{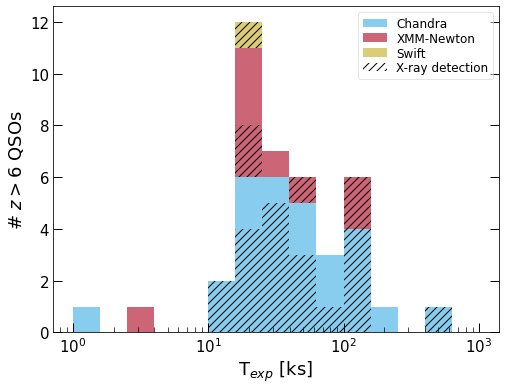}
    \caption{Histogram of the exposure times of published X-ray observations of $z>6$ QSOs (see Tab.~\ref{Tab_highz_QSOs}). Blue, red, and yellow colours refer to observations performed with the \chandra, \xmm, and \textit{Swift} observatories, respectively. Detected sources are marked with hatching patterns. }
    \label{fig_texp_hist}
\end{figure}

After their launch in 1999, \chandra and \xmm provided the required sensitivity to detect the X-ray emission from large samples of high-redshift QSOs, and, in turn, investigate their accretion properties. Starting from 2002, tens of $z>6$ QSOs have been either targeted by or serendipitously covered with X-ray observations, and their number is expected to increase in the next years.
Fig.~\ref{fig_texp_hist} presents the distribution of the exposure times of X-ray observations covering the 39 $z>6$ QSOs for which such data has been published (before February 2022, see Tab.~\ref{Tab_highz_QSOs}), including two QSOs pointed with \textit{Swift/XRT}. Most of the available datasets consist of tens of ks observations, and a few QSOs have been covered with $>100$ ks, up to the $\approx500$ ks coverage of J1030+0524 (\cite{nanni18}). In addition, the on-going {\it extended Roentgen Survey with an Imaging Telescope Array} (eROSITA\footnote{https://erosita.mpe.mpg.de/edr/}, \citep{2012SPIE.8443E..1RP,2012arXiv1209.3114M}) all-sky survey has already started contributing to increase the number of X-ray detected $z>6$ QSOs (\cite{Medveded20}).

\begin{table*}

	\caption{QSOs at $z\geq6$ covered with published X-ray observations, as of February 2022 to our knowledge.}
	\begin{tabular}{cccccccccc} 
		\hline
		\multicolumn{1}{c}{{ ID }} &
% 		\multicolumn{1}{c}{{ RA}} &
% 		\multicolumn{1}{c}{{ DEC }} &
		\multicolumn{1}{c}{{ $z$}} &
		\multicolumn{1}{c}{{ $M_{1450\ang}$}} &
		\multicolumn{1}{c}{{ Telescope}} &
		\multicolumn{1}{c}{{ $T_{exp}$}} &
		\multicolumn{1}{c}{{ Det.}} &
		\multicolumn{1}{c}{ Ref.} \\ 
		(1) & (2) & (3) & (4) & (5) & (6) & (7) \\
		\hline
VDES J002031.46$-$365341.8$^a$   & 6.834  & $-$26.47 & X   & 25   &N & \cite{Pons20}\\
PSO J006.1240+39.2219        & 6.6210 & $-$25.62 & C   & 20   &N & \cite{Wang21}\\ 
PSO PSOJ007.0273+04.9571$^b$ & 6.0008 & $-$26.34 & C   & 66   &N & \cite{Li21}\\
CFHQS J005006.67+344521.6    & 6.253  & $-$26.70 & C   & 34   &Y & \cite{Vito19a}\\
SDSSJ 010013.02+280225.9     & 6.3258 & $-$29.14 & X   & 45   &Y & \cite{Nanni17}\\
VIK J010953.13$-$304726.3    & 6.7909 & $-$25.64 & C   & 66   &N & \cite{Vito19a}\\
ATLAS J014243.73$-$332745.4  & 6.379  & $-$27.82 & S   & 21   &Y & \cite{Nanni17}\\
CFHQS J021013.19$-$045620.9  & 6.4323 & $-$24.53 & X   & 25   &N & \cite{Nanni17}\\
CFHQS J021627.81$-$045534.1  & 6.01   & $-$22.49 & X   & 29   &N & \cite{Nanni17}\\
VDES J02246.53$-$471129.4 & 6.5223 & $-$26.67 & C   & 18   & Y & \cite{Wang21}\\
PSO J036.5078+03.0498        & 6.541  & $-$27.33 & C   & 26   &Y & \cite{Vito19a}\\
VDES J02440102$-$5008537$^a$     & 6.724  & $-$26.61 & X   & 17   &N & \cite{Pons20}\\
SDSS J03033140$-$0019129     & 6.078  & $-$25.56 & C   & 2    &N & \cite{Nanni17}\\
VIK J03051692$-$3150559      & 6.6145 & $-$26.18 & C   & 50   &N & \cite{Vito19a}\\
PSO J030947.49+271757.31$^b$     & 6.10   & $-$25.1  & S   & 19   &Y & \cite{Moretti21}\\
UHS J043947.08+163415.7 & 6.5188 & $-$25.31 & X & 77 & Y &\cite{Yang22}\\
SDSS J08422943+121850.5      & 6.0763 & $-$26.91 & C   & 29   &Y & \cite{Vito19a}\\
SDSS J10302711+052455.0      & 6.308  & $-$26.99 & C   & 500  &Y & \cite{nanni18}\\
PSO J159.2257$-$02.5438$^a$        & 6.38   & $-$26.80 & X   & 20   &Y & \cite{Pons20}\\
VIK J10481909$-$010940.2     & 6.6759 & $-$26.03 & C   & 35   &N & \cite{Wang21}\\
SDSS J10484507+463718.5      & 6.2284 & $-$27.24 & C   & 15   &Y & \cite{Nanni17}\\
PSO J167.6415$–$13.4960      & 6.5148 & $-$25.57 & C   & 177  &N & \cite{Vito21}\\
ULAS J11200148+064124.3      & 7.0842 & $-$26.63 & X   & 152  &Y & \cite{Nanni17}\\
SDSS J11481665+525150.3      & 6.4189 & $-$27.62 & C   & 78   &Y & \cite{Nanni17}\\
SDSS J13060827+035626.3      & 6.0337 & $-$26.82 & C   & 126  &Y & \cite{Nanni17}\\
ULAS J13420827+092838.6      & 7.5413 & $-$26.76 & C   & 45   &Y & \cite{Vito19a}\\
CFHQS J14295217+544717.7     & 6.183  & $-$26.10 & X   & 20   &Y & \cite{Medvedev21}\\
CFHQS J15094178$-$174926.8   & 6.1225 & $-$27.14 & C   & 27   &Y & \cite{Vito19a}\\
PSO J231.6576$-$20.8335        & 6.5864 &  $-$27.14& C   & 140  &Y &  \cite{Connor20}\\
SDSS J16025398+422824.9      & 6.09   & $-$26.94 & C   & 13   &Y & \cite{Nanni17}\\
SDSS J16233181+311200.5      & 6.26   & $-$26.55 & C   & 17   &Y & \cite{Nanni17}\\
SDSS J16303390+401209.6      & 6.065  & $-$26.19 & C   & 27   &Y & \cite{Nanni17}\\
CFHQS J16412173+375520.1     & 6.047  & $-$25.67 & C   & 54   &Y & \cite{Vito19a}\\
PSO J308.0416$–$21.2339      & 6.24   & $-$26.35 & C   & 150  &Y &  \cite{Connor19}\\
PSO J323.1382+12.2986        & 6.5850 & $-$27.08 & C   & 18   &Y & \cite{Wang21}\\
HSC J221644.47$-$001650.1    & 6.10   & $-$23.62 & X   & 4    &N & \cite{Nanni17}\\
HSC J223212.03+001238.4$^c$  & 6.18   & $-$22.70 & C   & 17   &N & \cite{Li21}\\
PSO J338.2298+29.5089        & 6.666  & $-$26.14 & C   & 54   &Y & \cite{Vito19a}\\
SDSS J231038.89+185519.9     & 6.0031 & $-$27.80 & C   & 18   &Y & \cite{Vito19a}\\
VIK J23483335$-$305453.1     & 6.9018 & $-$25.84 & C   & 42   &N & \cite{Wang21}\\     
		\hline
	\end{tabular}\label{Tab_highz_QSOs}\\
\small	(1): QSO identification; (2): redshift; (3): absolute magnitude at rest-frame 1450 \AA; (4): telescope that performed the observation. C: \chandra; X: \xmm; S: \textit{Swift}. (5): exposure time in ks. For \xmm, we use the EPIC-PN exposure times excluding periods of background flaring, as reported in the papers which presented such observations. (6) X-ray detection; (7) reference for the X-ray observations. We refer to these works for complete descriptions of the objects, including the discovery papers. In case of an object observed with different instruments, here we report only the work with the most sensitive dataset. For simplicity, we refer to the collection of \cite{Nanni17} for all the observations published before 2017. We discarded two $z>6$ QSOs, HSC J084456.62+022640.5 and HSC J125437.08-001410.7, which fall at very large off-axis angles ($>10$ arcmin) in \chandra observations (\cite{Li21}). $^a$ See \cite{Pons21} for the $\aox$ values of these QSOs. $^b$ We assumed flux and luminosity of the quiescent state (see \cite{Moretti21}). $^c$ We have re-analyzed the data of these QSOs.
\end{table*}

\begin{figure*}
    \centering
    \includegraphics[width=\textwidth]{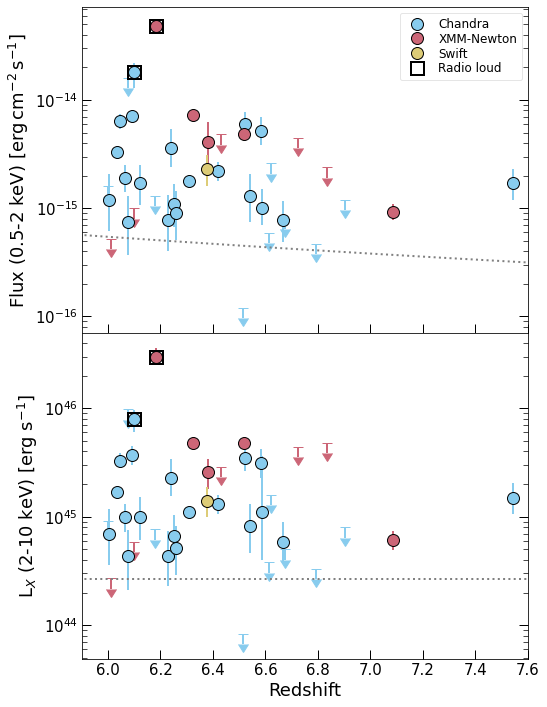}
    \caption{Soft-band flux (\textit{upper panel}) and 2--10 keV band luminosity (\textit{lower panel}) versus redshift for $z>6$ QSOs with published observations; see Fig.~\ref{fig_texp_hist} for references. The dotted line marks the X-ray flux and luminosity corresponding to the knee of the $z\approx6$ QSO UV luminosity function The conversion assumes UV power-law emission $f_{\lambda}\propto\lambda^{-0.3}$ (\cite{Banados2016}), the $\aox-L_{UV}$ relation of \cite{Just07}, and X-ray emission with $\Gamma=2$. }
    \label{fig_flux_z}
\end{figure*}

This significant observational effort led to the detection of 25 $z>6$ QSOs, typically spanning soft-band (i.e., 0.5-2 keV) fluxes of $\approx10^{-15}-10^{-14}\funits$ (Fig.~\ref{fig_flux_z}, upper panel). At $z>6$, such relatively faint fluxes are still associated with QSOs populating the bright end of the QSO luminosity function (Fig.~\ref{fig_flux_z}, lower panel). 
In support of this statement, the dotted lines in Fig.~\ref{fig_flux_z} mark the X-ray flux and luminosity corresponding to the break magnitude ($M_{1450\ang}=-24.9$; \cite{Matsuoka2018}). Therefore, current X-ray observations detected only luminous $z>6$ QSOs, that populate the bright end of the luminosity function. These objects typically are powered by the most massive or fastest accreting SMBHs. On the one hand, this implies that current X-ray investigations focus on the high-redshift QSOs which have grown, or are still growing, at the fastest rate at high redshift. Such objects are arguably the best to study in order to shed light on the physical processes responsible for the formation and fast mass building of SMBHs in the early Universe. On the other hand, luminous QSOs inevitably provide us with only a partial and biased view of the physical properties and evolution of accreting SMBHs.

It is worth noting that most of these expensive exposures aim at simply detecting the target, or at most constraining basic spectral parameters, such as the photon index, via rather crude X-ray spectral or colour (i.e., hardness ratio) analyses, often assuming a simple power-law model for the observed X-ray emission (e.g., \cite{Vito19a}). In fact, in addition to %the ``photon starvation" generally affecting X-ray astronomy and 
the large luminosity distances of high-redshift objects, X-ray observations of high-redshift QSOs sample very high rest-frame energies (i.e., $E\approx2.5-80$ keV at $z=6.0-7.5$), where the intrinsic rate of photon production in QSOs drops due to the power-law emission (i.e., $F(E)\propto E^{-\Gamma}$). Therefore, the outcome of X-ray observations of high-redshift QSOs is often limited to an estimate of the X-ray luminosity under reasonable assumptions on the spectral shape.
Typically, only the large statistical errors due to the Poisson nature of X-ray counts are reported for the X-ray flux and luminosity, and thus also derived quantities, of high-redshift QSOs, while the uncertainties associated with the assumed spectral shape are considered negligible (but see, e.g., \cite{Connor20}).
Only in some cases the target is bright enough, and the observations sufficiently sensitive, to constrain the  photon index and the level of possible absorption. 
Still, while the detailed analyses that can be performed on the X-ray spectra of low-redshift QSOs are precluded even for the most luminous objects at the cosmic dawn with currently available instrumentation (see Section \ref{sec:futureProspects} for future facilities), X-ray observations of high-redshift QSOs provide us with unique information on the physics of SMBH growth soon after the Big Bang.

\subsection{The X-ray view of accretion physics in high-z QSOs }
A common feature of the several proposed models of SMBH formation and early growth is the requirement of long periods of extremely efficient accretion on the first BH seeds (see Section 3) to explain the presence of $10^9-10^{10}\,M\odot$ BHs at $z\approx6-7$ (e.g., \cite{Onoue19}). Therefore, high-redshift QSOs are expected to be powered by SMBHs accreting at a significant fraction of the Eddington ratio ($\lambda_{Edd}\gtrsim0.1$).
The BH accretion physics is thought to be dependent on this parameter: standard models of geometrically thin accretion discs are no longer valid in the high-$\lambda_{Edd}$ regime, where different solutions have been proposed (e.g., \cite{ShakuraSunyaev1973, Abramowicz13}). This transition implies that the observational properties of QSOs should change as well approaching the Eddington limit. 
In spite of this, the typical rest-frame optical/UV continuum and emission line properties of QSOs do not appear to have undergone significant redshift evolution from the local Universe up to $z\approx7.5$ (e.g., \cite{Jiang2016, deRosa2014, Mazzucchelli2017,Shen2019}). Only recently, hints for an excess of weak emission-line QSOs (WLQs; e.g., \cite{Diamond-Stanic09}) among the population of $z\gtrsim6$ QSOs have been found by \cite{Shen2019}. Since WLQs are fast accreting systems (e.g., \cite{Marlar18}), the higher fraction of WLQs may indicate that $z>6$ QSOs are generally accreting at higher Eddington ratios than at lower redshift. Additionally, \cite{Meyer19} and \cite{Schindler2020} reported evidence for stronger C IV 1549\AA\ emission line blueshift in $z>6$ QSOs than at lower redshift, even considering luminosity-matched samples. C IV blueshift is generally associated with the presence of outflowing gas (e.g., \cite{Richards11} or even with accretion at high Eddington ratios (e.g., \cite{Coatman16}, but see also \cite{Schindler2020}).
While these observational findings can indicate widespread fast SMBH accretion at high redshift, as predicted and required by models of SMBH formation and growth, they can also be due to selection effects related to the typically high luminosity of known $z\gtrsim6$ QSOs. 

\begin{figure*}
    \centering
    \includegraphics[width=\textwidth]{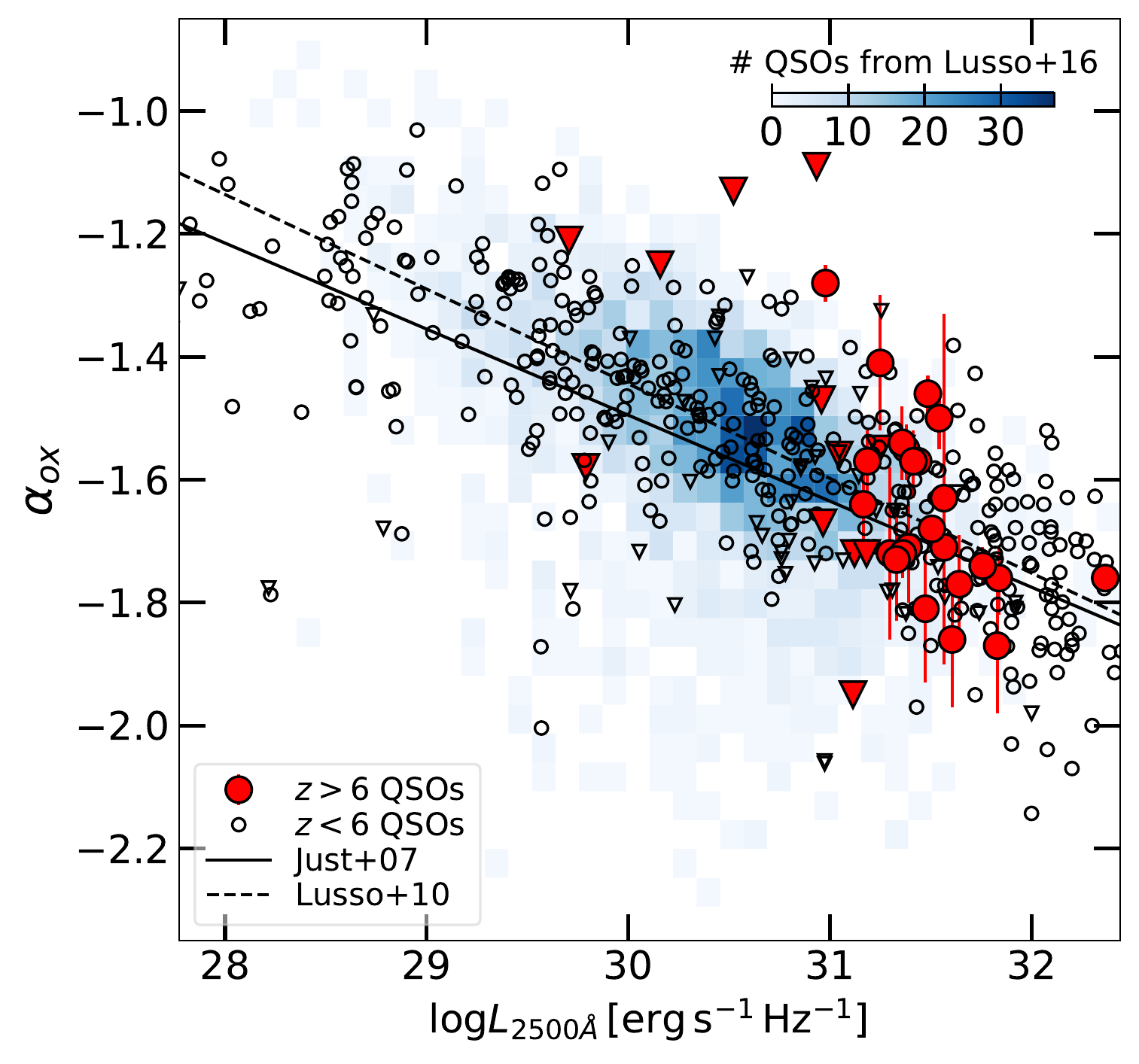}
    \caption{$\aox$ vs. UV luminosity of the high-redshift QSOs included in Tab~\ref{Tab_highz_QSOs} compared with the general QSO population (samples collected from \cite{Shemmer06,Steffen06,Just07,Lusso16,Nanni17,Salvestrini19}). Only radio-quiet QSOs are considered. Downward-pointing triangles are upper limits. The solid and dashed lines are the best-fitting relations of \cite{Just07} and \cite{Lusso10}.}
    \label{fig_aox_LUV}
\end{figure*}

\begin{figure*}
    \centering
    \includegraphics[width=\textwidth]{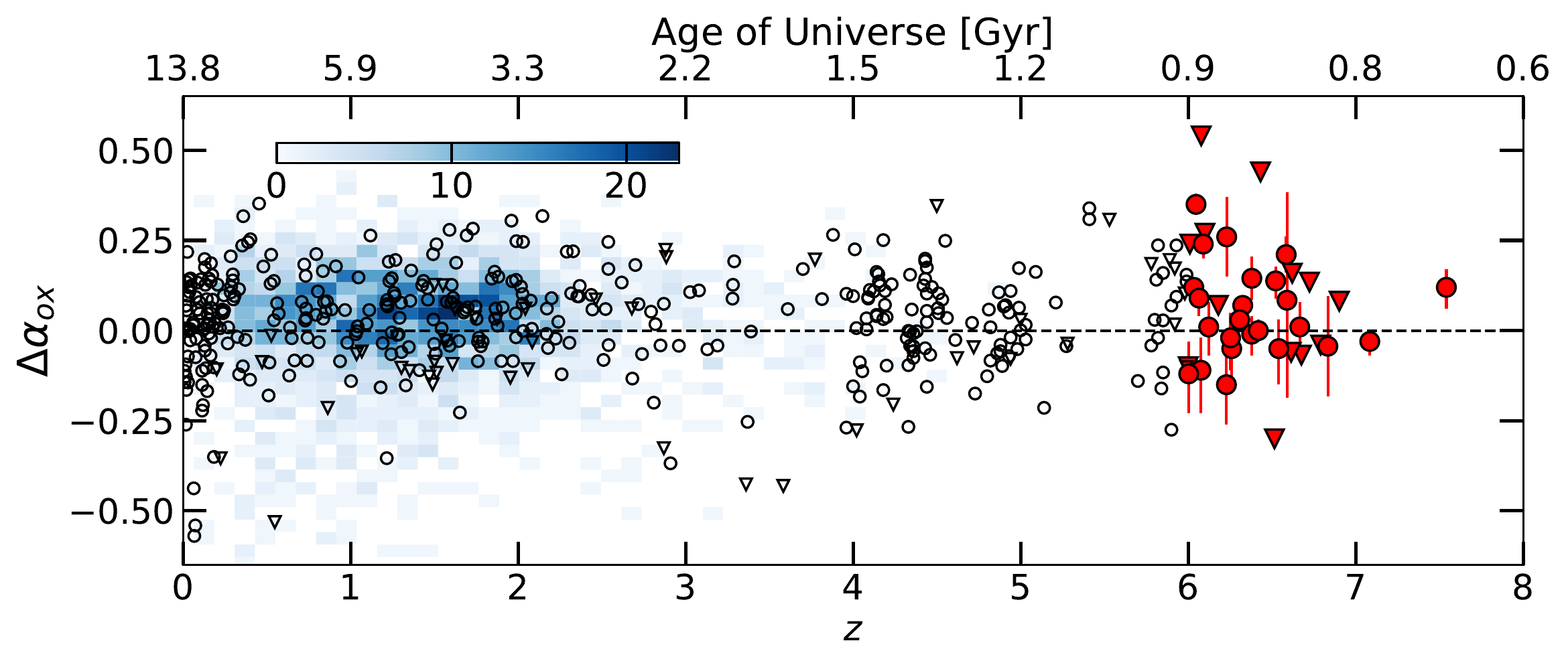}
    \caption{$\Delta\aox$ as a function of redshift . Symbols are as in Fig.~\ref{fig_aox_LUV}. The values of $\Delta\alpha_{ox}$ have been computed assuming the $\aox-L_{UV}$ relation of \cite{Just07}.}
    \label{fig_daox_z}
\end{figure*}

In this context, X-ray observations
can help us find possible intrinsic evolution of the properties of high-redshift QSOs.
For instance, the parameter $\aox=0.3838\times\mathrm{log}\frac{L_{2\,\mathrm{keV}}}{L_{2500\ang}}$ (see Chapters 1 and 2) measures the relative contribution of the rest-frame UV and X-ray emission to the total output of a QSO. Since UV and X-ray photons are produced in the accretion disc and hot corona, respectively, a change of the typical $\aox$ with parameters such as luminosity and redshift is a reliable proxy of a variation of the typical accretion physics. A well known anti-correlation exists between $\aox$ and $L_{UV}$, implying that the production of high-energy photons is more and more inhibited at increasing QSO luminosity. This behaviour indicates that the physics of the accretion disc/hot corona system changes at increasing QSO luminosity, and possibly with Eddington ratio, as expected considering models of BH accretion. Several works found that $z>6$ QSOs comply with the local $\aox-L_{UV}$ anticorrelation, implying that there is no appreciable redshift evolution of the QSO accretion physics over $\approx13$ Gyr of the Universe life (Fig.~\ref{fig_aox_LUV}; e.g., \cite{Nanni17, Banados18b, Vito19a, Wang21}).
This conclusion can be visualized better by factoring-out the $\aox$ dependence on $L_{UV}$ and defining $\Delta\aox=\aox^{obs}-\aox^{exp}$, where $\aox^{obs}$ is computed with the observed  UV and X-ray luminosities, while $\aox^{exp}$ is the $\aox$ value expected for a QSO with a given $L_{UV}$ according to best-fitting $\aox-L_{UV}$ relations (e.g., \cite{Just07,Lusso10}). Plotting $\Delta\aox$ as a function of redshift reveals that the distribution of $\Delta\aox$ is consistent with zero at all redshifts (Fig.~\ref{fig_daox_z}; see \cite{Vito19a,Wang21}), implying that the $\aox-L_{UV}$ relation, and thus the physical interplay between the accretion disc and hot corona, does not vary strongly with cosmic time.

\begin{figure*}
    \centering
    \includegraphics[width=\textwidth]{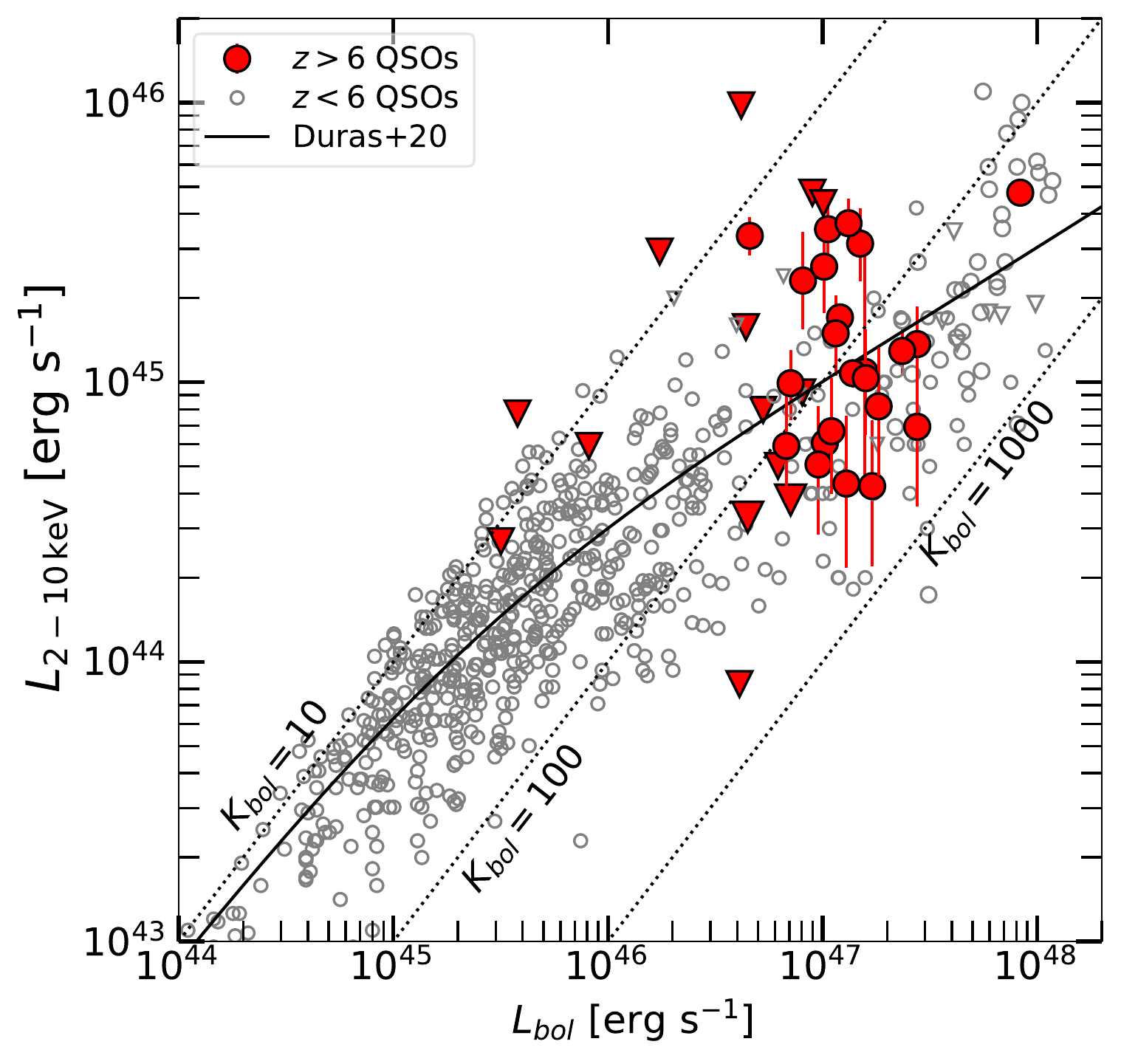}
    \caption{X-ray luminosity vs. bolometric output of $z>6$ QSOs (Tab.~\ref{Tab_highz_QSOs}) compared with QSOs at lower redshift collected from \cite{Lusso10, Nanni17,Salvestrini19,Martocchia17}. The black solid line is the best-fitting relation of \cite{Duras20}. The diagonal dotted lines marks loci of constant $K_{bol}$ as defined in the text.}
    \label{fig_Lx_Lbol}
\end{figure*}

While $\aox$ measures the relative strength of QSO emission in the UV and X-ray bands, the X-ray bolometric correction ($K_{bol}=\frac{L_{bol}}{L_X}$) measures the fraction of bolometric power emitted in the X-ray band. $\aox$ and $K_{bol}$ are clearly related: steeper values of $\aox$ correspond to higher values of $K_{bol}$, as the
contribution of the X-ray emission to the bolometric QSO luminosity decreases, and the QSO total power is increasingly dominated by the optical/UV band. 
Fig.~\ref{fig_Lx_Lbol} presents the $L_X-L_{bol}$ plane for QSOs. The relation between $L_x$ and $L_{bol}$ flattens at increasing bolometric luminosity (e.g., \cite{Duras20}), such that larger values of $K_{bol}$ are measured for more luminous QSOs. High-redshift QSOs are consistent with this general trend, once again implying that the accretion physics in QSOs changes with increasing luminosity, thus possibly Eddington ratio, but does not depend strongly on redshift, at least for what the disc-corona coupling is concerned. However, an important caveat regarding high-redshift QSOs is needed here: the spectral energy distributions (SEDs) of these systems is poorly sampled, and usually only a few rest-frame UV photometric points are available, in contrast to low-redshift QSOs. Therefore, for the vast majority of these objects, $L_{bol}$ is estimated by applying UV bolometric corrections (e.g., \cite{Decarli2018}), which have been calibrated at low redshift. Therefore, for high-redshift QSOs, the $L_X-L_{bol}$ relation is mostly a byproduct of the $\aox-L_{UV}$ anti-correlation.

\begin{figure*}
    \centering
    \includegraphics[width=\textwidth]{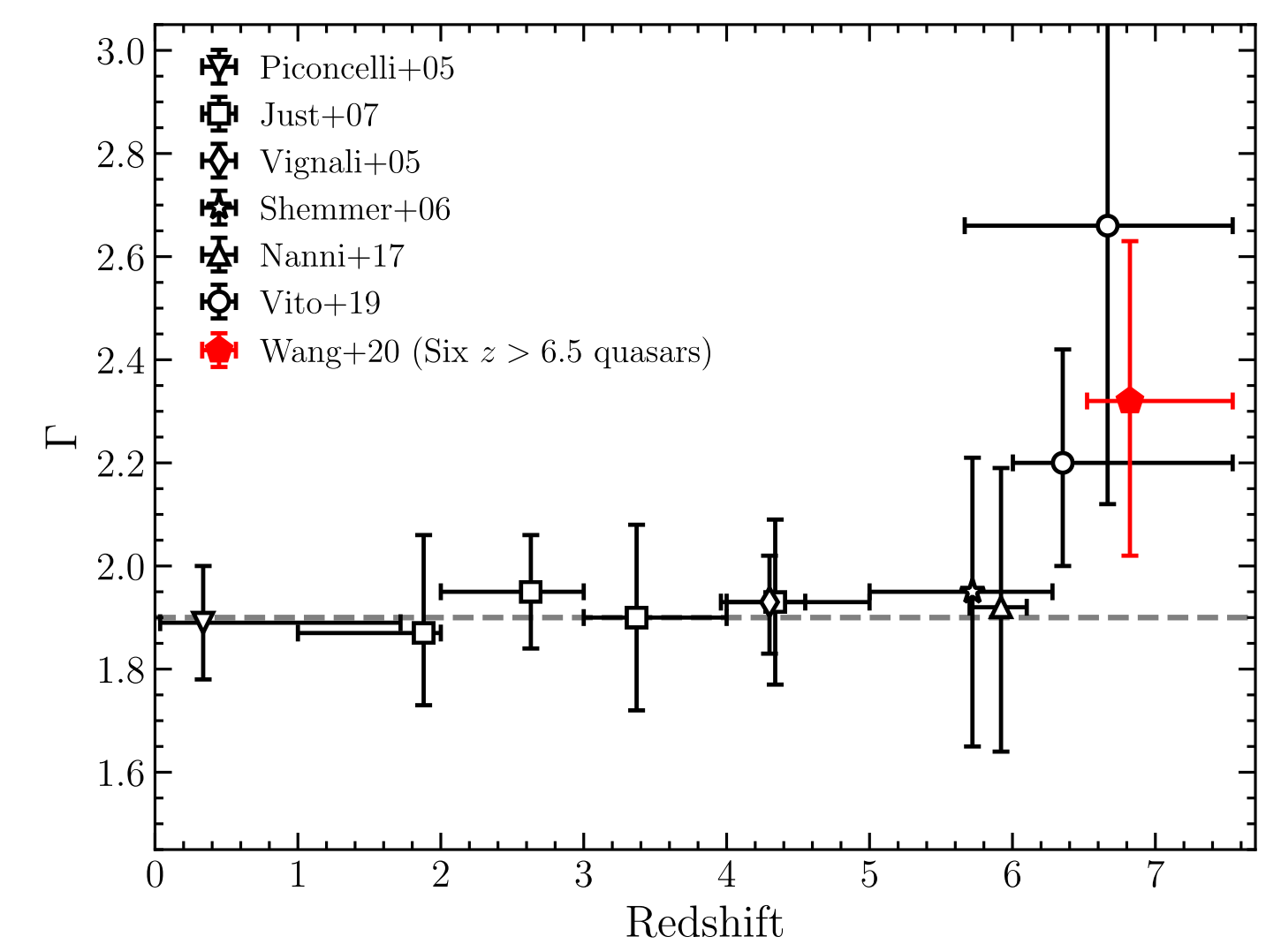}
    \caption{Average photon index of optically selected QSOs as a function of redshift. From \cite{Wang21}. $\copyright$ AAS. Reproduced with permission.}
    \label{fig_Gamma_z}
\end{figure*}

In addition to $\aox$ and $K_{bol}$, the intrinsic photon index $\Gamma$ of the X-ray power-law emission also carries information about the physical interplay between the accretion disc and the hot corona, and it is indeed often considered a proxy of the Eddington ratio (e.g., \cite{Brightman13}, but see also \cite{Trakhtenbrot17}). The average value of $\Gamma$ for optically selected QSOs has been found by several different authors (e.g., \cite{Nanni17} and references therein) to be remarkably constant ($\Gamma\approx1.9$) up to $z\approx6$ (Fig.~\ref{fig_Gamma_z}). Evidence for a possible steepening at earlier cosmic times has been presented by \cite{Vito19a,Wang21}.
However, the typically steep (i.e., $\Gamma\approx2.3$, \cite{Wang21}) X-ray spectra derived at $z=6.5-7$ do not necessarily imply a fundamental change of the physics of the hot corona in these objects, but is, once again, most probably an effect related to the high luminosities, and hence possibly high Eddington ratios, of the high-redshift QSOs which were followed-up in the X-ray band. In fact, \cite{Wang21} showed that the measured value for $z>6.5$ QSOs is in agreement with the $\Gamma-\lambda_{Edd}$ relation of \cite{Brightman13} derived at $z<2$.

\section{Quasars as cosmological probes}
\label{Quasars as cosmological probes}
%(prologue) Several techniques have employed quasars in cosmology. I will provide a (almost) comprehensive list of several methodologies in addition to the $\Lo-\Lx$ relation.
%I will make a parallel with supernovae Ia as well, since it is a well-established cosmological probe. Highlight that quasars are sampling a cosmological epoch that it is hard to probe with other astrophysical objects (apart from GRBs and surely not with the same statistics, but I will mention them as well). Quasars are complementary to other cosmological probes and provide an independent way to test cosmological models.
The potential of QSOs as cosmological probes was evident since the identification of the first QSO redshifts in 1963 \citep{s1963,oke1963,hms1963}. 
%%%%%%%%%%%%%%%%%%%%%%%%%%%%%%%%%%
Over the last four decades, several techniques employing empirical correlations between various QSO properties have been proposed to gauge cosmological parameters. Striking examples include the relations between the luminosity and spectral properties such as the emission-line equivalent width \citep{baldwin1978}, the broad-line region radius \citep{watson2011}, and the X-ray variability amplitude \citep{lafranca2014}. Nonetheless, all these correlations are either affected by a large observational dispersion (up to 0.6 dex) or applicable over a limited redshift range with the current facilities. Other methods \citep[e.g.][]{ek2002,wang2013,ms2014} could be promising, but are still more a proof of concept than a real cosmological tool. For these reasons, QSOs are not yet competitive against standard probes like SNe Ia.
%%%%%%%%%%%%%%%%%%%%%%%%%%%%%%%%%%
Recently, a new proposed method leverages on the observed non-linear relation between the QSO ultraviolet (at the rest frame 2500 \AA, $L_{\rm UV}$) and X-ray (at the rest frame 2 keV, $L_{\rm X}$) emission  (e.g., \citep{avnitananbaum79}), %,zamorani81,avnitananbaum82}, 
parametrised as $L_{\rm X}\propto L_{\rm UV}^\gamma$, with an observed slope $\gamma\simeq0.6$. 
This relation allows us an independent measurement of the QSO distances (provided an external calibration) and makes possible to extend the distance modulus--redshift relation (or the so-called Hubble-Lema\^itre diagram) of SNe Ia to a redshift range that is still poorly explored \cite[$z>2$;][]{rl15}. 
The applicability of this methodology is based on two fundamental points. Firstly, the understanding that most of the observed dispersion in the $L_{\rm X}-L_{\rm UV}$ relation is not intrinsic to the relation itself but rather due to observational issues (e.g., gas absorption in the X-rays, dust extinction in the UV, calibration uncertainties in the X-rays (e.g., \cite{lusso2019a}), variability, selection biases). In fact, with an optimal selection of sources where the \emph{intrinsic} UV and X-ray QSO emission can be measured, the observed dispersion drops from previously observed 0.4 dex to $\simeq$0.2 dex \citep{lr16,lr17}. As shown in the previous section, the slope of the $L_{\rm X}-L_{\rm UV}$ relation does not evolve with redshift up to the highest redshift where the source statistics is currently sufficient to verify any possible dependence of the slope with distance ($z\simeq4$).
A major implication of the results above mentioned is that the $L_{\rm X}-L_{\rm UV}$ relation must be the manifestation of a universal mechanism at work in the QSO engines. Nonetheless, the details on the physical process at the origin of this relation are still unknown (e.g., \cite{nicastro2000,2003MNRAS.341.1051M,lr17,arcodia2019}, and references therein). In the following, we will focus on the details and the results obtained with this technique.

\subsection{How to build a quasar Hubble diagram: the technique}
%Discuss how the distance modulus is estimated. Present the Bayesian MCMC fit of the data: likelihood and priors  . Present the cosmographic technique. Cosmological fits of the Hubble diagram
To build the Hubble diagram of QSOs, two quantities are required: the distance modulus and the redshift for a statistical significant sample of sources. The distance modulus is obtained from the luminosity distance (e.g., \cite{rl15,rl19}) defined as:
\begin{equation}
\label{dl}
\log d_{\rm L} = \frac{\left[\log F_{\rm X} -\beta -\gamma(\log F_{\rm UV}+27.5) \right]}{2(\gamma-1)}-\frac{1}{2}\log(4\pi) + 28.5,
\end{equation}
where $F_{\rm X}$ and $F_{\rm UV}$ are the flux densities (in erg s$^{-1}$ cm$^{-2}$ Hz$^{-1}$). $F_{\rm UV}$ is normalised to the (logarithmic) value of 27.5 in the equation above, whilst $d_{\rm L}$ is in units of cm and is normalised to 28.5 (in logarithm)\footnote{The normalisation values depend upon the luminosity range probed by the considered SO sample and should be tailored accordingly.}. 
The slope of the $F_{\rm X}-F_{\rm UV}$ relation, $\gamma$, is a free parameter, and so is the intercept $\beta$\footnote{The intercept $\beta$ of the $L_{\rm X}-L_{\rm UV}$ relation is related to the one of the $F_{\rm X}-F_{\rm UV}$, $\hat\beta$, as $\hat\beta(z)=2(\gamma-1)\log d_{\rm L}(z) + (\gamma-1)\log 4\pi + \beta$.}. The distance modulus, $DM$, thus follows the standard definition:
\begin{equation}
DM = 5 \log d_{\rm L} - 5 \log (10\,{\rm pc}),
\end{equation}
whilst its uncertainty, $d DM$, can be obtained through the propagation of the error on the measured quantities (assuming that all the parameters are independent):
\begin{equation}
\label{dDM}
\begin{aligned}
d DM = \frac{5}{2(\gamma-1)} \left[ \left(d\log F_{\rm X}\right)^2 + \left(\gamma d\log F_{\rm UV}\right)^2 + \left(d\beta\right)^2 + \right.\\
\left.
\left( \frac{d\gamma \left[\beta+\log F_{\rm UV}+27.5-\log F_{\rm X}\right]}{\gamma-1} \right)^2\right]^{1/2},
\end{aligned}
\end{equation}
where $d\log F_{\rm X}$ and $d\log F_{\rm UV}$ are the logarithmic uncertainties on $F_{\rm X}$ and $F_{\rm UV}$, respectively.

The likelihood function, $LF$, that is then utilised to fit the Hubble diagram is then defined as:
\begin{equation}
\label{lf}
\ln LF = - \frac{1}{2} \sum_i^N\left(\frac{(y_i-\psi_i)^2}{s_i^2} - \ln s^2_i\right)
\end{equation}
where $N$ is the number of sources, $s_i^2 = d y_i^2 +\gamma^2 d x_i^2 + \exp(2\ln\delta)$ takes into account the uncertainties on both the $x_i$ ($\log F_{\rm UV}$) and $y_i$ ($\log F_{\rm X}$) parameters of the fitted relation. The parameter $\delta$ represents what is left in the dispersion of the relation once it is marginalised over all the parameters and thus it can be considered a proxy of the \textit{intrinsic} dispersion\footnote{$\delta=0$ means that all the observed dispersion is intrinsic.}. The variable $\psi$ is the modelled X-ray monochromatic flux ($F_{\rm X,\,mod}$), defined as:
\begin{equation}
\label{model}
\psi = \log F_{\rm X,\,mod} = \beta + \gamma(\log F_{\rm UV}+27.5) +2(\gamma-1)(\log d_{\rm L,\,mod} -28.5),
\end{equation}
and is dependent upon the data, the redshift and the model (cosmological or parametric, see, e.g.,\citep{bargiacchi2021} for further details) assumed for the distances (e.g., $\Lambda$CDM, $w$CDM or a polynomial function). 
Additionally, the QSO distances obtained through the method described above are not absolute and should be calibrated by making use of the distance ladder through, for instance, already calibrated SNe Ia. Such a cross-calibration parameter $k$ could be obtained by a simultaneous fit of SNe Ia and QSOs.
Finally, it should be noted that this technique does not provide an estimate of the Hubble constant $H_0$ as it is degenerate with $k$, so it can assume any arbitrary value (e.g., \cite{lusso19b,lusso2020,bargiacchi2021}).

\subsection{How to build a quasar Hubble diagram: required measurements and sample selection.}
\label{measurements and sample selection}
\begin{figure}
    \includegraphics[width=\columnwidth]{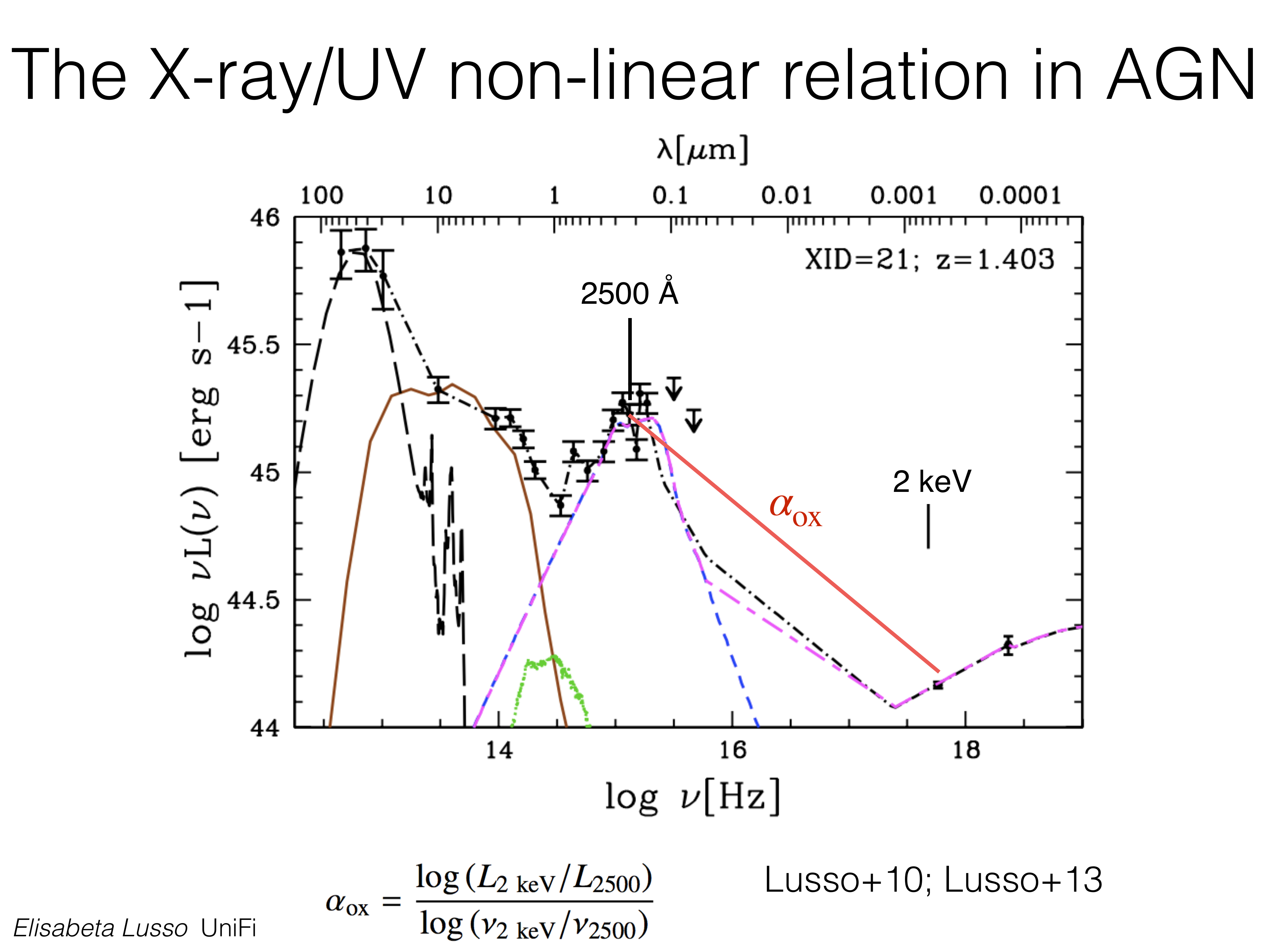}
    \caption{Example of full AGN SED from far-IR to X-rays at redshift 1.4 from the XMM-COSMOS survey. The rest-frame data, used to construct the total SED (shown with a black dot-dashed line), are represented with black points. Various lines represent the different models adopted to describe the data. Specifically, the black long-dashed, brown solid, green dotted, and blue dashed lines correspond to the best-fit starburst, hot-dust from reprocessed AGN emission, host-galaxy, and the empirical disc templates, respectively. The magenta short-long dashed line is the best-fit empirical disc template with the inclusion of X-rays.
    The rest-frame 2500 \AA\ and 2 keV are marked, as well as a representation of the $\aox$ index for this object. This figure is adapted from \cite{lusso2013}. }
    \label{fig:sed}
\end{figure}

%Describe how to compute the needed parameters from observations.
To construct a Hubble diagram for QSOs, the required observational parameters are the rest-frame flux density at both 2500 \AA\ and 2 keV (see equation~\ref{dl}) and the redshift.
A detailed spectroscopic analysis at both the UV and X-ray energies is preferred but can be carried out only for a relatively small number of sources. Therefore, the currently published QSO sample still heavily relies on broadband photometry at both UV and X-rays, and so the computation of both $F_{\rm X}$ and $F_{\rm UV}$ is through their photometric rest-frame SED.  
To compile the QSO SEDs, multiwavelength data from radio to UV should be considered, where Galactic reddening must be taken into account in the optical/UV by utilising the selective attenuation of the stellar continuum (e.g., \cite{F99}), along with the relative Galactic extinction (e.g., \cite{schlegel98}) for each object. Most of the relevant broadband information, as well as the spectroscopic redshifts, are compiled in the online QSO catalogues of several surveys such as SDSS \citep{lyke2020}. The reader may refer to \cite{richards2006,elvis2012,lusso2013} for a detailed description of how a spectral energy distribution can be built from broad band photometry. An example of a full broad band SED (from far-infrared to the X-rays) for an AGN in the XMM-COSMOS survey is presented in Figure~\ref{fig:sed} (see also Figure~1 in Chapter 1). The various lines represent the different models adopted to describe the data. Specifically, the best-fit starburst in the far-infrared (with a peak emission at $\lambda\simeq50-80\,\mu$m), hot-dust from reprocessed AGN emission ($\lambda\simeq 10\,\mu$m), host-galaxy  ($\lambda\simeq 1\,\mu$m), and the empirical disc templates ($\lambda\simeq 0.2-0.6\,\mu$m), respectively (see \cite{lusso2013} for further details on the SED modelling). The rest-frame 2500 \AA\ and 2 keV are marked, as well as a representation of the $\aox$ index for this object.

% 2500 Å luminosity
By compiling a broad photometric coverage, the rest-frame luminosity at 2500 \AA\ can be computed via interpolation for the majority of the QSOs whenever the reference frequency is covered by the SED. Otherwise, the value can be extrapolated by considering the slope between the luminosity values at the closest frequencies, for instance.
% 2 keV luminosity
To obtain the rest frame luminosities at 2 keV, a detailed X-ray spectral analysis of large samples of QSOs is, again, impractical. Therefore, to compute the rest-frame 2 keV monochromatic flux, a photometric approach is a viable solution for large number of sources. Briefly, the rest-frame 2 keV flux densities and the relative (photometric) photon indices, $\Gamma_X$ (along with their 1$\sigma$ uncertainties), can be derived from the 0.5--2 keV (soft) and the 2--10 keV (hard) fluxes which are tabulated in many online X-ray catalogues (i.e. XMM-\textit{Newton}\footnote{http://xmmssc.irap.omp.eu/Catalogue/4XMM-DR11/4XMM\_DR11.html} and \textit{Chandra}\footnote{https://cxc.cfa.harvard.edu/csc/}). The interested reader should refer to \cite{rl19,lusso2020} for further details on the computation of these quantities.

%%%%%%%%%%%%%%%%%%%% sample selection
A QSO sample that can be employed for cosmological measurements should comply with a series of criteria that are aimed at reducing possible contaminants/systematics that may prevent the determination of the \emph{intrinsic} X-ray (corona) and UV (disc) luminosity and bias the final sample. Whilst the interested reader should refer to 
\cite{lusso2020} for a complete description and implementation of the filters (see also \citep{rl15,lr16,rl19}) aimed at obtaining the final ``best" sample for a cosmological analysis, a brief overview is provided below.
The main possible sources of contamination that may affect the flux measurements are dust reddening and host-galaxy contamination in the optical/UV and gas absorption in the X-rays. Moreover, any flux-limited sample is biased towards brighter sources at high redshifts and this should be more relevant to the X-rays, since the relative observed flux range is narrower than in the optical/UV.
Specifically, AGN with an average X-ray intensity close to the flux limit of the observation will be observed only in case of a positive fluctuation. This introduces a systematic, redshift-dependent bias towards high fluxes, known as {\it Eddington bias}, which has the effect to flatten the $F_{\rm X}-F_{\rm UV}$ relation.
%Samples where only detected sources are included might thus be affected such a bias.
\EL{Samples that include only quasars with a detection at X-ray energies might thus be affected such a bias. This is true even when variability is accounted for, in the case detections only are considered.} One possibility is to include censored data in the analysis. Yet, the investigation of both the $F_{\rm X}-F_{\rm UV}$ and the distance modulus--redshift relations is far from trivial, since it strongly depends upon the weights assumed in the fitting algorithm. Therefore, one needs to find an alternative method to obtain an (almost) unbiased sample (see, e.g., Appendix~A in \citep{lr16}).
Finally, QSOs with bright radio emission and/or broad absorption lines (BALs) should also be neglected. Synchrotron emission in a radio QSO can also include a bright X-ray component that adds to the one of the corona, whilst the strong absorption features observed in BALs, and usually attributed to winds/outflows, hamper a robust measurement of the QSO continuum in the UV.

The most up-to-date broad-line QSO sample (available online) considered for cosmological purposes is composed by 2,421 QSOs spanning a redshift interval $0.009\leq z\leq7.52$, with a mean redshift of 1.442 \cite[][]{lusso2020}.

\subsection{Cosmological constraints from the quasar Hubble diagram}
% discuss the tension between the fit of the quasar Hubble diagram and the standard concordance model ($\Lambda$CDM)
%-------------------------------------- 
   \begin{figure}[t!]
   \centering
   \includegraphics[width=\columnwidth]{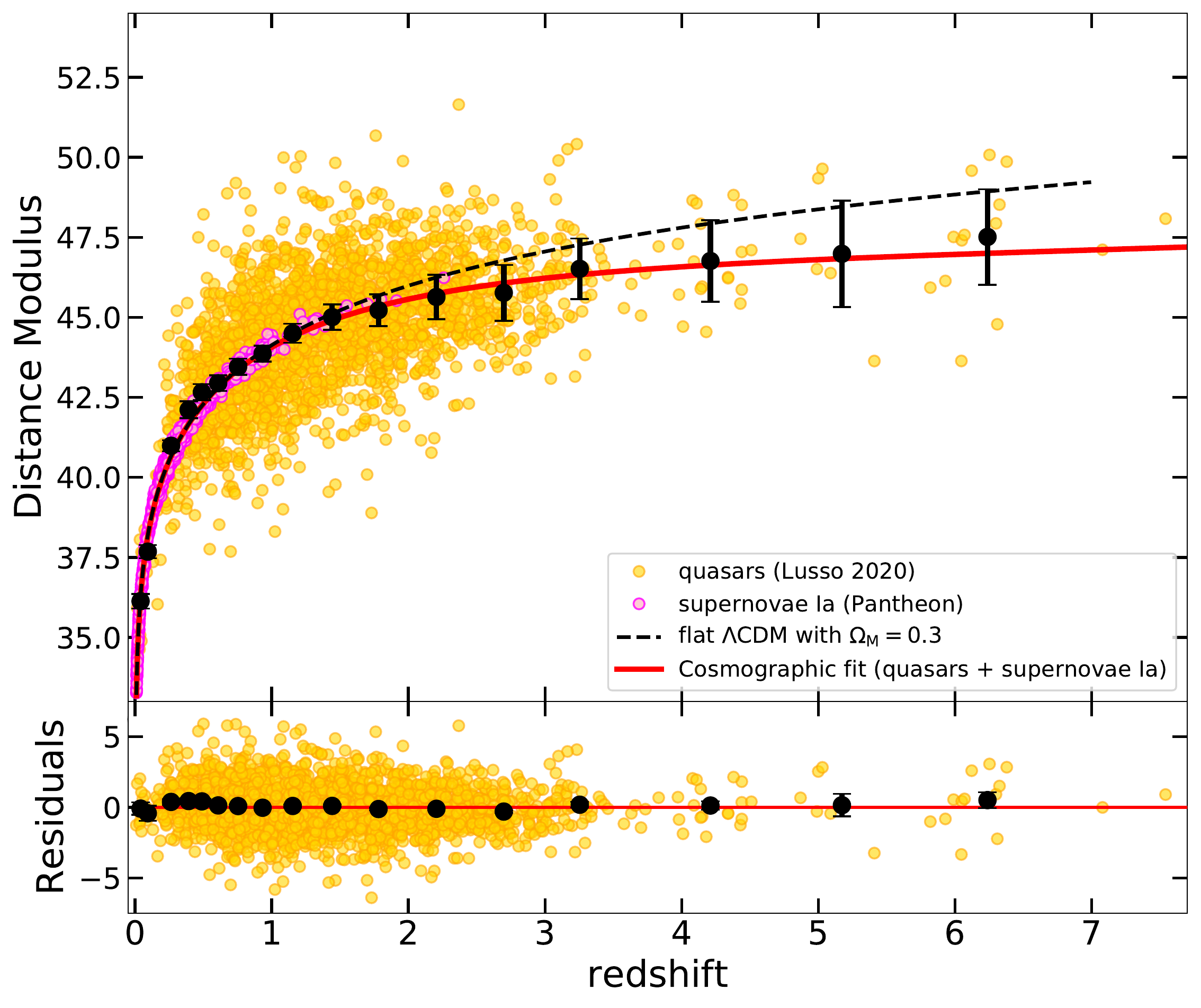}   
   \caption{Hubble diagram of QSOs \citep{lusso2020} and SNe Ia \citep[\textit{Pantheon}][magenta points]{scolnic2018}. The red line represents a fifth order cosmographic fit of the data, whilst the black points are averages (along with their uncertainties) of the distance moduli in narrow (logarithmic) redshift intervals. The dashed black line shows a flat $\Lambda$CDM model fit with $\Omega_{\rm M}=0.3$. The bottom panel shows the residuals with respect to the cosmographic fit and the black points are the averages of the residuals over the same redshift intervals. Figure adapted from \cite{lusso2020}.}
              \label{hubbleclean}
    \end{figure}
%-------------------------------------- 
Measurements of the expansion rate of the Universe based on a Hubble diagram of QSOs combined with SNe Ia show a deviation from the concordance model at high redshifts ($z>1.4$), with a statistical significance of $\sim3-4\sigma$ \citep{rl19}. Figure~\ref{hubbleclean} presents the Hubble diagram for the most up-to-date samples of QSOs \citep{lusso2020} and SNe Ia from the {\it Pantheon} survey \citep{scolnic2018}. 
The best cosmographic fit (performed through a Monte Carlo Markov Chain analysis) is shown with the red line \citep[see][for more details on the cosmographic technique]{bargiacchi2021}, whilst black points are the means (along with the uncertainty on the mean) of the distance modulus in narrow (logarithmic) redshift intervals, plotted for clarity purposes only. 
It is evident from this figure the potential of QSOs to extend the distance modulus-redshift relation in a cosmic epoch ($z>2$) that it is poorly sampled by any other cosmological probe. QSOs are indeed key to discriminate amongst various cosmological models that are instead degenerate at low redshifts, and to improve the constraints on a possible evolution of the dark energy parameter with time.
Additional data are required at high redshifts where very few QSOs are available with both X-ray and UV observations that fulfil the selection criteria discussed in section~\ref{measurements and sample selection}. 

\begin{figure}[t!]
   \centering
   \includegraphics[width=0.8\columnwidth]{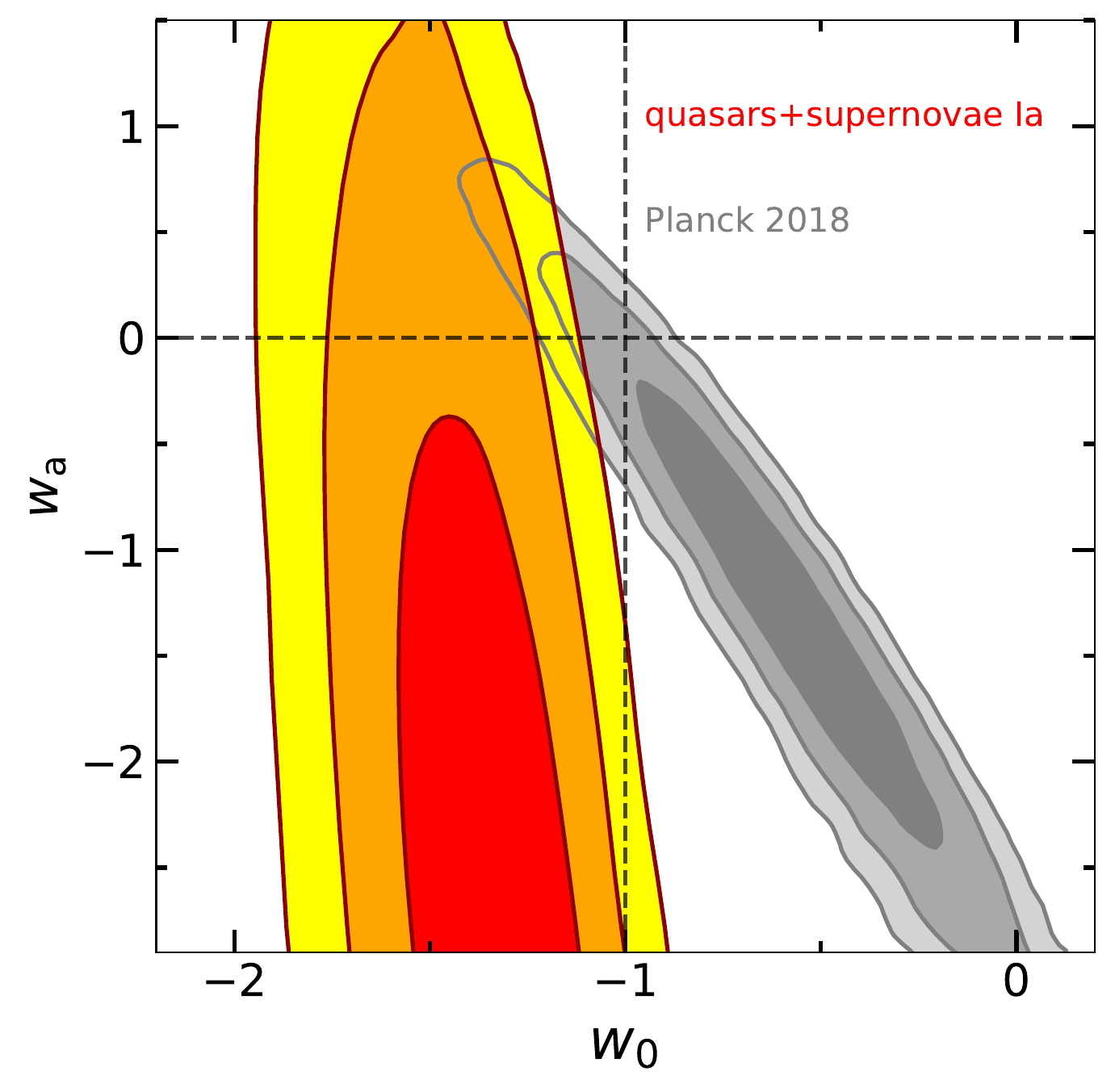}   
   \caption{Marginalized posterior distributions (1, 2 and 3$\sigma$) of the ($w_0$,$w_{\rm a}$) parameters for the combined QSOs \citep{lusso2020} and SNe Ia \citep{scolnic2018} samples (red-orange-yellow contours). The constraints from the combination of \textit{Planck} TT,TE,EE+lowE+lowl (i.e. baseline parameter analysis) with the inclusion of baryonic acoustic oscillations are also shown \citep[grey contours][]{planck2018}. The dashed lines mark the point corresponding to the $\Lambda$CDM model. The resulting ($w_0$,$w_{\rm a}$) for the combined QSOs + SNe are statistically consistent with the phantom regime ($w>-1$) and at variance with the $\Lambda$CDM model at more than the $3\sigma$ statistical level. Image adapted from \cite{2022arXiv220107241M}. }%copyright by .... \textbf{that depends upon who publishes first...}}
              \label{w0wa}
    \end{figure}

The resulting constraints on the cosmological parameters that represent the evolution of the equation of state of dark energy in a $w_z$CDM cosmological model, namely $w_0$ and $w_{\rm a}$, are shown in Figure~\ref{w0wa} for the combined latest QSO and SNe samples. The statistical contours for the analysis that jointly considers QSOs and SNe is consistent with the phantom regime ($w>-1$) and is at variance with the $\Lambda$CDM model at more than the $3\sigma$ statistical level. The constraints from the combination of \textit{Planck} TT,TE,EE+lowE+lowl (i.e. baseline parameter analysis\footnote{https://wiki.cosmos.esa.int/planck-legacy-archive/index.php/CMB\_spectrum\_\%26\_Likelihood\_Code}) with the inclusion of baryonic acoustic oscillations are also shown \citep{planck2018}. The detailed discussion of the cosmological implications of this deviation and its statistical significance is discussed at length by \cite[][]{rl19,lusso19b}.
The cosmological information given by these samples may shed new light on the possible evolution of the dark energy with time and, if confirmed independently, would suggest a profound revision of our current understanding of cosmic evolution. 
% NOTE add Bargiacchi et al 2021, submitted, once it really is...

%-------------------------------------- 
%-------------------------------------- 

\section{The unexplored black hole Universe} \label{sec:unexploredBHs}
%\FV{Possiamo provare a unire questa sezione con quella che sotto chiamo "The missing QSO population" (o qualcosa del genere)? E.g., fare una sezione unica in cui parliamo di tutti quegli oggetti che dovrebbero esserci e non vediamo (ancora). Oppure meglio lasciare separata la parte di BH mancanti di "bassa massa" (questa section) e "bassa luminosita/alto oscuramento" (la mia sezione)?}\RV{si potrebbe essere una buona idea. Quando avrò finito di scrivere e sistemare uesta sezione ne discutiamo alla luce di quello che ci inserisco ;-)}
Theoretically, the mass spectrum of astrophysical BHs is potentially unlimited. To date, however, observational evidences of their existence are limited to two flavours, the stellar-mass BHs ($\lesssim 100 \msun$) and the supermassive ($\gtrsim 10^5 \msun$). 
%%%% PARAGRAFO ELIMINATO DA RV il 15/11/2021
%The observation of $\sim 10^9 \, \rm M_\odot$ SMBHs at $z>6$ suggests that we may expect (and observe) a population of BHs with a much lower mass, forming at even higher redshift $(z>10-20)$, from which the giants have grown. In addition, only a small fraction ($<20\%$) of all the seeds formed at high redshift are expected to later grow into SMBHs by $z\sim 6$ (e.g. \cite{Valiante2016, Sassano2021}. Some of these may remain almost unaltered (ungrown due to scarce gas accretion and limited number of interactions with other BHs) for long periods of time, thus being observable at lower redshifts (e.g. in dwarf galaxies).

Currently, the faint-end tail of the AGN luminosity function at $z\sim 6-7$ has been sampled down to absolute magnitude of $M_{1450}=-22$ mag \citep{Matsuoka2018}. However, the whole population of high-redshift (accreting) seed BHs ($10^2-10^6 ~\rm M_\odot$) is still elusive to current electromagnetic (EM) facilities and remain undetected, probably because \textit{(ii)} they are too faint for current EM facilities (below their sensitivity) and/or \textit{(ii)} their active fraction at high redshift is low ($\sim 0.1\%$ at $z>7$; e.g., \cite{Pezzulli2017a}) and/or \textit{(iii)} their number density is relatively low, as in the case of heavy seeds (e.g., 
%\cite{Habouzit2016c, Valiante2016, valiante2017, Cowie2020}, 
\cite{Habouzit2016c, Valiante2016, Cowie2020}, but see \cite{Wise2019}).
%The non-detection of faint high-z AGN may be a consequence of their low active fraction \citep[$\sim 0.1\%$ at $z>7$][]{Pezzulli17} and/or of their relatively low number density \citep[][but see \citealt{Wise19}]{Habouzit16, Valiante16,Cowie20}.
An additional complication may be represented by obscuration from gas and dust (observed to be abundant even at high redshift) that may prevent the discovery of accreting BHs (see Section \ref{sec:missingQSOs}), as they may evolve buried in dense dusty gas clouds.
On the other hand, the LIGO/VIRGO collaboration extended the exploration of stellar-mass BHs detecting the gravitational signal from merging binaries up to a total remnant mass of $\sim 150 \, \rm M_\odot$ (GW190521; \cite{LigoVirgo2020}), opening a new window on the BH Universe.
Thus, observationally, the nature of the earliest seed BHs still remain unproved. %unconstrained
Note that the whole population of BHs with mass in the range $10^2-10^6 \, \rm M_\odot$ is often refereed as Intermediate-Mass Black Holes (IMBHs).%, the ones filling the gap between stellar-mass ($10^2 \, \rm M_\odot$) and supermassive BHs ($>10^6 \, \rm M_\odot$).

To date, the only insights on IMBHs of $\sim 10^5 \, \rm M_\odot$ can be inferred at lower redshift, observing BHs hosted in local dwarf galaxies %\cite{Reines2015, Baldassare2015, Mezcua2016, Mezcua2018}. 
(see, e.g., \cite{ReinesComastri2016} for a review). 
These galaxies are expected to evolve almost unaltered by mergers and accretion processes (e.g., \cite{vanWassenhove2010}). Thus, their central BHs could be the (ungrown) relics of the earliest population of seed BHs, as mentioned above, still showing strong observational signatures of seed formation (e.g., \cite{Habouzit2016c}).

Detailed observations of a large sample of the highest redshift QSOs and their fainter/less massive counterparts are necessary to test the theoretical models reviewed in Section~\ref{sec:seedGrowth} and to improve our understanding of SMBH formation and evolution. 
For a detailed discussion on seed BHs observational diagnostics and IMBHs searches, we refer the reader to the recent reviews by \cite{inayoshi2020} and \cite{Greene2020}.

\subsection{The missing QSO population} \label{sec:missingQSOs}

As discussed in Section \ref{Sec_selection}, known samples of high-redshift QSOs are limited to luminous, optically selected systems. Based on our knowledge of the population of accreting SMBHs at later cosmic times, luminous type-1 QSO represent only a small fraction of the overall AGN population. The bulk of the AGN population, which is comprised of low-luminosity (i.e., $L<L_*$, where $L_*$ is the break luminosity of the AGN X-ray luminosity function, XLF) and often obscured systems, is currently completely precluded to observational investigations. This bias leaves a big hole in our knowledge of accreting SMBHs at the cosmic dawn, and hampers a fair comparison with results from theoretical models and numerical simulations, which instead suggest that the early phases of SMBH growth could occur in heavily obscured conditions, and with a total radiative output significantly lower than the typical luminosity of known $z\gtrsim6$ QSOs (e.g., \cite{Pacucci2015, Valiante2018a}). Only candidate type-2 QSOs (i.e., objects with narrow emission lines and weak or undetected rest-frame UV continuum) have been reported at $z\gtrsim6$ in the SHELLQ survey down to $M_{1450\ang}\approx-22$ (e.g., \cite{Matsuoka18,Matsuoka2019b}), but their nature (i.e., type-2 QSO or bright Ly$\alpha$ emitting galaxies) has been not be determined clearly yet (\cite{Onoue21}). Detection of strong X-ray emission from these objects would confirm their type-2 nature with high confidence. Recently, two hard X-ray source candidates, indicative of heavy obscurations, at $z>6$ have been reported in the literature.
Both of them have been identified with satellite galaxies interacting with optically selected QSOs.
\cite{Vito19b} presented the tentative detection of a faint hard X-ray source consistent with the position of a satellite galaxy close to a $z=6.5$ QSO detected with ALMA and HST, but subsequent \chandra follow-up did not confirm the detection (\cite{Vito21}). \cite{Connor19} presented \chandra observations of a QSO/galaxy merging system, and discussed the possible faint detection of the merging galaxy in the hard X-ray band.

While luminous and rare QSOs provide useful insights into the physics of extremely fast and efficient accretion required to form $10^8-10^9\,M_\odot$ SMBHs in a few hundred million years, low-luminosity AGN represent a less extreme, but much more abundant, population of accreting SMBH. Therefore, they carry important information about the more common physical processes responsible for the formation of most BHs in the early Universe. In particular, measuring the space density of low-luminosity AGN at high redshift would provide observational constraints to cosmological simulations which assume different prescriptions for BH seeding (e.g., \cite{Habouzit2017}).

Deep X-ray surveys, like the 7 Ms \textit{Chandra} Deep Field-South \cite{Luo17}, the 2 Ms \textit{Chandra} Deep Field-North (e.g., \cite{Xue16}) and COSMOS-Legacy (e.g., \cite{Civano16}) are arguably the best tools to detect and study low-luminosity AGN at high redshift. In fact, relatively strong ($L_{X}\gtrsim10^{42}\,\mathrm{erg\,s^{-1}}$) X-ray emission is a reliable indication of on-going nuclear activity, as it does not suffer from dilution by host-galaxy emission (e.g., \cite{Xue17}). Moreover, in contrast to optical/UV observations, the large penetrating power of X-ray photons allows AGN to be detected even when obscured by substantial hydrogen column densities, especially at high-redshift, as high-energy, and thus more penetrating, X-ray photons are redshifted into the bandpass of current X-ray telescopes. However, the redshift confirmation of X-ray selected sources still requires the spectroscopic identification of the optical counterpart. This task is particularly difficult at high redshift, due to the faint optical magnitudes of X-ray sources. Therefore, it is not surprising that sizeable samples of X-ray selected AGN are available only up to $z\approx5$ even in the deepest X-ray surveys covered with sensitive multiwavelength observations (e.g., \cite{Vito14, Marchesi16, Vito18,Cowie2020}.

\begin{figure*}
    \centering
    \includegraphics[width=\textwidth]{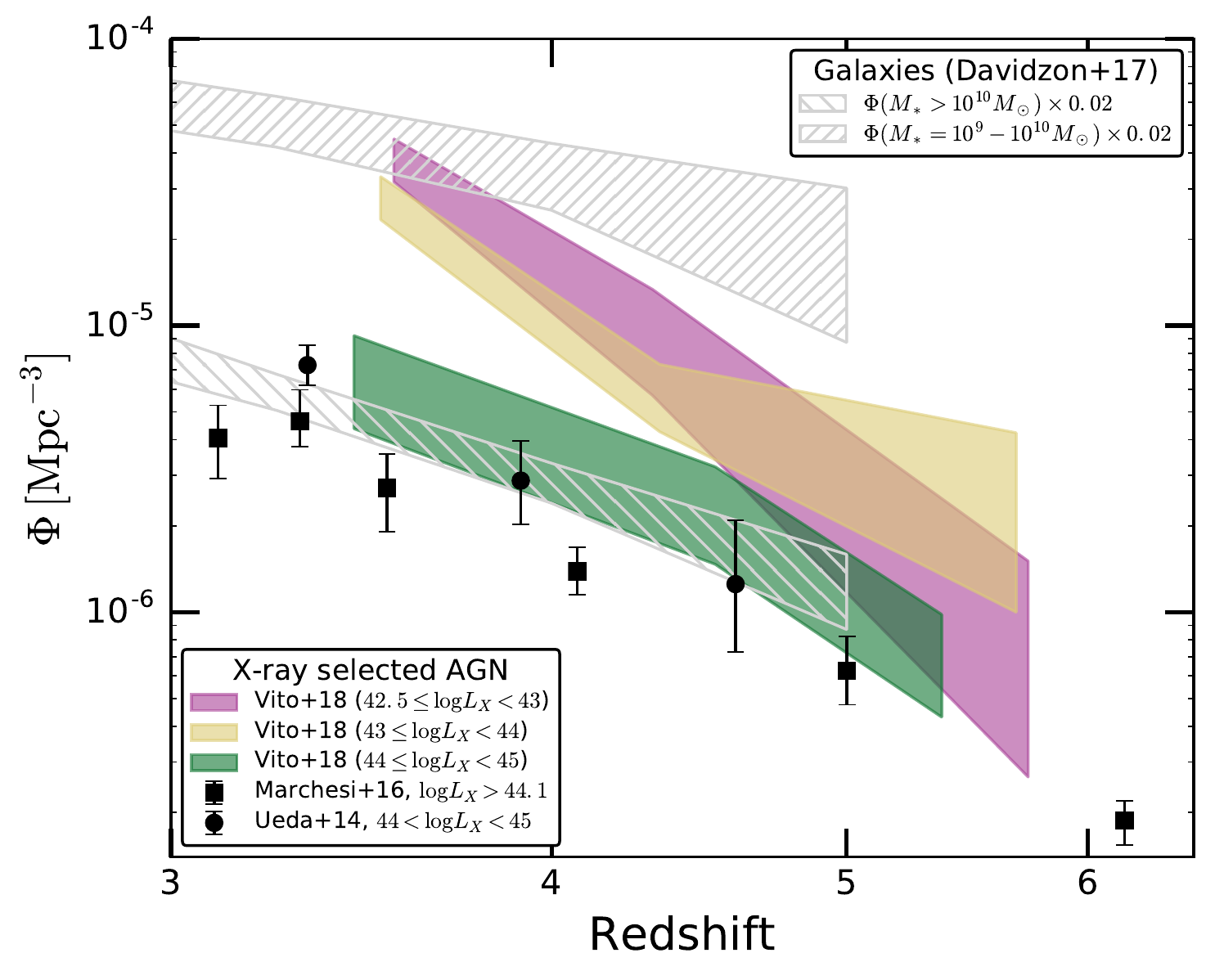}
    \caption{Evolution of the space density of X-ray selected AGN with different luminosities at $z=3-6$ (from \cite{Ueda14, Marchesi16, Vito18}), compared with the space density of galaxies (from \cite{Davidzon17}) with different stellar masses (rescaled by a factor of 0.02). }
    \label{fig_spden}
\end{figure*}

Using these limited samples, dedicated studies found a strong (i.e., a factor of $\approx10$) decline of the AGN space density from $z\approx3$ to at least $z\approx5$ (Fig.~\ref{fig_spden}; e.g., \cite{Vito18,Marchesi16} and references therein): not only X-ray selected AGN are difficult to be spectroscopically identified, but their intrinsic number is also decreasing noticeably at increasing redshift. For $L\gtrsim L_*$ AGN, the space density decline appears to be driven by the general evolution of the parent galaxy population (\cite{Vito18}; see the green and grey stripes and the black points in Fig.~\ref{fig_spden}): relatively luminous AGN are thought to be preferentially hosted by massive galaxies, which become rarer at earlier cosmic times. The density evolution of the low-luminosity population is still not well constrained, mostly due to the small sizes and low identification completeness of such faint objects. Evidence for a flattening of the faint-end of the AGN XLF at $z>3$ (i.e., a steeper decline of the space density of low-luminosity AGN compared with that of more luminous objects) has been reported by, e.g.,  \cite{Georgakakis15} and \cite{Vito18} (see the magenta and yellow stripes in Fig.~\ref{fig_spden}). However, other investigations involving X-ray and UV selected AGN samples do not find such a flattening (e.g., \cite{Giallongo19}). Constraining the number density of low-luminosity AGN up to high-redshift ($z=6$ and possibly beyond) is one of the main goals of future X-ray telescopes (e.g., \cite{Aird13}), which will provide us with deeper and wider X-ray surveys than those currently available. {\bf Recently, \cite{Wolf21} and \cite{Barlow-Hall22} presented the first estimates of the bright-end of the AGN XLF at $z\approx6$, via X-ray detection of one high-redshift AGN in wide-field {\it eROSITA} and {\it SWIFT} surveys, respectively.}

The cosmic evolution of the obscured AGN population has also been determined with deep X-ray surveys up to $z\approx5$. The fraction of obscured systems has been found to increase steadily from the local Universe ($\approx10-30\%$) up to $z\approx3-5$ ($\approx70-80\%$), especially in the high-luminosity regime (e.g., \cite{Vito14, Buchner15, Vito18}). Such evolution can be possibly caused by an increasing contribution  of the interstellar medium (ISM) in the host galaxies to the nuclear obscuration at high redshift, where galaxies are typically smaller and more gas-rich than at later cosmic times (e.g., \cite{DAmato20}).

\begin{figure*}
    \centering
    \includegraphics[height=\textwidth]{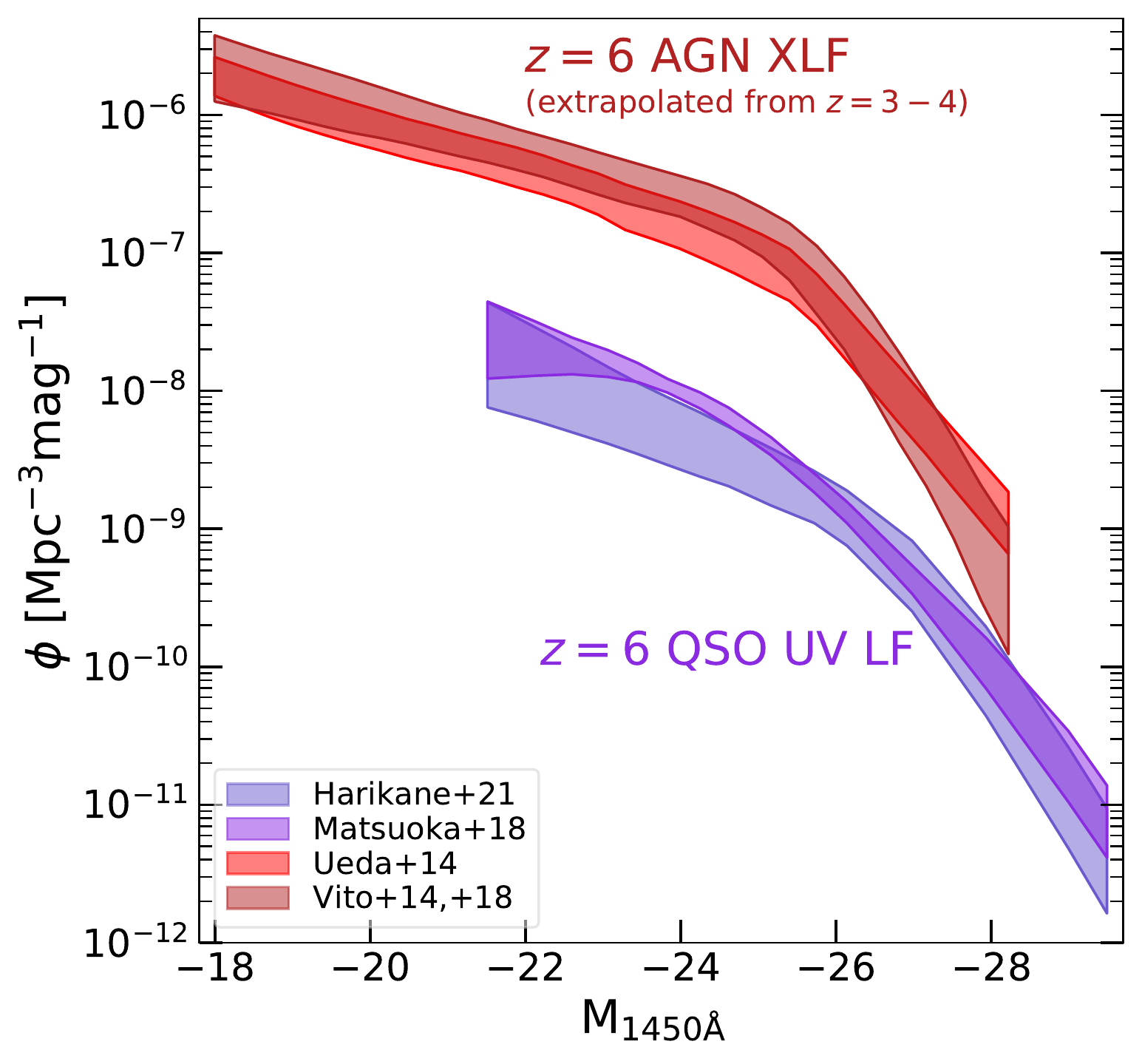}
    \caption{Comparison between the UVLF (from \cite{Matsuoka18, Harikane21}) and XLF (from \cite{Ueda14, Vito14}, estrapolated and transformed into UVLF as described in the text) of $z\approx6$ QSOs. The different normalizations suggest the existence of $\approx90\%$ QSOs missed by UV selection, most probably due to obscuration.}
    \label{fig_XLF_UVLF}
\end{figure*}

It is worth noting that X-ray observation of obscured AGN at $z\gtrsim3$ 
sample rest-frame energies corresponding to the peak of the X-ray emission. Therefore, deep X-ray surveys are thought to be highly complete for what the obscured AGN population at high redshift is concerned, with the only possible exception of the most heavily obscured, Compton-thick systems (i.e., those with column densities $>10^{24}~\mathrm{cm^{-2}}$). Therefore, an estimate of the fraction of missed $z>6$ QSOs due to obscuration can be attempted by comparing the $z\approx6$ QSO UV luminosity function (UVLF) to the extrapolation of the analytical form of the AGN XLF from $z\approx4$ to $z\approx6$.
In fact, the existence of a significant number of high-redshift QSOs  missed by currently used selection techniques based on the rest-frame UV colours should produce a higher normalization of the AGN XLF than that of the UVLF. 
Fig.~\ref{fig_XLF_UVLF} presents such a comparison, where the analytical forms of the AGN XLF from \cite{Ueda14} and \cite{Vito14} have been extrapolated from $z\approx4$ to $z=6$ and converted into UVLF assuming the $\aox-L_{UV}$ relation of \cite{Just07}, $\Gamma=2$ and UV emission in the form $f_{\lambda}\propto\lambda^{-0.3}$ (e.g., \cite{Banados2016}). The larger space density of $z\approx6$ QSOs suggested by the XLF points toward a fraction of $\approx90\%$ of high-redshift QSOs missed by standard UV selection, most probably due to widespread obscuration that suppresses the UV emission. Such fraction is consistent with measurements at $z\approx4$ (e.g., \cite{Vito18}).  The uncovered population of obscured QSOs at high redshift represents a wide discovery space for future multi-wavelength facilities, and next-generation X-ray telescopes likely will play a leading role in the identification of such objects.

\section{Future prospects}\label{sec:futureProspects}
%Overcoming current limitations - the landscape of upcoming, studieds and proposed X-ray missions/surveys (Athena, Lynx, Axis) to understand BH formation (at high-z) and to constrain high-z QSOs/galaxy formation models
%\begin{itemize}
%    \item The detectability of X-rays signatures of high-z (massive) BH seeds (Athena, Lynx, Axis...) and synergy with the next generation of GW observatories (e.g. LISA, ET) 
%    \item Populating the low-mass/faint end of the AGN mass/luminosity function with Athena (and Lynx and AXIS) 
%\end{itemize}
%%% RV
As mentioned in Section \ref{sec:unexploredBHs}, the existence and the nature of the earliest (seed) BHs is still observationally unconstrained.
The advent of the next-generation EM facilities and GW interferometers will provide unique insights on the yet unexplored BH Universe, especially at high redshift. The first will see little patches of the deep Universe, unveiling the dawn of galaxies and accreting BHs, while GW detectors will witness the dawn of black hole binaries. 

\begin{figure*}[!]
    \includegraphics[width=\columnwidth]{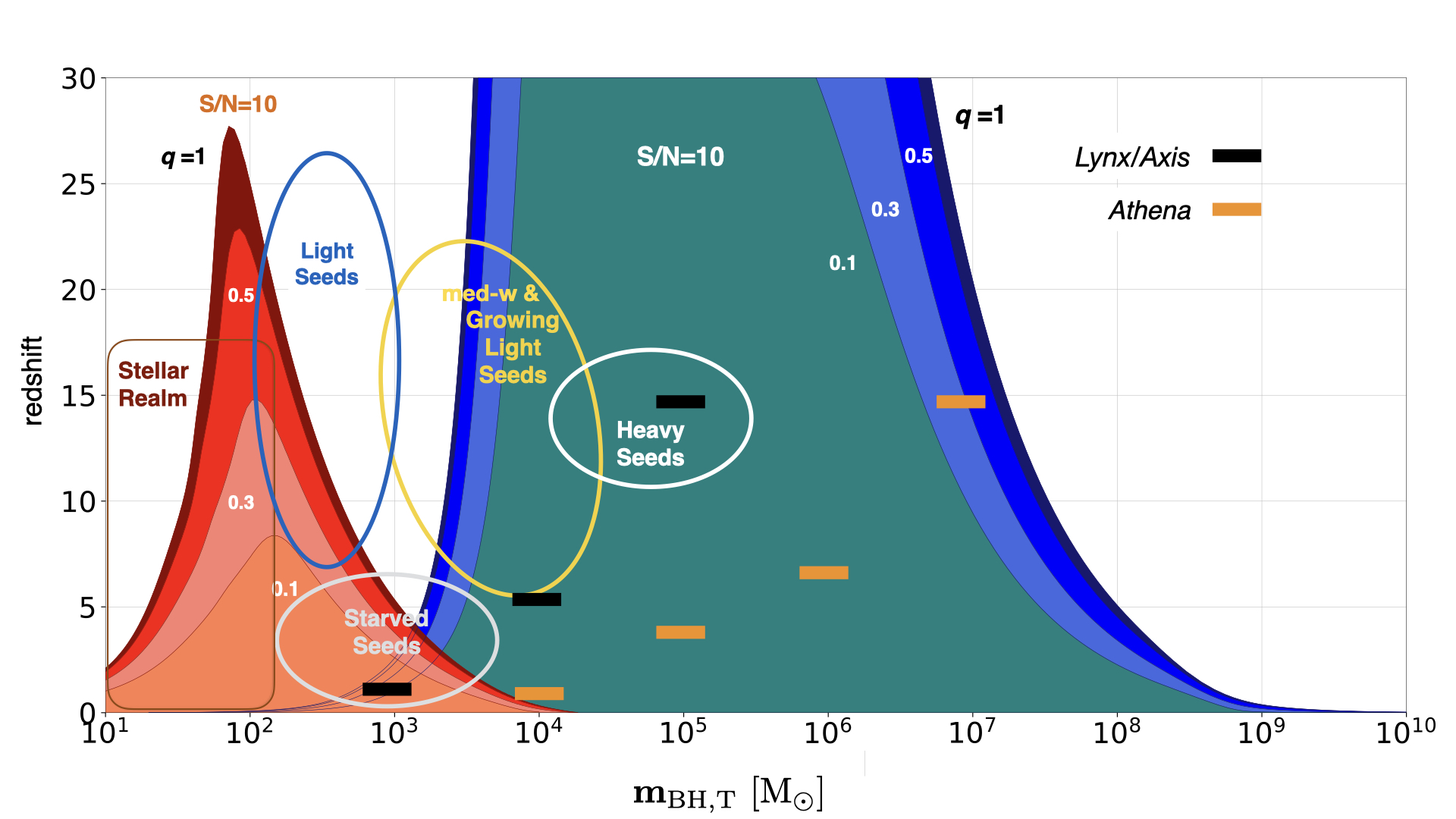}
    \caption{Summary of the GW and EM landscape. 
    The coloured curves are the sensitivity limits (also known as waterfall plots) showing the average GW horizon computed for $\rm S/N=10$ and mass ratios $q=1,0.5,0.3,0.1$ both in the ET (red) and LISA (blue) bandwidth 
    %\cite{Hild2011, Santamaria2010, Robson2019}. 
    \cite{Hild2011, Robson2019}.
    Thick horizontal bars indicate the sensitivity of the deepest pointing, in the $[0.5-2]$ keV observed band, of the missions {\it Athena} (orange) and of the NASA concepts {\it Lynx/AXIS} (black). %given the limiting fluxes of $2.4\times 10^{-17}$ \citep[][]{2013arXiv1306.2325A} and $10^{-19}$ erg s$^{-1}$ cm$^{-2}$ \cite[see https://www.lynxobservatory.com/ and][]{2019BAAS...51g.107M}, respectively. %The upper limits are inferred assuming that BHs are emitting at the Eddington limit and adopting a bolometric correction ($L_{\rm X}/L_{\rm bol})$ of $10\%$. 
    Ellipses show where stellar-mass BHs (red), light (blue) medium-weight (yellow) and heavy (white) seeds are expected to form. In the yellow region newly formed medium-weight seeds and light seeds that grow in time via accretion and mergers are expected to co-exist. The transit to the SMBH domain covers the entire LISA area and EM observations are key to discover the high-mass tail of the SMBH distribution. The light-grey ellipse at $z\leq 5$ marks the population of long-living ``starved" merging seeds (i.e. seeds whose growth is stunted by several processes) that could be
    detectable at late times when merging with other SMBHs at $z\leq 5$ (\cite{Valiante2021}). The gap in the deci-Hz window, between the ET and LISA domain, is the territory of decihertz GW observatories/programs like DECIGO and AION.
    Figure adapted from \cite{Valiante2021}.}.
    \label{fig:GW-EM}
\end{figure*}

Observatories like the Square Kilometer Array (SKA) in the radio band, the James Webb Space Telescope ({\it JWST}) space mission and the ground-based Extremely Large Telescope in the optical and near-infrared, the Advanced Telescopes for High Energy Astrophysics {\it Athena} and the mission-concepts {\it Lynx} and {\it Axis} in the X-rays, will enable deep EM observations of galaxies and active BHs up to $z\sim 8-10$ (see Figure~\ref{fig:GW-EM}), providing new information on the faintest AGN, the earliest accreting BHs and extending the search for binary/multiple AGN in interacting systems. Nonetheless, it will be possible to identify the electromagnetic counterparts of gravitational wave sources (merging massive BHs, that are among the loudest sources of GWs). %\cite{Dalcanton2019,SesanaNature2020}. 
%
%Observational signatures of seed BHs must be searched close to their formation epoch, i.e. at $z>10$ when the seed and its host galaxies still evolve almost undisturbed (the system do not experience significant mergers) and the characteristic features/properties related to a specific seeding scenario (BH accretion rate, star formation rate, ISM metallicity, total stellar mass etc.) are still distinguishable (e.g. \cite{Valiante2018a}).
%%%%At lower redshift, these differences are progressively erased, as the systems experience minor or major mergers and every trace of the BH origin gets lost.

Theoretical studies suggest that deep X-ray observations may be used to discriminate among different BH seeding and growth mechanisms as these processes should leave distinct imprints in the high-redshift BH occupation fraction and luminosity functions, although uniquely disentangling the electromagnetic signatures of seeds of different origin will be challenging 
%\cite[e.g.][]{Pacucci2015, Natarajan2017, Volonteri2017, Valiante2018b, Ricarte2018}, Trinca et al. in prep.
(e.g., \cite{Natarajan2017, Valiante2018b, Ricarte2018}).
%%%As an example, Ricarte & Natarajan (2018) showed that light seeds produce a significantly higher number of low luminosity (.1042 ergs/s) AGN and low mass (.104 M ) BH mergers compared to heavy seeds at z &6. 
%% PARAGRAFO ELIMINATO DA RV IL 15/11/2021
%Observational strategies for discriminating BH seeding/growth models are widely discussed in \cite{Pezzulli2017a, Valiante2018a, Natarajan2017, Valiante2018b, Ricarte2018}.

The imprints of different seeding mechanisms may be even more pronounced in the observed BH-BH merger rates, potentially detectable through GWs. If light seeds were the dominant seed formation channel at high redshift, a higher merger rate (as a consequence of the higher predicted occupation fraction), with respect to that expected from heavy seeds, should be observed (e.g., \cite{Huang2020, Sesana2007, Bonetti2019, Ricarte2018}).%, Bhowmick2021}.

Ground-based gravitational wave observatories with frequency sensitivities down to a few Hz, like the Einstein Telescope (ET; \cite[][]{ET2012}) and Cosmic Explorer (CE; \citep[][]{CE2017} will discover millions of coalescing stellar binary BHs ($\sim 10^2 \msun$) out to $z\sim 10-15$ (see Figure~\ref{fig:GW-EM}), thus probing the existence of light seeds that are expected to be to faint for EM facilities (e.g., \cite{Valiante2021} and references therein). %Nonetheless, ET will be the only instrument that will let us discover light BH seeds forming at cosmic dawn, as accreting BHs of similar mass are expected to be too faint to be detected at $z>5$ even with the most sensitive EM facilities like Lynx and Axis (e.g. \cite{Valiante2021} and references therein). 
Complementary, Decihertz Observatories, such as the proposed Deci-hertz Interferometer Gravitational wave Observatory (DECIGO) and the Atom interferometer Observatory and Network (AION), will be sensitive to $\sim 10-10^4 \, \rm M_\odot$ binary BHs at $z>12$ %\cite[e.g.][]{Sato2017, Voyage2019, Badurina2020}.
(e.g., \cite{Badurina2020}, and references therein).

In the low-frequency domain ($100\mu$Hz and 100 mHz), space-based interferometers such as the Laser Interferometer Space Antenna (LISA; \cite{LISA2017}), the interferometer TianQin (under design \cite{Luo2016}) and the proposed Taiji program (\cite{Ruan2018}) will instead detect the GW signals from massive binary BH coalescences (in the range $\sim 10^4-10^7 \, \rm M_\odot$), across cosmic ages (e.g., \cite{Dayal2019, Valiante2021} and references therein). 
%Together, LISA and ET, will provide the first ever census of the population of BHs that are expected to form in the aftermath of galaxy collisions, detecting signals of BHs in the yet unexplored mass range ($10^2-10^6 \, \rm M_\odot$), thus filling the gap between the stellar-mass ($\lesssim 10^2 \, \rm M_\odot$) and supermassive BHs ($\gtrsim 10^6 \, \rm M_\odot$) and shedding light into the seed BH formation process (\cite{Colpi2019, Valiante2021}).
%%%%%As light/medium-weight seeds evolve via accretion and mergers, they will transit across the LISA bandwidth and the match between ET and LISA events will statistically shed light into the seeding mechanism. LISA has also the potential to detect the rare heavy seeds in their transit to become supermassive. The lack of events on the right side of ET waterfall plot could be an indication that only heavy seeds are the progenitor of the SMBHs or that light seeds grow at a very fast (super-Eddington) rate, following their formation without experiencing cosmologically driven mergers.
%
%The highest redshift light and medium-weiht seeds (as well as the growing light seeds) mergers would be captured in the deci-Hz frequency domain, thus detectable with a Decihertz Observatory such as the proposed Deci-hertz Interferometer Gravitational wave Observatory (DECIGO) and the Atom interferometer Observatory and Network (AION) that will designed to be sensitive to $\sim 10-10^4 \, \rm M_\odot$ binary BHs at $z>12$ (see e.g. \citep{Sato2017, Voyage2019, Badurina2020}).\\

Finally, facilities and experiments like SKA, the Pulsar Timing Array (PTA) and the concept PIXIE (Primordial Inflation Explorer) will enable to probe the existence of primordial BHs.
%(Byrnes et al., 2019; Inomata & Nakama, 2019; Kalaja et al., 2019; Gow et al., 2020).
\newline
\newline

With currently operating facilities, dedicated observations of well-selected high redshift QSOs will greatly improve the test of the cosmological model and the study of the dispersion of the $\Lx-\Lo$ relation. 
{\it eROSITA}, flagship instrument of the ongoing Russian {\it Spektrum-Roentgen-Gamma} (SRG) mission, will represent an extremely powerful and versatile X-ray observatory in the next decade. The sky of eROSITA will be dominated by the AGN population, with $\sim$3 million AGN at a median redshift of $z\sim1$ expected by the end of the nominal 4-year all-sky survey at the sensitivity of $F_{0.5-2\,\rm keV} \simeq 10^{-14}$ erg s$^{-1}$ cm$^{-2}$, for which extensive multiwavelength follow-up is already planned. 
Regarding the constraints on the cosmological parameters (such as $\om$, $\ol$, and $w$) through the Hubble diagram of QSOs, we predict that the 4-year eROSITA all-sky survey alone, complemented by redshift and broadband photometric information, will supply the largest QSO sample at $z<2$ (average redshift $z\simeq1$, see \citep{liu2021}), but a relatively small population should survive the Eddington bias cut at higher redshifts (see, e.g., \citep{medvedev2020}), thus being available for cosmology \citep[][]{lusso2020} as eROSITA will sample the brighter end of the X-ray luminosity function \citep[][]{comparat2020}.
None the less, the large number of eROSITA QSOs at $z\simeq1$ will be pivotal for both a better cross-calibration of the QSO Hubble diagram with SNe and a more robust determination of $\ol$, which is  sensitive to the shape of the low redshift part of the distance modulus--redshift relation.

In the mid/long term, surveys from {\it Euclid} and {\it Vera C. Rubin Observatory} (LSST) in the optical/UV, and {\it Athena} in the X-rays, will also provide samples of millions of QSOs. 
With these samples it will be possible to obtain constraints on the observed deviations from the standard cosmological model, which will rival and complement those available from the other cosmological probes.
\newline
\newline

\noindent
From the theoretical perspective, further improvements in both SAMs and simulations are required to fully understand the physical mechanisms regulating the formation of high-redshift QSOs.
Self-consistent models for seed BH formation and growth need to be implemented in future simulations, possibly connecting large-scale cosmological studies with higher-resolutions, (non-cosmological) simulations (those devoted to study single halos and/or suited to resolve the typical physical scales, close or below the Bondi radius, of BH accretion and dynamics).
More refined prescriptions for BH dynamics (down to the GW driven domain) are required, in both SAMs and simulations, when predicting the statistical properties of BH populations (e.g., their merger rate) in a cosmological context, %These should be included in SAMs and simulations, %adopting prescriptions motivated by simulations,
alongside the hierarchical assembly of galaxies. %\cite[e.g.][]{SesanaGair2011, Klein2016, Ricarte2018, Bonetti2019, Dayal2019, Katz2020, Volonteri2020}. 
In addition, efforts should also be devoted to consistently link the assembly history of SMBHs (BH dynamics, seed formation and growth mechanisms) to the chemical evolution history of their host galaxies. %\cite[e.g.][]{Valiante2018a, Valiante2021}.
%%% RV - END

%[NOTES]: Mention possible future developments on the topic related to X-ray instruments such as Athena...

% \bibliographystyle{spbasic}
\bibliography{sectionXII_chapter4biblio}

\end{document}